\documentclass[neutralpreprint]{imsart}
\RequirePackage{amsthm,amsmath,amsfonts,amssymb,array,placeins}
\RequirePackage{algorithm}
\RequirePackage{xcolor}
\RequirePackage[authoryear,round,sort]{natbib}
\RequirePackage[colorlinks,citecolor=blue,urlcolor=blue,linkcolor=blue,bookmarksopen=true,bookmarksnumbered=true]{hyperref}
\makeatletter
\def\common@pub@types@hook{\def\betal##1{et~al.\space}\let\bseriesno\bvolume}
\def\toc@size{\small}
\makeatother
\renewcommand{\contentsname}{Table of contents}
\startlocaldefs
\numberwithin{equation}{section}
\newtheorem{theorem}{Theorem}[section]
\newtheorem{proposition}[theorem]{Proposition}
\newtheorem{lemma}[theorem]{Lemma}
\newtheorem{corollary}[theorem]{Corollary}
\theoremstyle{definition}
\newtheorem{assumption}[theorem]{Assumption}

\DeclareMathOperator{\Tr}{Tr}
\DeclareMathOperator{\Rea}{Re}
\DeclareMathOperator{\Ima}{Im}
\DeclareMathOperator{\TV}{TV}
\DeclareMathOperator{\diag}{diag}
\newcommand{\R}{\mathbb R}
\newcommand{\E}{\mathbb E}
\newcommand{\Prb}{\mathbb P}
\newcommand{\Id}{\mathbf 1}
\newcommand{\norm}[1]{\left\lVert#1\right\rVert}
\newcommand{\abs}[1]{\left\lvert#1\right\rvert}
\newcommand{\ip}[2]{\langle #1,#2\rangle}
\newcommand{\op}{\mathrm{op}}
\newcommand{\HS}{\mathrm{HS}}
\newcommand{\cQ}{\mathcal Q}
\newcommand{\cG}{\mathcal G}
\DeclareRobustCommand{\limitationtext}[1]{#1}

\endlocaldefs
\begin{document}
\begin{frontmatter}
\title{Constructive finite-sample collective quantum inference}
\runtitle{Finite-sample collective quantum inference}
\begin{aug}
\author[A]{\fnms{Ning}~\snm{Ning}\thanksref{cor}\ead[label=e1]{patning@tamu.edu}}
\thankstext{cor}{Corresponding author.}
\address[A]{Department of Statistics, Texas A\&M University, College Station, Texas, USA\printead[presep={,\ }]{e1}}
\end{aug}
\begin{abstract}
We develop a constructive framework for finite-sample quantum inference that connects an explicit collective measurement with confidence coverage, testing power, and estimation risk. For faithful finite-dimensional quantum models with nonsingular symmetric logarithmic derivative information, calibrated observables and a finite ancillary system define an exactly normalized measurement near a known nominal parameter. Assuming two continuous derivatives of the density map, we establish explicit density and total variation bounds for the measurement output relative to a smoothed Gaussian shift, uniformly over a local parameter ball. A feasible point of the Holevo optimization determines the comparison covariance, and its optimization gap remains visible in the finite-sample risk certificate. The distributional comparison yields confidence regions, local tests, projected quadratic-risk bounds, radius-dependent shrinkage, and a finite-readout version of the procedure. A transverse qubit example makes the ancillary construction and measurement disturbance explicit and illustrates the actual output law numerically. Detailed supporting results cover inference after specified measurements, pilot adaptation, finite-radius channel comparison, and an optional quadratic likelihood surrogate. Together, the construction and its error analysis give a self-contained route from quantum observables to statistical guarantees for a specified finite number of systems.
\end{abstract}
\end{frontmatter}

\tableofcontents
\clearpage

\section{Introduction}\label{sec:intro}
In quantum statistics, the data available for inference depend on the
measurement chosen. A measurement converts information in a quantum state
into classical outcomes, from which one constructs estimates, confidence
regions and tests. When observables for different parameters do not commute,
they generally cannot all be measured jointly without changing their outcome
laws. Measurement choice is therefore part of statistical design, and
achievable precision depends on both the measurement and the inference based
on its outcomes~\citep{GillGuta2013,Holevo2011}.

Local Gaussian approximation provides a powerful framework for understanding
this interaction. Quantum local asymptotic normality (qLAN) describes how
qubit and finite-dimensional experiments approach Gaussian models under
suitable regularity and spectral assumptions
\citep{GJK2008,GutaJencova2007,GK2006,KG2009}. The likelihood formulation of
\citet{YFG2013} establishes asymptotic attainability
of the Holevo bound, while subsequent work develops quantum contiguity and
representation of limiting measurement distributions~\citep{FY2020,FY2023}.
Related Gaussian comparison methods also apply beyond finite-dimensional
models. \citet{ButuceaGutaNussbaum2018}
establish local asymptotic equivalence to quantum Gaussian white noise for
nonparametric pure-state models, and \citet{LahiryNussbaum2022} obtain sharp asymptotic minimax and adaptive
estimation results in that setting. These theories explain how local model
geometry and measurement constraints determine asymptotic precision.

Finite-sample theory brings this perspective closer to inference with a
specified number of systems. Existing work provides reliable tomography and
confidence regions~\citep{ChristandlRenner2012,Sugiyama2014,WangScholzRenner2019}
and metrological guarantees for prescribed accuracy~\citep{Meyer2025}.
\citet{Yang2019} obtain polynomial-rate local
output approximation and a global estimation procedure under their
hypotheses. Recent work by \citet{KLY26} develops
asymptotic testing efficiency, including local power bounds and adaptive
scalar procedures. These developments connect measurement design with an
increasing range of inferential objectives.

For a specified measurement, carrying this connection through at a fixed
sample size requires quantitative control of its actual output distribution.
A limiting covariance alone does not determine the coverage of a confidence
region or the power of a test at that sample size. Quadratic-risk analysis
also has to account for the size of the parameter neighborhood, the use of
projection, and the possibility of shrinkage. Our question is therefore:
\emph{how can an explicitly specified collective measurement provide
simultaneous finite-sample guarantees for estimation, confidence regions,
and testing over a common local parameter neighborhood?}

We formulate this problem for a family of density matrices
$\rho_\theta$, indexed by a parameter $\theta$ in an open subset of $\R^p$.
Assume a faithful finite-dimensional model whose density map is twice
continuously differentiable on a neighborhood of a known nominal center
$\theta_0$, with nonsingular symmetric logarithmic derivative (SLD) information
there. Faithfulness means positive definiteness and permits repeated
positive eigenvalues. For $n$ independently prepared copies, write
$h=\sqrt n(\theta-\theta_0)$ and
$K_r=\{h\in\R^p:\norm h\le r\}$, where $\norm{\cdot}$ is the Euclidean norm
and $r\ge0$ is chosen so that the corresponding parameters remain in that
neighborhood. The state is
$\rho_{n,h}=\rho_{\theta_0+h/\sqrt n}^{\otimes n}$.
The statistical problem is to construct a measurement and reporting rule,
independent of the unknown $h$, with explicit guarantees uniformly over
$K_r$.

The local covariance target is specified by the Holevo optimization.
Put $\rho=\rho_{\theta_0}$, fix a real symmetric $p\times p$ loss weight $W\succ0$, and
let $\mathcal X_{\theta_0}$ be the set of lists $X=(X_1,\ldots,X_p)$ of
Hermitian observables satisfying
\begin{equation}\label{eq:intro-calibration}
 \Tr\rho X_i=0
 \quad\text{and}\quad \Tr(\partial_j\rho_{\theta_0})X_i=\delta_{ij},\quad 1\le i,j\le p.
\end{equation}
Here $\partial_j$ denotes the derivative in parameter coordinate $j$, and
$\delta_{ij}$ is the Kronecker delta. For each list $X$, write $Z=Z(X)$
for the ordered moment matrix with entries $Z_{ij}=\Tr\rho X_iX_j$.
The optimization problem is
\begin{equation}\label{eq:intro-holevo}
 C_W=\min_{X\in\mathcal X_{\theta_0}}\Phi_W(Z(X))
 \quad\text{and}\quad \Phi_W(Z)=\Tr(W\Rea Z)+\norm{W^{1/2}(\Ima Z)W^{1/2}}_1.
\end{equation}
Real and imaginary parts are taken entrywise, $\Tr$ denotes the trace, and
$\norm{\cdot}_1$ is the trace norm which is the sum of singular values. This
established optimization supplies the target for our construction. A feasible
list $X$ has gap $\epsilon_{\mathrm{opt}}=\Phi_W(Z)-C_W\ge0$ and determines the
real covariance $V_X$ in~\eqref{eq:Vstar}, with
$\Tr(WV_X)=\Phi_W(Z)$.

Building on the ordered Gaussian-root kernels in \citet[Appendix A.5, Lemma A.1]{YFGSupp}, we combine these observables and a finite ancillary system
to construct an exactly normalized continuous collective measurement.
Let $P_{n,h}^{\eta}$ be its outcome law, where $\eta>0$ is the smoothing
parameter, and let $I_p$ be the $p\times p$ identity matrix.
Theorem~\ref{thm:collective} gives the uniform comparison
\begin{equation}\label{eq:intro-output}
 \sup_{h\in K_r}\TV\!\left(P_{n,h}^{\eta},N(h,V_X+\eta^2I_p)\right)
 \le\varepsilon_n(r,\eta,R_{\rm out}),
\end{equation}
where $N(h,V)$ denotes a Gaussian law with mean $h$ and covariance $V$,
$\TV(P,Q)=\sup_B|P(B)-Q(B)|$ is total variation distance over measurable
events $B$, and $R_{\rm out}>r$ is an auxiliary output cutoff used to bound TV.
The error bound $\varepsilon_n$ is explicit in
\eqref{eq:collective-TV-budget}. It retains the dependence on density
derivatives, observable norms, radius and smoothing. The fluctuation
argument in Theorem~\ref{thm:local-char} uses the trace-zero density remainder
to gain an additional factor $n^{-1/2}$, allowing two density derivatives
to suffice.

The comparison~\eqref{eq:intro-output} controls confidence coverage and
testing power through the same error bound. It also gives a direct link
between the optimization problem and finite-sample risk. If $Y$ is the
measurement outcome and $\widehat h=\Pi_r^W(Y)$ is its projection onto $K_r$
in the norm $\norm v_W^2=v^TWv$, then
\begin{equation}\label{eq:intro-risk}
 \sup_{h\in K_r}\E_h\norm{\widehat h-h}_W^2
 \le C_W+\epsilon_{\mathrm{opt}}+\eta^2\Tr W
       +4\norm W_\op r^2\varepsilon_n(r,\eta,R_{\rm out}).
\end{equation}
Here $\E_h$ is expectation under the constructed outcome law and
$\norm W_\op$ is the operator norm. The four terms distinguish the Holevo
covariance target, the optimization gap, the smoothing cost and the
finite-sample approximation cost. Radius-dependent shrinkage further
improves the available risk certificate under the conditions of
Corollary~\ref{cor:collective-shrinkage}. These are guarantees for a specified
procedure near its known nominal center.
Algorithm~\ref{alg:collective-inference} summarizes the construction and
reporting rule; Theorem~\ref{thm:collective} and
Corollary~\ref{cor:collective-shrinkage} give its assumptions and guarantees.

\begin{algorithm}[!t]
\caption{Constructive collective measurement and local inference}\label{alg:collective-inference}
\small
{\raggedright\noindent
\textbf{Inputs.} The known density model $\rho_\theta$ with $p$ real parameters
and nominal center $\theta_0$; $n$ signal copies; local ball
$K_r=\{h\in\R^p:\norm h\le r\}$ with $r\ge0$ such that
$\theta_0+K_r/\sqrt n$ lies in the neighborhood of the density derivative
bounds $D_1,D_2$ in~\eqref{eq:trace-regularity}; real symmetric positive-definite
loss weight $W$; smoothing parameter $\eta>0$; and error-bound cutoff
$R_{\rm out}>r$. The unknown signal parameter is
$\theta=\theta_0+h/\sqrt n$ with $h\in K_r$.
\par}
\begin{enumerate}
\item \textbf{Choose calibrated observables.} Choose centered Hermitian
$X_1,\ldots,X_p$ satisfying~\eqref{eq:intro-calibration}.
Optionally minimize the Holevo objective defined in~\eqref{eq:intro-holevo}.
\item \textbf{Construct the ancilla.} Compute the moment matrix
$Z_{ij}=\Tr\rho_{\theta_0}X_iX_j$, the target covariance $V_X$ in~\eqref{eq:Vstar},
and $C=V_X-Z$. Factor $C=B^*B$ and let $l_i$ be the columns of $B$.
Use Section~\ref{sec:holevo-ancilla} to form the ancillary Hilbert space
$\mathcal H_a$, and~\eqref{eq:ancilla} to define its fixed state $\omega$,
ancillary observables $A_i$, and signal--ancilla observables $U_i$.
\item \textbf{Form the collective observables.} Attach one ancilla in
state $\omega$ to each signal copy and set
$B_{n,i}=n^{-1/2}\sum_{k=1}^nU_i^{(k)}$, where $U_i^{(k)}$ acts as $U_i$
on the $k$th signal--ancilla pair and as the identity on every other pair.
\item \textbf{Calculate the inference constants.} Evaluate the uniform
total variation error bound $\varepsilon=\varepsilon_n(r,\eta,R_{\rm out})$
using~\eqref{eq:collective-TV-budget} and set
\[
 A_\eta=\Tr W(V_X+\eta^2I_p),\quad
 B_r=\norm W_\op r^2
 \quad\text{and}\quad c_*=\frac{B_r}{A_\eta+B_r}.
\]
Here $I_p$ is the $p\times p$ identity and $\norm W_\op$ is the largest
eigenvalue of $W$; $c_*$ is the shrinkage coefficient.
\item \textbf{Measure and record.} For $i=1,\ldots,p$, apply the
Gaussian instrument with operator $g_{\eta,y_i}(B_{n,i})$, obtained by
spectral calculus from the kernel~\eqref{eq:gaussian-kernel}, and record
$Y_i=y_i$. These stages act in this order on the same collective
signal--ancilla system, with joint operator $K_y$ in~\eqref{eq:smooth-povm}.
Retain $Y=(Y_1,\ldots,Y_p)$.
\item \textbf{Select the estimator.} Set
\[
 \widehat h=
 \begin{cases}
  \Pi_r^W(c_*Y),
  &\displaystyle \frac{A_\eta B_r}{A_\eta+B_r}+4B_r\varepsilon<B_r,\\[4pt]
  0,&\text{otherwise}.
 \end{cases}
\]
Here $\Pi_r^W(y)=\operatorname*{arg\,min}_{z\in K_r}(z-y)^TW(z-y)$
is projection onto $K_r$ in the $W$ metric.
\end{enumerate}
{\raggedright\noindent\textbf{Outputs.} Return
$\widehat\theta=\theta_0+\widehat h/\sqrt n$ and the local $W$-weighted
quadratic-risk bound~\eqref{eq:collective-selected-risk}; divide this bound
by $n$ for the quadratic risk of $\widehat\theta$. Retain the unprocessed $Y$
for confidence regions and tests in Corollary~\ref{cor:collective-inference}.
\par}
\end{algorithm}

The guarantees distinguish the statistical task from the covariance target.
The Holevo objective specifies the covariance cost of a calibrated observable
list, whereas the actual finite-sample procedure also incurs smoothing and
output-approximation costs. The local radius enters the bounded-loss transfer
and the shrinkage coefficient explicitly. Thus the same output comparison
supports several reporting rules, with a separate guarantee for each. The
finite-readout result additionally converts the continuous construction into
finitely many recorded outcomes while tracking the discretization cost.

The qubit illustration gives a concrete view of this mechanism. It includes
faithful states with repeated eigenvalues, observables that remain
noncommuting after ancillary cancellation, and exact first moments showing
how the measurement order changes the finite-sample output. The numerical
calculations examine the actual ordered output density as well as the
analytic error certificate. This connects the operator construction to the
coverage, approximation, and loss criteria used to assess a statistical
procedure.

These results provide a constructive framework that treats measurement
design and statistical inference as parts of the same finite-sample problem.
Explicit control of the constructed measurement's output law gives a common
basis for estimation, confidence regions and hypothesis tests. Under the
stated regularity and information assumptions, the construction applies to
smooth, faithful finite-dimensional models near a known nominal parameter
and extends beyond the qubit illustration. Its significance is the explicit
connection between the selected quantum observables, the classical readout
law, and the inferential guarantees available from finitely many systems.

The appendices develop complementary inference tools in detail.
Inference after a specified POVM is developed
through score concentration, localization, and Fisher and Wilks expansions.
Pilot-based procedures are analyzed conditionally, including an affine
scalar model with an exact risk identity. Separate channel assumptions give
experiment-level risk comparisons, and a quadratic likelihood surrogate
provides an alternative route to the same collective-output guarantee.
These supporting results state their own hypotheses and show precisely
which additional controls are needed for their respective tasks.

The rest of the paper proceeds as follows.
Section~\ref{sec:setup} introduces states, measurements and calibration, explains
the local parameter scale, and connects classical scores to the SLD.
Section~\ref{sec:collective} develops the covariance and measurement
construction, establishes the output bound, and derives its statistical
consequences. Section~\ref{sec:examples} works through these calculations
for the transverse qubit model. Section~\ref{sec:conclusion} summarizes the findings.
Appendix~\ref{app:background} gives detailed background;
Appendices~\ref{sec:measured}--\ref{app:polynomial-support} develop the
supporting inference and comparison results. The remaining appendices
supply concentration arguments, instrument moments, collective corollary
proofs, and Pauli coverage calculations and numerical illustrations.

\FloatBarrier
\section{Model, measurements and information}\label{sec:setup}
The density operators $\rho_\theta$ introduced in Section~\ref{sec:intro}
act on $\mathcal H=\mathbb C^d$, equipped with the Hermitian inner product
$\langle u,v\rangle=u^*v$, where $*$ denotes conjugate transpose.
For real or complex vectors, $\norm v=(v^*v)^{1/2}$ is the Euclidean norm.
Let $\Theta\subset\R^p$ denote their open
parameter domain, where $p$ is the parameter dimension and $d$ is the
system dimension. Fixing an
orthonormal basis identifies each operator with its matrix
in $\mathbb C^{d\times d}$.
We retain the nominal center $\theta_0$ and reference state $\rho=\rho_{\theta_0}$.
For an operator $A$ on $\mathcal H$ and $1\le k\le n$, let $A^{(k)}$
act as $A$ on the $k$th factor of $\mathcal H^{\otimes n}$ and as
the identity on every other factor. Derivatives of matrix-valued maps
are taken entrywise; in particular,
$\partial_j\rho_{\theta_0}=\left.\partial\rho_\theta/\partial\theta_j\right|_{\theta=\theta_0}$.
For a smooth matrix-valued map $f$, $D^jf(u)[v_1,\ldots,v_j]$ denotes its
$j$th derivative evaluated on the listed directions. A Hermitian matrix
satisfies $A=A^*$. For Hermitian matrices of the same size, $A\succeq B$
means $v^*(A-B)v\ge0$ for every $v$, and $A\succ0$ means $v^*Av>0$
for every nonzero $v$. We write $[A,B]=AB-BA$ and $\{A,B\}=AB+BA$;
in complex expressions, $i=\sqrt{-1}$. Recall the local coordinate
$h=\sqrt n(\theta-\theta_0)$ and closed parameter ball
$K_r=\{h\in\R^p:\norm h\le r\}$ of radius $r\ge0$.
The associated local experiment is
\begin{equation}\label{eq:local-model}
 \cQ_{n,r}=\left\{\rho_{\theta_0+h/\sqrt n}^{\otimes n}:\norm h\le r\right\}\quad\text{\normalfont and}\quad \rho_{n,0}=\rho^{\otimes n}.
\end{equation}
Unless stated otherwise, $n,p,d$ are positive integers and $r\ge0$. Inverse-radius statements require $r>0$. The scalar $x>0$ is a tail-probability tuning parameter; for example, the measured likelihood event in Theorem~\ref{thm:measured} has failure probability at most $e^{-x}$, so $x=\log(1/\alpha)$ gives failure tolerance $\alpha$, where $0<\alpha<1$. The reference center and one-copy constants are fixed; a triangular array requires explicit uniform control. Constants may depend on the fixed model, local neighborhood, dimension, loss weight and chosen observables unless stated otherwise. Repeated auxiliary symbols are local to their stated result and do not identify constants across theorems.

\subsection{Experiments, distances and decisions}\label{sec:distances}
A quantum state determines the probabilities of measurement outcomes.
Its density matrix satisfies
\[
 \rho=\rho^*,\quad \rho\succeq0
 \quad\text{and}\quad \Tr\rho=1.
\]
The trace is the sum of diagonal entries, and the eigenvalues of $\rho$
are nonnegative and sum to one. We write $\Id$ for the identity on the
relevant Hilbert space and $I_q$ for the $q\times q$ identity matrix,
in particular $I_p$ on the real parameter space. As in
Section~\ref{sec:intro}, the state is \emph{faithful} when every eigenvalue
is positive, equivalently when $\rho\succ0$ or its rank equals $d$.
For a qubit, a system with $d=2$, both $\diag(0.7,0.3)$ and $\Id/2$
are faithful, whereas $\diag(1,0)$ is not. Faithfulness permits repeated
positive eigenvalues: the maximally mixed state $\Id/d$ is faithful.

Matrix functions are defined by spectral calculus. If a Hermitian matrix
has decomposition $A=V\diag(a_1,\ldots,a_d)V^*$ with $V$ unitary, then
\[
 f(A)=V\diag\bigl(f(a_1),\ldots,f(a_d)\bigr)V^*
\]
whenever the scalar values exist. In particular, $A^{1/2}$ is the positive
square root when $A\succeq0$, and $\log A$ is Hermitian when $A\succ0$.
Faithfulness therefore ensures that $\rho^{-1}$ and $\log\rho$ exist as
finite matrices. The same calculus defines the Gaussian functions of
collective observables used in Section~\ref{sec:collective-output}.

A measurement effect is an operator $0\preceq E\preceq\Id$ representing
an event with probability $\Tr\rho E$. At a faithful state, every
nonzero effect has positive probability, since
\[
 \Tr\rho E\ge\lambda_{\min}(\rho)\Tr E>0,
\]
where $\lambda_{\min}(\rho)$ is the smallest eigenvalue. This observation
ensures positivity of the probabilities used to define measured scores
in Section~\ref{sec:likelihood-setup}, after zero effects are discarded.

A Hermitian observable has spectral decomposition $X=\sum_a aP_a$,
where $a$ runs over its distinct real eigenvalues and the orthogonal
spectral projections $P_a$ sum to $\Id$. The Born
rule specifies the distribution of its scalar measurement outcome $Y_X$:
\[
 \Prb_\theta\{Y_X=a\}=\Tr\rho_\theta P_a
 \quad\text{and}\quad
 \E_\theta Y_X=\sum_a a\Tr\rho_\theta P_a=\Tr\rho_\theta X.
\]
Applying the same identity to the squared outcome gives
\[
 \operatorname{Var}_\theta(Y_X)
 =\Tr\rho_\theta X^2-(\Tr\rho_\theta X)^2.
\]
Thus trace means and variances are ordinary moments of a specified
classical outcome. The symbols $\Prb_\theta$ and $\E_\theta$ refer to
the measurement performed on $\rho_\theta$, with the measurement held
fixed as $\theta$ varies.
Each Hermitian observable specifies such a scalar outcome law. Listing
several observables gives their separate spectral means; it does not yet
specify one measurement that records all of them together.

A positive operator-valued measure (POVM) $M$ on a measurable outcome space $(\mathcal Y,\mathcal A)$ assigns positive operators to sets in $\mathcal A$, is countably additive in the weak operator sense, and satisfies $M(\mathcal Y)=\Id$. Its classical outcome law is
\[
 P_\theta^M(B)=\Tr\rho_\theta M(B).
\]
For finite outcomes, $M_y\succeq0$ and $\sum_yM_y=\Id$, giving
$p_\theta(y)=\Tr\rho_\theta M_y$. A real statistic $g(Y)$ has mean
$\E_\theta g(Y)=\sum_y g(y)p_\theta(y)$; spectral measurement is the
special case $M_a=P_a$ and $g(a)=a$. When a POVM has an operator
density relative to a fixed measure $\nu$, it satisfies
\[
 M(B)=\int_Bm(y)\nu(dy),\quad m(y)\succeq0
 \quad\text{and}\quad \int_{\mathcal Y}m(y)\nu(dy)=\Id.
\]
Sums of outcome probabilities and means then become integrals. The
constructed measurement in Section~\ref{sec:collective-output} uses
Lebesgue measure, so $M(dy)=m(y)dy$. Write
$P_\sigma^M(B)=\Tr\sigma M(B)$ when a state $\sigma$, rather than a
parameter, indexes the law. These conventions are standard in quantum
statistics~\citep{Holevo2011}.

Statistical procedures apply a measurable reporting rule
$\widehat\theta(y)\in\R^p$ to the outcome. The reported estimate is also
called an \emph{action}. For a finite POVM, grouping outcomes with the
same reported action gives, for each measurable $B\subset\R^p$,
\[
 \widetilde M(B)=\sum_{y:\widehat\theta(y)\in B}M_y
 \quad\text{and}\quad
 \Prb_\theta\{\widehat\theta(Y)\in B\}=\Tr\rho_\theta\widetilde M(B).
\]
The positive effects $\widetilde M(B)$ sum to $\Id$ over any countable
measurable partition of $\R^p$. This is an \emph{action-valued POVM}:
its outcomes are reported estimates, while its effects act on the
quantum state space. It represents the measurement and reporting rule
together when formulating risk.

For a finite POVM with reported displacement
$x_i(y)=\widehat\theta_i(y)-\theta_{0,i}$, the associated first-moment
operator makes the connection between numerical estimates and quantum
observables explicit:
\[
 X_i=\sum_y x_i(y)M_y
 \quad\text{and}\quad
 \E_\theta[\widehat\theta_i(Y)-\theta_{0,i}]
 =\sum_y x_i(y)\Tr\rho_\theta M_y=\Tr\rho_\theta X_i.
\]
The real reporting values make $X_i$ Hermitian. This operator encodes the
mean of the report, although the implemented procedure measures $M$ and
then applies the reporting rule. Both choices are held fixed when the
parameter varies; different procedures can have the same first-moment
operators and different outcome distributions.

For a fixed list $X$, write
$\mu(\theta)=(\Tr\rho_\theta X_1,\ldots,\Tr\rho_\theta X_p)^T$.
Under the $C^2$ density conditions stated below, the calibration
constraints~\eqref{eq:intro-calibration} give
\[
 \mu(\theta_0)=0
 \quad\text{and}\quad D\mu(\theta_0)=I_p,
 \quad\text{so}\quad
 \mu(\theta_0+u)=u+O(\norm u^2).
\]
Centering fixes the mean at the nominal point. The identity Jacobian
requires slope one in the coordinate being estimated and slope zero in
every other coordinate, preventing a change in another parameter from
being interpreted as a first-order displacement in that coordinate.
For first-moment operators of a reported estimate, these conditions say
that its bias and the first derivatives of its bias vanish at $\theta_0$.
They impose local calibration, rather than exact unbiasedness throughout
a neighborhood or exact unbiasedness of the finite-sample collective
readout constructed later.

This calibration also explains the local scale in~\eqref{eq:local-model}.
Put $X_i^{[n]}=n^{-1/2}\sum_{k=1}^nX_i^{(k)}$ for a calibrated list.
The product state and linearity of the trace give
\[
 \Tr\rho_{n,h}X_i^{[n]}
 =\sqrt n\,\mu_i(\theta_0+h/\sqrt n)
 =h_i+O(\norm h^2/\sqrt n).
\]
The remainder is uniform on a fixed local ball when the second density
derivatives are bounded on the corresponding physical neighborhood.
At the reference state, independence and centering make the spectral
variance of $X_i^{[n]}$ equal to $\Tr\rho X_i^2$, independently of $n$.
Thus physical displacements of size $n^{-1/2}$ produce mean changes of
order one on the same scale as the fluctuations. The vector $h$ records
these changes. Constructing one classical readout with the required joint
behavior remains a separate measurement problem.

For arbitrary matrices, the trace norm introduced in Section~\ref{sec:intro}
can be expressed through the matrix absolute value. We also use the
Hilbert--Schmidt (HS) and state-weighted norms:
\[
 |A|=(A^*A)^{1/2},\quad \norm A_1=\Tr|A|,\quad
 \norm A_{\HS}^2=\Tr A^*A
 \quad\text{and}\quad \norm A_{2,\rho}=\norm{A\sqrt\rho}_{\HS}.
\]
The quantity $\norm{\cdot}_{2,\rho}$ is a norm when $\rho$ is
faithful and a seminorm when $\rho$ is singular; the latter case
may occur after adjoining a pure ancilla. The notation is retained in
both cases, and all uses of Cauchy--Schwarz remain valid.
The HS inner product is $\langle A,B\rangle_{\HS}=\Tr A^*B$. For real matrices we also write $\norm A_{\mathrm F}=(\sum_{ij}A_{ij}^2)^{1/2}$ for this Frobenius norm. The operator norm $\norm A_\op$ is the largest singular value. Matrix transpose is denoted by $A^T$, adjoint by $A^*$, and entrywise complex conjugation by $\overline A$; real and imaginary matrix parts are taken entrywise.

 Only for self-adjoint $A$ do we use the spectral probability notation
$\Prb_\rho\{|A|>t\}=\Tr\rho\Id_{\{|A|>t\}}$. More generally, $\Id_E(A)$ is the spectral projection of $A$ onto a Borel set $E$.
The trace distance between states and the density representation of the
TV distance introduced in Section~\ref{sec:intro} are
\[
 d_{\rm tr}(\rho,\sigma)=\tfrac12\norm{\rho-\sigma}_1\quad\text{\normalfont and}\quad \TV(P,Q)=\sup_B|P(B)-Q(B)|=\tfrac12\int|p-q|.
\]
The last equality uses densities with respect to the same dominating measure, which may always be chosen as $P+Q$. A quantum channel $\mathcal T$ is a parameter-independent completely positive trace-preserving linear map. Its adjoint satisfies $\Tr\mathcal T(\rho)A=\Tr\rho\mathcal T^*(A)$ and is positive and unital, hence pulls POVMs back to POVMs. An ancilla is an additional system in a known state, independent of the unknown parameter. In our construction, an ordered product $K_y$ of positive functions of noncommuting observables specifies both effects $K_y^*K_y$ and a particular measurement operation. The effects alone do not specify all post-measurement states.
The distance-contraction and bounded-loss transfer inequalities used below
are collected in Lemma~\ref{lem:decision-transfer} in
Appendix~\ref{app:bg-operations}.

For a fixed POVM with outcome density $p_\theta^M$, its classical
Fisher information is
\[
 F(\theta;M)=\E_\theta\left[\nabla_\theta\log p_\theta^M(Y)
                       \nabla_\theta\log p_\theta^M(Y)^T\right].
\]
The gradient is in physical parameter coordinates; the measurement is held
fixed when differentiating. The expectation $\E_\theta$ refers to its
classical outcome law. The local coordinate remains
$h=\sqrt n(\theta-\theta_0)$ throughout the paper.
For the weight $W$, the norm $\norm v_W^2=v^TWv$ used in~\eqref{eq:intro-risk}
is induced by the inner product $\langle u,v\rangle_W=u^TWv$.
The projection $\Pi_r^W$ introduced there is taken in this inner product.
It is generally not radial Euclidean clipping. The projection variational inequality
$\langle y-\Pi_r^W y,h-\Pi_r^W y\rangle_W\le0$ for $h\in K_r$ gives
\begin{equation}\label{eq:projection-facts}
\begin{gathered}
 \norm{\Pi_r^W y-h}_W\le\norm{y-h}_W,\quad
 \norm{\Pi_r^W y-\Pi_r^W z}_W\le\norm{y-z}_W,\\
 \text{\normalfont and}\quad\norm{a-h}_W^2\le4\norm W_\op r^2\quad(a,h\in K_r).
\end{gathered}
\end{equation}
The nonexpansiveness inequality in~\eqref{eq:projection-facts} follows by applying the projection variational inequality to $y$ and $z$, adding the two inequalities and using Cauchy--Schwarz. The bound on $\norm{a-h}_W^2$ follows from $\norm{a-h}\le2r$. Local quadratic risk equals $n$ times physical-parameter quadratic risk for $\widehat\theta=\theta_0+\widehat h/\sqrt n$.

\subsection{Scores and density regularity}\label{sec:likelihood-setup}
The classical score provides a useful interpretation of the SLD. For a
fixed finite POVM, discard zero effects; faithfulness makes the remaining
probabilities positive, as shown in Section~\ref{sec:distances}.
Abbreviate them by $p_\theta(y)$. Let
$s_j(y)=\left.\partial_j\log p_\theta(y)\right|_{\theta=\theta_0}$.
Write $s(y)=(s_1(y),\ldots,s_p(y))^T$ for the score vector.
For any fixed real statistic $g(y)$, differentiation of the finite sum
shows that
\[
 \left.\partial_jp_\theta(y)\right|_{\theta=\theta_0}
 =p_{\theta_0}(y)s_j(y)
 \quad\text{and}\quad
 \left.\partial_j\E_\theta g(Y)\right|_{\theta=\theta_0}
 =\E_{\theta_0}[g(Y)s_j(Y)].
\]
Normalization gives $\E_{\theta_0}s(Y)=0$. Thus the Fisher matrix is
the score covariance, $F=\E_{\theta_0}[s(Y)s(Y)^T]$.
If $F\succ0$, the displacement report $x(y)=F^{-1}s(y)$ satisfies
\[
 \E_{\theta_0}x(Y)=0
 \quad\text{and}\quad
 \left.D_\theta\E_\theta x(Y)\right|_{\theta=\theta_0}
 =F^{-1}\E_{\theta_0}[s(Y)s(Y)^T]=I_p.
\]
Its reference covariance is $F^{-1}FF^{-1}=F^{-1}$.
Inverse information therefore converts the fixed score into a locally
calibrated report with the familiar inverse-information covariance.
The score in this report is evaluated at the known nominal point;
the distribution used to take its expectation varies with $\theta$.
This distinction lets a single procedure respond to nearby parameters
without inserting the unknown parameter into its definition.
The SLD performs the corresponding derivative construction at the level
of the state, before any POVM has been chosen.

At the faithful reference state, the SLD $L_j$ is the unique Hermitian solution of
\begin{equation}\label{eq:sld}
 \partial_j\rho_{\theta_0}=\tfrac12(\rho L_j+L_j\rho),
 \quad 1\le j\le p.
\end{equation}
The symmetric product is needed because the derivative of a density
matrix is Hermitian, while $\rho L_j$ need not be. If $\rho$ and $L_j$
commute, this average reduces to the classical product form.
For any fixed Hermitian observable $X_i$, the SLD equation gives the
analogue of the score identity:
\[
 \left.\partial_j\Tr\rho_\theta X_i\right|_{\theta=\theta_0}
 =\tfrac12\Tr\rho(X_iL_j+L_jX_i)
 =\Rea\Tr\rho X_iL_j.
\]
Thus symmetrized state moments determine the local mean slopes that
calibration must match.
The linear map $A\mapsto(\rho A+A\rho)/2$ is invertible on Hermitian
matrices: in an eigenbasis of $\rho$ it multiplies entry $(a,b)$ by
$(\lambda_a+\lambda_b)/2>0$. Taking traces in~\eqref{eq:sld} gives
$\Tr\rho L_j=0$. Put $L=(L_1,\ldots,L_p)$ and
\begin{equation}\label{eq:derivatives}
 J_{ij}=\Tr\rho L_iL_j,\quad \mathbb{I}=\Rea J
 \quad\text{and}\quad \Omega=\Ima J.
\end{equation}
Thus $\mathbb{I}$ is real symmetric and $\Omega$ is real antisymmetric; $\mathbb{I}$ is
information, whereas $I_p$ denotes the identity matrix. We assume $\mathbb{I}\succ0$.
For real $v\in\R^p$, write $L[v]=\sum_i v_iL_i$. The normalized score list is $\Delta_{n,i}=n^{-1/2}\sum_{k=1}^nL_i^{(k)}$. The SLD identity implies $\Tr(\partial_j\rho_{\theta_0})L_i=\mathbb{I}_{ij}$,
so $X_i=\sum_k(\mathbb{I}^{-1})_{ik}L_k$ is a feasible calibrated list:
\[
 \Tr\rho X_i=0
 \quad\text{and}\quad
 \Tr(\partial_j\rho_{\theta_0})X_i
 =\sum_k(\mathbb{I}^{-1})_{ik}\mathbb{I}_{kj}=\delta_{ij}.
\]
Its symmetrized moment matrix is $\mathbb{I}^{-1}$, just as the
classical inverse-score report has covariance $F^{-1}$. The SLD
construction supplies observables before a joint measurement is specified.
Their ordered moments can have an imaginary part, which records expected
commutators and affects the covariance achievable by one classical
reporting rule. Section~\ref{sec:holevo-ancilla} develops this additional
step through the Holevo objective and the ancillary construction.
Further details of the score analogy and the statistical meaning of
calibration are given in Appendix~\ref{app:bg-unbiased}.

For the constructive result the density map is $C^2$ on a neighborhood
of $\theta_0+U$, where $a>0$, $U=\{u\in\R^p:\norm u\le a\}$ and
$\theta_0+U\subset\Theta$. A faithful reference state remains uniformly
faithful on a sufficiently small such ball: continuity of the density
map and of the smallest eigenvalue permits shrinking $a$ so that, for
some $\kappa>0$,
\[
 \rho_{\theta_0+u}\succeq\kappa\Id
 \quad\text{for every }u\in U.
\]
Thus local faithfulness supplies a positive spectral lower bound for
nearby states without requiring distinct eigenvalues. The first two directional derivatives have
finite trace-norm bounds $D_1,D_2$ defined in~\eqref{eq:trace-regularity}.
The signal states are faithful there, the center is known, and all chosen
observables and tuning parameters are independent of the unknown $h$.
A pure ancillary state need not be faithful: the ordered fluctuation
argument uses density derivatives and remains valid on that enlarged space.
No third density derivative is required. All constants refer to the fixed
model and neighborhood. Stronger regularity conditions for the supporting
likelihood results are stated separately in
Assumptions~\ref{ass:meas} and~\ref{ass:regular}.
Table~\ref{tab:inference-routes} summarizes the supporting inference routes
and their assumptions. In particular, the first row shows how feasible
observables and $C^2$ density bounds, with a known center, lead to confidence
regions, tests and bounded-radius risk guarantees for a constructed readout.

\begin{table}[htbp]
\caption{Assumptions and guarantees of the inference routes. Each row has its own inputs.}
\label{tab:inference-routes}
\small
\setlength{\tabcolsep}{3pt}
\renewcommand{\arraystretch}{1.2}
\begin{tabular}{|>{\raggedright\arraybackslash}p{.18\textwidth}|>{\raggedright\arraybackslash}p{.38\textwidth}|>{\raggedright\arraybackslash}p{.37\textwidth}|}
\hline
Route & Required inputs & Resulting guarantee \\
\hline
Direct collective output & Known center, \(C^2\) density bounds, feasible observables & Constructed readout law; confidence regions and tests; projected bounded-radius risk \\
\hline
Fixed measured likelihood & Specified finite POVM, positive measured information, standardized score/Hessian/third-derivative bounds & Localization, Fisher/Wilks and observed-information bounds \\
\hline
Pilot adaptation & Fresh second-stage copies; uniform conditional bounds and a pilot localization guarantee, or an exact affine identity & Conditional and unconditional adaptive inference \\
\hline
Full Gaussian comparison & Separately supplied parameter-independent channels & Experiment-level bounded-loss and minimax comparisons \\
\hline
Quadratic likelihood surrogate & Cubic one-copy likelihood remainder and exponent bounds & State comparison on the same Hilbert space; transfer of the selected collective procedure \\
\hline
\end{tabular}
\end{table}
\FloatBarrier

\section{Constructive collective inference}\label{sec:collective}
We analyze a measurement on the original copies at a known, fixed nominal center $\theta_0$, with unknown local parameter $h\in K_r$. Its ordered Gaussian kernels are those used in Lemma A.1 of Appendix A.5 in \citet{YFGSupp}. Locally unbiased observables determine the target covariance; an ancilla cancels its imaginary part, and a continuous version of the ordered kernels produces an exactly normalized classical output. The direct finite-sample comparison with a Gaussian law uses only two density derivatives.
The argument has three steps. An ordered product of one-copy exponentials gives a finite-$n$ fluctuation bound. Ancillary covariance cancellation makes its comparison law classical, and Gaussian Fourier integration converts that bound into an output-density estimate. Normalization then gives a TV bound, from which confidence and bounded-risk statements follow. The following subsections carry out these steps in that order.

\subsection{Local alternatives and Fourier tests}\label{sec:ordered-tests}
For this section it is enough that the density map is $C^2$ on
a neighborhood of $\theta_0+U$, using the closed ball $U$ and radius
$a>0$ from Section~\ref{sec:likelihood-setup}, with the following finite
bounds for $j=1,2$:
\begin{equation}\label{eq:trace-regularity}
 D_j=\sup_{u\in U,\norm v=1}\norm{D^j\rho_{\theta_0+u}[v,\ldots,v]}_1<\infty.
\end{equation}
The constants $D_1,D_2$ in~\eqref{eq:trace-regularity} are finite by continuity of the derivatives on the compact neighborhood. Theorem~\ref{thm:local-char} also applies to a density model with a fixed pure ancilla, without any faithfulness assumption on that enlarged model. For $S,r\ge0$, define
\begin{gather}
 Q(r,S)=S^2/2+D_1rS\nonumber\\
 \text{\normalfont and}\quad b_n(r,S)=\min\left\{2,\frac{S^3/6+D_1rS^2/2+D_2r^2S/2}{\sqrt n}
       +\frac{Q(r,S)^2}{2n}\right\}.\label{eq:local-char-budget}
\end{gather}
Dependence on $D_1,D_2$ is suppressed only in this notation.

For an integer $m\ge1$ and centered self-adjoint matrices
$A_1,\ldots,A_m$, define the ordered fluctuation quantities
\begin{gather*}
 X_{n,t}=n^{-1/2}\sum_{k=1}^n A_t^{(k)},\quad S=\sum_t\norm{A_t}_\op,\\*
 \text{and}\quad q_0=-\tfrac12\sum_t\Tr\rho A_t^2-\sum_{t<s}\Tr\rho A_tA_s.
\end{gather*}
Products below have the displayed left-to-right order.

\begin{theorem}[Ordered fluctuations under local alternatives]\label{thm:local-char}
Under~\eqref{eq:trace-regularity}, suppose $\Tr\rho A_t=0$ for each
$t$ and $r/\sqrt n\le a$. Then, uniformly for $\norm h\le r$,
\begin{equation}\label{eq:local-char}
 \left|\Tr\rho_{n,h}\prod_{t=1}^m e^{iX_{n,t}}
 -\exp\left\{q_0+i\Tr\left(D\rho_{\theta_0}[h]\sum_tA_t\right)\right\}\right|
 \le b_n(r,S).
\end{equation}
\end{theorem}
\begin{proof}
We first expand the ordered one-copy product and its expectation under a local alternative, then tensorize the scalar comparison.
For real $z$, write $U(z)=\prod_t e^{izA_t}$. The product rule and unitarity give, for $k=1,2,3$,
\[
 \sup_{z\in\R}\norm{\frac{d^k}{dz^k}U(z)}_\op
 \le\sum_{k_1+\cdots+k_m=k}\frac{k!}{k_1!\cdots k_m!}
          \prod_t\norm{A_t}_\op^{k_t}=S^k.
\]
Taylor's integral formula along the real segment therefore bounds the third-order remainder at $n^{-1/2}$ by $S^3/(6n^{3/2})$. In particular,
\[
 \norm{U(z)-\Id}_\op\le |z|S\quad\text{\normalfont and}\quad
 \norm{U(z)-\Id-iz\sum_t A_t}_\op\le z^2S^2/2.
\]
To track the three remainders separately, put $t=n^{-1/2}$,
$H_h=D\rho_{\theta_0}[h]$, $A_\Sigma=\sum_jA_j$, and
\[
 B_U=-\tfrac12\sum_j A_j^2-\sum_{j<k}A_jA_k\quad\text{\normalfont and}\quad U(t)=\Id+itA_\Sigma+t^2B_U+R_3(t).
\]
In particular $\Tr\rho B_U=q_0$; the order $A_jA_k$ is inherited from the ordered product. Also write $$R_2(t)=U(t)-\Id-itA_\Sigma\quad \text{and}\quad
\rho_{\theta_0+th}=\rho+tH_h+E,$$ where
$\Tr H_h=\Tr E=0$, $\norm{H_h}_1\le D_1r$, and
$\norm E_1\le D_2r^2t^2/2$. With $q_h=q_0+i\Tr H_hA_\Sigma$, expanding the trace gives
\begin{gather*}
 z_n:=\Tr\rho_{\theta_0+th}U(t)=1+t^2q_h+e_n\\
 \text{\normalfont and}\quad e_n=\Tr\rho R_3(t)+t\Tr H_hR_2(t)
                     +\Tr E\bigl\{U(t)-\Id\bigr\}.
\end{gather*}
Using $|\Tr AB|\le\norm A_1\norm B_\op$ in these three terms gives
\[
 |e_n|\le t^3\bigl\{S^3/6+D_1rS^2/2+D_2r^2S/2\bigr\}.
\]
The trace-zero property of $E$ is essential here: pairing it with $U-\Id$ supplies one extra factor $t$, rather than an order-$t^2$ error. Moreover $|q_h|\le Q(r,S)$. Hermiticity and cyclicity of trace give
\[
 2\Rea\Tr\rho A_jA_k=\Tr\rho(A_jA_k+A_kA_j).
\]
Consequently $\Rea q_h=-\Tr\rho A_\Sigma^2/2\le0$; the linear shift is purely imaginary because $H_h,A_\Sigma$ are Hermitian.

Operators on different tensor sites commute, so regrouping by site preserves the within-site order and gives
\[
 \prod_j e^{iX_{n,j}}=U(t)^{\otimes n}\quad\text{\normalfont and}\quad
 \Tr\rho_{n,h}\prod_j e^{iX_{n,j}}=z_n^n.
\]
Both $|z_n|$ and $|e^{q_h/n}|$ are at most one. Thus
$|z_n^n-e^{q_h}|\le n|z_n-e^{q_h/n}|$. For $\Rea z\le0$,
\[
 e^z-1-z=z^2\int_0^1(1-t)e^{tz}\,dt\quad\text{\normalfont and}\quad
 |e^z-1-z|\le |z|^2/2.
\]
Apply this with $z=q_h/n$ to obtain the remaining term, without exponential frequency factors. Both expectations have modulus at most one, which gives the cap two.
\end{proof}
For centered observables $X_1,\ldots,X_q$, put
\begin{equation}\label{eq:extended-covariance}
 \Sigma_{ij}=\Tr\rho X_iX_j\quad\text{\normalfont and}\quad T_{ij}=\Tr(\partial_j\rho_{\theta_0})X_i.
\end{equation}
For coefficient vectors $v_t\in\R^q$, with $v_t^TX=\sum_i v_{t,i}X_i$, taking $A_t=v_t^TX$ identifies the target in~\eqref{eq:local-char} as
\[
 \exp\left\{i\sum_tv_t^TTh-\tfrac12\sum_tv_t^T\Sigma v_t
                   -\sum_{t<s}v_t^T\Sigma v_s\right\}.
\]
This is an ordered Gaussian test with covariance convention $X_iX_j$. Write $\sigma_x,\sigma_y,\sigma_z$ for the Pauli matrices, and $\sigma=(\sigma_x,\sigma_y,\sigma_z)$. Thus a qubit with Bloch vector $r\in\R^3$, $\norm r\le1$, has density $(\Id+r\cdot\sigma)/2$, where $r\cdot\sigma=r_x\sigma_x+r_y\sigma_y+r_z\sigma_z$; it is faithful exactly when $\norm r<1$. For example, for $|s|<1$ and real $a,b$, the choices $\rho=(\Id+s\sigma_z)/2$, $A_1=a\sigma_x$ and $A_2=b\sigma_y$ give $q_0=-(a^2+b^2)/2-iabs$; reversing their order changes the sign of this imaginary term. If $\Sigma$ is real, then it is symmetric and the exponent becomes
\[
 i\left(\sum_t v_t\right)^TTh
 -\tfrac12\left(\sum_t v_t\right)^T\Sigma\left(\sum_t v_t\right).
\]
This is $\E\exp\{i\sum_t v_t^TG\}$ for a single $G\sim N(Th,\Sigma)$, including repeated or oppositely signed coordinates. The same $G$ appears in every factor.

\subsection{The Holevo target and a finite ancilla}\label{sec:holevo-ancilla}
For the faithful reference model and weight $W\succ0$, choose
$X\in\mathcal X_{\theta_0}$. Recall the centering and derivative calibration
conditions from~\eqref{eq:intro-calibration}:
\begin{equation}\label{eq:local-unbiased-observables}
 \Tr\rho X_i=0\quad\text{\normalfont and}\quad \Tr(\partial_j\rho_{\theta_0})X_i=\delta_{ij}.
\end{equation}
The local-mean interpretation in Section~\ref{sec:distances} explains these constraints; Appendix~\ref{app:bg-unbiased} gives further details and a two-parameter example. Positive definite SLD information makes the constraints~\eqref{eq:local-unbiased-observables} feasible. For this list, the ordered moment matrix $Z=Z(X)$ and Holevo objective introduced in Section~\ref{sec:intro} are
\[
 Z_{ij}=\Tr\rho X_iX_j\quad\text{\normalfont and}\quad
 \Phi_W(Z)=\Tr W\Rea Z+\norm{W^{1/2}(\Ima Z)W^{1/2}}_1.
\]
The matrix $Z$ records ordered products of observables, so its entries
need not be real. For noncommuting $X_i$ and $X_j$, it is generally not
the covariance of two jointly observed classical outcomes. Nevertheless,
$Z_{ji}=\overline{Z_{ij}}$ and, for any $c\in\mathbb C^p$,
\[
 c^*Zc=\Tr\rho\left(\sum_i c_iX_i\right)^*
                    \left(\sum_j c_jX_j\right)\ge0.
\]
Thus $Z$ is a positive semidefinite Hermitian Gram matrix. Its real and
imaginary parts have distinct interpretations. Writing $Z=G+i\Gamma$ gives
\begin{equation}\label{eq:main-moment-parts}
 G_{ij}=\tfrac12\Tr\rho(X_iX_j+X_jX_i)
 \quad\text{and}\quad \Gamma_{ij}=\frac1{2i}\Tr\rho[X_i,X_j].
\end{equation}
Here $G$ is real symmetric and $\Gamma$ is real antisymmetric. The diagonal
$G_{ii}$ is the spectral outcome variance of $X_i$, and $a^TGa$ is the
variance of the single observable $\sum_i a_iX_i$ for real $a$. The matrix
$\Gamma$ records expected commutators. Its vanishing alone does not imply
that the observables commute or have a common spectral measurement.

To relate these moments to statistical error, consider a real reported
estimate $\widehat\theta$ with finite second moments. Let
$b=\E_{\theta_0}\widehat\theta-\theta_0$ be its bias and let $V$ be
its ordinary covariance matrix at $\theta_0$. Centering the error about
its expectation removes the cross term and gives
\[
 \E_{\theta_0}\norm{\widehat\theta-\theta_0}_W^2
 =\Tr WV+b^TWb.
\]
For diagonal $W$, the covariance term weights each coordinate variance;
for a general positive definite $W$, it also includes cross-coordinate
covariances. The second term measures squared bias in the same geometry.
At a locally unbiased reference point $b=0$, so minimizing weighted
error amounts to minimizing $\Tr WV$. For a biased estimator, both terms
matter. In particular, the later finite-radius shrinkage rule trades a
controlled bias against a reduction in variance. This identity explains
why a local unbiased covariance benchmark and the risk of an estimator
using a known parameter ball answer different statistical questions.

The covariance constraint behind the Holevo objective can now be
understood through an actual reporting rule. For a finite POVM $(M_y)$, let
$x(y)=\widehat\theta(y)-\theta_0$ be real displacement reports with
$\E_{\theta_0}x(Y)=0$. Define their first-moment observables
$X_i^M=\sum_yx_i(y)M_y$ and put $Z^M=Z(X^M)$. If the reporting rule is
locally unbiased, these observables satisfy
\eqref{eq:local-unbiased-observables}. The covariance of the reported vector is
\[
 V^M_{ij}=\sum_y x_i(y)x_j(y)\Tr\rho M_y
 \quad\text{and}\quad V^M-Z^M\succeq0.
\]
For a brief verification, fix $c\in\mathbb C^p$, put
$b_y=\sum_i c_ix_i(y)$ and $B_c=\sum_yb_yM_y$, and expand the positive
operator $\sum_y(b_y\Id-B_c)^*M_y(b_y\Id-B_c)$.
It equals $\sum_y|b_y|^2M_y-B_c^*B_c$; taking its trace against $\rho$
gives $c^*(V^M-Z^M)c\ge0$. This explains why the observed covariance
can exceed the operator moments. At $\theta_0$, the weighted risk of the
locally unbiased reports is $\Tr WV^M$. The same argument applies to
continuous outcomes with finite second moments.

For a fixed feasible list $X$, minimize this necessary covariance
constraint over real symmetric matrices. The result is
\begin{equation}\label{eq:main-holevo-variational}
 \min_{\substack{V\in\R^{p\times p},\ V=V^T\\ V-Z\succeq0}}\Tr WV
 =\Tr WG+\norm{W^{1/2}\Gamma W^{1/2}}_1=\Phi_W(Z).
\end{equation}
The first term is the weighted symmetrized variance. The trace-norm term
is the additional cost imposed by the imaginary moments when the
reported covariance must be real. To see the minimization, put
$A=W^{1/2}\Gamma W^{1/2}$ and $D=W^{1/2}(V-G)W^{1/2}$, so the constraint
is $D-iA\succeq0$. In an orthogonal basis decomposing $A$ into blocks
$\left(\begin{smallmatrix}0&a_k\\-a_k&0\end{smallmatrix}\right)$,
positivity of each corresponding principal block gives
\[
 D_{2k-1,2k-1}+D_{2k,2k}
 \ge2\sqrt{D_{2k-1,2k}^2+a_k^2}\ge2|a_k|.
\]
Summing yields $\Tr D\ge\Tr|A|$, and equality holds at $D=|A|$ because
$|A|-iA\succeq0$. The minimization gives a covariance benchmark;
it does not assert single-copy attainability. Appendix~\ref{app:bg-holevo}
gives the full expansion proving the covariance constraint for
locally unbiased estimators. In the construction below, the minimizing covariance
becomes the target of the Gaussian output comparison.

The corresponding optimal value and optimization gap from the introduction satisfy
\begin{equation}\label{eq:holevo-finite}
 C_W=\min_{X\in\mathcal X_{\theta_0}}\Phi_W(Z(X))
 \quad\text{\normalfont and}\quad \epsilon_{\mathrm{opt}}=\Phi_W(Z)-C_W.
\end{equation}
The covariance $V_X$ used in~\eqref{eq:intro-output} has the explicit expression
\begin{equation}\label{eq:Vstar}
 V_X=\Rea Z+W^{-1/2}\left|W^{1/2}(\Ima Z)W^{1/2}\right|W^{-1/2}.
\end{equation}
The matrix absolute value convention in Section~\ref{sec:distances} applies also to real antisymmetric matrices. The finite-dimensional minimum exists: faithfulness and $W\succ0$ make the quadratic real part coercive on the closed affine feasible space. Any feasible $X$ supplies the computable objective $\Phi_W(Z)$. A claimed numerical upper bound on $\epsilon_{\mathrm{opt}}$ additionally requires a certified lower bound for $C_W$ (for example a dual certificate). A feasible point therefore suffices to report the risk bound in terms of $\Phi_W(Z)$. Theorem B.1 of \citet{YFGSupp} reduces this known Holevo optimization to a real matrix problem on a $D_\rho$-invariant extension of the SLD span. Here $D_\rho$ is a real linear map on Hermitian operators: for each Hermitian $A$, $D_\rho(A)$ is the unique Hermitian solution of $\rho D_\rho(A)+D_\rho(A)\rho=i[\rho,A]$. Invariance means that this map sends the chosen real observable span into itself. The covariance objective and cancellation principle are established ingredients used in the finite-sample analysis.

For any feasible $X$, write $X=\mathbb{I}^{-1}L+R$. The constraints give $\Rea\Tr\rho R_iL_j=0$, and hence
\begin{equation}\label{eq:real-gram-lower}
 \Rea Z=\mathbb{I}^{-1}+(\Rea\Tr\rho R_iR_j)_{ij}\succeq \mathbb{I}^{-1}\succ0.
\end{equation}
In particular $V_X\succ0$. Put $C=V_X-Z$ and $A=W^{1/2}(\Ima Z)W^{1/2}$. The Hermitian matrix $iA$ has absolute value $|A|$, so diagonalization gives $|A|-iA\succeq0$ and therefore $C\succeq0$. Its imaginary part satisfies $\Ima C=-\Gamma$, while its real part
$\Rea C=V_X-G$ is the covariance added to make the target real.
A known ancillary state will supply these moments while preserving the
mean and derivative calibration of the signal observables. Set
$
 r_C := \operatorname{rank}(C).
$
Choose a factorization $C=B^*B$ with $B\in\mathbb C^{r_C\times p}$. Let $l_i\in\mathbb C^{r_C}$ be the columns of $B$. In Dirac notation, $|v\rangle$ is a column vector and $\langle v|$ is its adjoint, so $|u\rangle\langle v|$ is their outer product. Choose a unit reference vector $|0\rangle$ orthogonal to $\mathbb C^{r_C}$. On $\mathcal H_a=\mathbb C|0\rangle\oplus\mathbb C^{r_C}$ set
\begin{equation}\label{eq:ancilla}
 \omega=|0\rangle\langle0|,\quad A_i=|l_i\rangle\langle0|+|0\rangle\langle l_i|,\quad 
 \text{\normalfont and}\quad U_i=X_i\otimes\Id+\Id\otimes A_i.
\end{equation}
Here $Z,V_X,C$ are $p\times p$ parameter matrices, while $A_i$ acts on the ancillary space and $U_i$ on the signal--ancilla space. Since $A_i|0\rangle=|l_i\rangle$,
\[
 \langle0|A_iA_j|0\rangle=\langle l_i,l_j\rangle=C_{ij}\quad\text{\normalfont and}\quad \langle0|A_i|0\rangle=0.
\]
The nonzero eigenvalues of $A_i$ are $\pm\norm{l_i}$ when $l_i\ne0$, so $\norm{A_i}_\op=\sqrt{C_{ii}}$. Expanding $U_iU_j$ under the product state gives $Z_{ij}+C_{ij}$: both mixed terms vanish because the ancillary means are zero. Consequently, writing $\bar\rho_\theta=\rho_\theta\otimes\omega$ and $\bar\rho=\bar\rho_{\theta_0}$,
\begin{equation}\label{eq:cancelled-cov}
 \Tr\bar\rho U_iU_j=(V_X)_{ij},\quad
 \Tr(\partial_j\bar\rho_{\theta_0})U_i=\delta_{ij},\quad 
 \text{\normalfont and}\quad \Tr WV_X=C_W+\epsilon_{\mathrm{opt}}.
\end{equation}
The derivative constants~\eqref{eq:trace-regularity} are unchanged by adjoining $\omega$. Real covariance means zero expected commutators; it does not mean that the $U_i$ commute as operators.

\begin{proposition}[Ancillary dimension of the Gram construction]\label{prop:ancilla-rank}
The above realization satisfies
\[
 r_C=\tfrac12\operatorname{rank}(\Ima Z)\quad\text{\normalfont and}\quad
 \dim\mathcal H_a=1+r_C\le1+\lfloor p/2\rfloor.
\]
\end{proposition}
\begin{proof}
Recall $A=W^{1/2}(\Ima Z)W^{1/2}$. The definition of $C$ gives
\[
 C=W^{-1/2}(|A|-iA)W^{-1/2}
 \quad\text{and}\quad
 \operatorname{rank}A=\operatorname{rank}(\Ima Z).
\]
A real orthogonal change of basis decomposes $A$ into nonzero blocks
$A_j=\left(\begin{smallmatrix}0&a_j\\-a_j&0\end{smallmatrix}\right)$
and a zero block. Each $A_j$ has rank two, whereas $|A_j|-iA_j$ has
 eigenvalues $0,2|a_j|$ and hence rank one. Invertible congruence preserves
rank, so
\[
 r_C=\operatorname{rank}(|A|-iA)
 =\tfrac12\operatorname{rank}A
 =\tfrac12\operatorname{rank}(\Ima Z)\le\lfloor p/2\rfloor.
\]
The Gram realization adds the reference vector $|0\rangle$ to an
$r_C$-dimensional space, giving $\dim\mathcal H_a=1+r_C$.
If $C=0$, the ancilla is one-dimensional and contributes no observable.
\end{proof}
Proposition~\ref{prop:ancilla-rank} bounds the ancillary dimension of the stated Gram realization. Finite covariance realization and ancillary cancellation have antecedents in \citet[Appendix A.5, Theorem B.7 and Lemma B.9]{YFGSupp}. To specify the procedure, solve the SLD equations, choose feasible centered Hermitian $X$, compute $Z,V_X,C$, factor $C$ and record the resulting norms of $U_i$ and the density constants.

\subsection{A normalized POVM and its risk}\label{sec:collective-output}
On $(\mathcal H\otimes\mathcal H_a)^{\otimes n}$, let
\[
 U_i^{(k)}=\Id^{\otimes(k-1)}\otimes U_i\otimes\Id^{\otimes(n-k)}
 \quad\text{and}\quad B_{n,i}=n^{-1/2}\sum_{k=1}^nU_i^{(k)}.
\]
Here $U_i$ is the signal--ancilla observable in~\eqref{eq:ancilla}, and
$U_i^{(k)}$ acts as $U_i$ on the $k$th signal--ancilla pair and as the
identity on every other pair, for $1\le i\le p$ and $1\le k\le n$.
For $\eta>0$ and $y\in\R^p$, the scalar Gaussian kernel is
\begin{equation}\label{eq:gaussian-kernel}
 g_{\eta,y_i}(x)=(2\pi\eta^2)^{-1/4}\exp\left\{-(x-y_i)^2/(4\eta^2)\right\}.
\end{equation}
The kernel gives a measurement with Gaussian resolution. To see its
statistical meaning, take one Hermitian observable $B=\sum_b bP_b$ and
an input state $\sigma$. Spectral calculus gives the outcome density
\[
 p_\sigma(y)=\Tr\sigma g_{\eta,y}(B)^2
 =\sum_b\Tr(\sigma P_b)\frac{e^{-(y-b)^2/(2\eta^2)}}{\sqrt{2\pi\eta^2}}.
\]
Thus an eigenstate with eigenvalue $b$ produces $N(b,\eta^2)$, and a
general state produces a mixture with the Born probabilities as weights.
For this one stage, the Gaussian noise leaves the mean unchanged and
adds $\eta^2$ to the spectral outcome variance:
\[
 \E_\sigma Y=\Tr\sigma B
 \quad\text{and}\quad
 \operatorname{Var}_\sigma(Y)=\Tr\sigma B^2-(\Tr\sigma B)^2+\eta^2.
\]
These identities follow by taking the first two moments of the mixture.
They explain the smoothing scale in the construction. For several
observables, the input to each successive stage is the state produced by
the preceding stages, so the joint law must also account for their
measurement operations.

The ordered product of the kernels defines that operation and its outcome measure:
\begin{equation}\label{eq:smooth-povm}
 K_y=g_{\eta,y_p}(B_{n,p})\cdots g_{\eta,y_1}(B_{n,1})
 \quad\text{\normalfont and}\quad M_{n,\eta}(dy)=K_y^*K_y\,dy.
\end{equation}
The palindromic effect $K_y^*K_y$ is the continuous counterpart of the ordered Gaussian-root construction in Lemma A.1 of Appendix A.5 in \citet{YFGSupp}, with $\eta^2=1/q$ in that lemma's notation. We use exact spectral integration in place of its finite-grid normalization and derive explicit finite-$n$ error budgets below.
Regroup the tensor factors as signal systems followed by ancillas and set $J_0\psi=\psi\otimes|0\rangle^{\otimes n}$. The actual signal POVM is
\begin{equation}\label{eq:signal-povm}
 M_{n,\eta}^{\rm signal}(dy)=J_0^*K_y^*K_yJ_0\,dy.
\end{equation}
Put $\bar\rho_{n,h}=\bar\rho_{\theta_0+h/\sqrt n}^{\otimes n}$ on the enlarged space. The same classical outcome law can be written in either representation:
\[
 P_{n,h}^{\eta}(dy)=\Tr\bar\rho_{n,h}M_{n,\eta}(dy)
 =\Tr\rho_{n,h}M_{n,\eta}^{\rm signal}(dy).
\]
The procedure uses $n$ known ancillas of dimension $1+r_C$, and no additional unknown copies. Its parameters depend on the known nominal center, model, $W$, and chosen tuning, not on the unknown $h$.

The rightmost factor in $K_y$ acts first. At stage $i$, the map
$\sigma\mapsto g_{\eta,y_i}(B_{n,i})\sigma g_{\eta,y_i}(B_{n,i})\,dy_i$
is the unnormalized state update. Iterating these updates gives
$K_y\sigma K_y^*\,dy$, whose trace is $\Tr\sigma K_y^*K_y\,dy$.
For an outcome with positive density, the conditional state after the
whole sequence is
\[
 \sigma_y=\frac{K_y\sigma K_y^*}{\Tr\sigma K_y^*K_y}.
\]
The effects $K_y^*K_y$ specify probabilities, while the chosen operators
$K_y$ also specify the state change. Within the sequence, each stage acts
on the state left by the preceding stages. This explains why the order matters when the
$B_{n,i}$ do not commute. Section~\ref{sec:collective-qubit} computes the
resulting change in the second output mean for the transverse qubit
model, using exact identities for the measurement operation.

To specify the error budgets for $\eta>0$, put
\[
 \ell_U=\max_i\norm{U_i}_\op,\quad a_\eta=(2\pi\eta^2)^{-1/4},\quad\text{\normalfont and}\quad
 v_\eta=\norm{V_X+\eta^2I_p}_\op.
\]
Let $w_1,\ldots,w_{2p}$ be the norms of the observables in the density word. Their order is
\[
 (U_1,\ldots,U_p,U_p,\ldots,U_1).
\]
For independent $Z_1,\ldots,Z_{2p}$ with density $\eta e^{-\eta^2z^2}/\sqrt\pi$, define $\mathfrak M_k=\E(\sum_{j=1}^{2p}w_j|Z_j|)^k$ for $k=1,2,3,4$. Define the density-error budget by
\begin{equation}\label{eq:collective-density-budget}
 D_n(r,\eta)=a_\eta^{2p}\biggl\{
 \frac{\mathfrak M_3/6+D_1r\mathfrak M_2/2+D_2r^2\mathfrak M_1/2}{\sqrt n}+\frac{\mathfrak M_4/4+D_1r\mathfrak M_3+D_1^2r^2\mathfrak M_2}{2n}\biggr\}.
\end{equation}
For $R_{\rm out}>r$, the corresponding TV budget is
\begin{equation}\label{eq:collective-TV-budget}
 \varepsilon_n(r,\eta,R_{\rm out})=\min\left\{1,(2R_{\rm out})^pD_n(r,\eta)
       +2p e^{-(R_{\rm out}-r)^2/(2v_\eta)}\right\}.
\end{equation}
Let $\Gamma$ denote the gamma function and $[t^k]f(t)$ the coefficient
of $t^k$ in a polynomial or power series.

\begin{lemma}[Fourier moments]\label{lem:fourier-moments}
For the independent frequencies and weights above, put
$\mu_k=\Gamma((k+1)/2)/(\sqrt\pi\eta^k)$, so $\mu_0=1$.
For $1\le k\le4$,
\begin{equation}\label{eq:fourier-moments}
 \mathfrak M_k=k![t^k]\prod_{j=1}^{2p}\left(\sum_{a=0}^4\frac{w_j^a\mu_a}{a!}t^a\right)
 \quad\text{\normalfont and}\quad \mathfrak M_k\le\left(\sum_jw_j\right)^k\mu_k.
\end{equation}
\end{lemma}
\begin{proof}
Integration against the frequency density gives $\mu_k=\E|Z_j|^k$;
in particular,
\[
 \mu_1=\frac1{\sqrt\pi\eta},\quad
 \mu_2=\frac1{2\eta^2},\quad
 \mu_3=\frac1{\sqrt\pi\eta^3},\quad\text{\normalfont and}\quad
 \mu_4=\frac3{4\eta^4}.
\]
The multinomial expansion of $(\sum_jw_j|Z_j|)^k$, followed by
independence, proves the coefficient formula. For example,
\[
 \mathfrak M_1=\frac{\sum_jw_j}{\sqrt\pi\eta}\quad\text{\normalfont and}\quad
 \mathfrak M_2=\frac{\sum_jw_j^2}{2\eta^2}
            +\frac{2\sum_{j<k}w_jw_k}{\pi\eta^2}.
\]
If $\sum_jw_j>0$, Jensen's inequality with weights
$w_j/\sum_iw_i$ gives the moment bound. If all weights vanish, both
sides are zero.
\end{proof}

\begin{theorem}[Constructive local output and risk bounds]\label{thm:collective}
Use the $C^2$ signal model on the compact neighborhood
$\theta_0+U$ of Section~\ref{sec:likelihood-setup}, with faithful signal
states throughout that neighborhood, positive definite reference SLD
information, a real symmetric weight $W\succ0$, and the construction
of Section~\ref{sec:holevo-ancilla}. Suppose
\eqref{eq:trace-regularity} and~\eqref{eq:local-unbiased-observables}
hold, $r/\sqrt n\le a$, $\eta>0$ and $R_{\rm out}>r$.
The measure~\eqref{eq:smooth-povm} is normalized at every $n$, and
its outcome law obeys
\begin{equation}\label{eq:collective-TV}
 \sup_{\norm h\le r}\TV\left\{P_{n,h}^{\eta},N(h,V_X+\eta^2I_p)\right\}
 \le\varepsilon_n(r,\eta,R_{\rm out}).
\end{equation}
The projected estimator $\widehat h=\Pi_r^W(Y)$ from~\eqref{eq:intro-risk}
has local quadratic risk bounded by
\begin{equation}\label{eq:collective-risk}
 \sup_{\norm h\le r}\E_h\norm{\widehat h-h}_W^2
 \le C_W+\epsilon_{\mathrm{opt}}+\eta^2\Tr W
              +4\norm W_\op r^2\varepsilon_n(r,\eta,R_{\rm out}).
\end{equation}
This is an upper bound for the specified procedure, with radius $r$ and tuning parameters $\eta,R_{\rm out}$ explicit.
\end{theorem}
\begin{proof}
We prove normalization, integrate the ordered Fourier bound, and then use the resulting TV bound to control quadratic risk.
For any finite-dimensional self-adjoint $B$, spectral calculus gives $\int g_{\eta,y}(B)^2dy=\Id$. For example, when $p=2$, abbreviating $g_{\eta,y_i}(B_{n,i})$ by $g_i$,
\[
 K_y^*K_y=g_1g_2^2g_1,\quad
 \int K_y^*K_y\,dy_2=g_1^2,\quad\text{\normalfont and}\quad
 \int g_1^2\,dy_1=\Id.
\]
For general $p$, the two factors involving $B_{n,p}$ are likewise adjacent. Integrate $y_p$ first, then $y_{p-1}$, down to $y_1$. Tonelli applies to each nonnegative quadratic form; finite dimensionality then identifies the operator integral as $\int K_y^*K_y\,dy=\Id$. Thus $P_{n,h}^{\eta}$ has a nonnegative density $p_{n,h}^{\eta}$ with integral one.

Use the Fourier convention $\widehat f(t)=\int_{\R}e^{-itx}f(x)\,dx$ and $f(x)=\int_{\R}e^{itx}\widehat f(t)\,dt/(2\pi)$. The Fourier measure for $g_{\eta,y}$ has density
\[
 a_\eta\frac{\eta}{\sqrt\pi}e^{-\eta^2t^2}e^{-ity}.
\]
Its variation mass is $a_\eta$, independently of $y$. The density is an ordered word of $2p$ such functions. After normalizing the product variation measure, its frequencies are the independent $Z_j$ above, and the norm sum in Theorem~\ref{thm:local-char} is $S=\sum_jw_j|Z_j|$, whose moments are given by Lemma~\ref{lem:fourier-moments}.
Apply Theorem~\ref{thm:local-char} inside the full Fourier integral, dropping its cap two. The polynomial bound is integrable because these moments are finite, and
\[
 \E Q(r,S)^2=\tfrac14\mathfrak M_4+D_1r\mathfrak M_3+D_1^2r^2\mathfrak M_2.
\]
Together with the variation mass $a_\eta^{2p}$, this gives exactly~\eqref{eq:collective-density-budget}. Bounded spectral calculus and finite variation justify the integral, and the integrable discrepancy justifies taking its absolute value.

By~\eqref{eq:cancelled-cov} the comparison word is classical. If $\varphi_{\eta^2I_p}$ denotes the centered Gaussian density with covariance $\eta^2I_p$, then for $G\sim N(h,V_X)$,
\[
 \E\prod_{i=1}^p g_{\eta,y_i}(G_i)^2
   =\E\varphi_{\eta^2I_p}(y-G)=q_h^\eta(y).
\]
This is the density of $G+\xi$ for independent $\xi\sim N(0,\eta^2I_p)$, explaining the smoothed covariance $V_X+\eta^2I_p$. Hence
\begin{equation}\label{eq:collective-density}
 \sup_y\left|p_{n,h}^{\eta}(y)-q_h^\eta(y)\right|\le D_n(r,\eta).
\end{equation}
Let $Q=N(h,V_X+\eta^2I_p)$ be the comparison law. Apply Lemma~\ref{lem:decision-transfer} to $A=[-R_{\rm out},R_{\rm out}]^p$. The compact integral is at most $(2R_{\rm out})^pD_n$; a coordinate union bound for the ordinary Gaussian gives $$Q(A^c)\le2p\exp\bigl\{-(R_{\rm out}-r)^2/(2v_\eta)\bigr\}.$$ This proves~\eqref{eq:collective-TV} using normalization, without a separate tail assumption on the quantum output.

Projection onto the closed convex ball in the $W$ inner product cannot increase distance to any $h\in K_r$. Under the Gaussian law its risk is thus at most $\Tr W(V_X+\eta^2I_p)$. After projection, loss lies in $[0,4\norm W_\op r^2]$, so its expectation changes by at most this range times the TV distance. Equation~\eqref{eq:cancelled-cov} proves~\eqref{eq:collective-risk}.
\end{proof}

\subsection{From output approximation to statistical guarantees}\label{sec:output-inference}
Theorem~\ref{thm:collective} compares the entire distribution of the
recorded vector, so its consequences apply to different reporting rules
computed from the same $Y$. Fix $h\in K_r$ and write
$P_h=P_{n,h}^{\eta}$, $Q_h=N(h,V_X+\eta^2I_p)$ and
$\varepsilon=\varepsilon_n(r,\eta,R_{\rm out})$. The definition of TV
ensures that every measurable outcome event has probabilities within
$\varepsilon$ under these two laws. For a proposed confidence region of nominal level $1-\alpha$, with
$0<\alpha<1$, the event $A_h=\{y:\theta_0+h/\sqrt n\in\mathcal C(y)\}$ is such an
event, even though its definition involves the fixed parameter $h$.
Consequently,
\[
 Q_h(A_h)=1-\alpha
 \quad\Longrightarrow\quad
 1-\alpha-\varepsilon\le P_h(A_h)\le1-\alpha+\varepsilon.
\]
A test's rejection region is another outcome event, so the same
comparison controls its size under the null and its power at each local
alternative. Uniformity of the output bound over $K_r$ makes these
statements uniform over that same parameter set.

Risk is an expectation rather than an event probability.
The bounded-loss inequality in Lemma~\ref{lem:decision-transfer} gives,
for each loss $0\le\ell(y,h)\le B_\ell<\infty$,
\[
 \left|\E_{P_h}\ell(Y,h)-\E_{Q_h}\ell(Y,h)\right|
 \le B_\ell\varepsilon.
\]
For quadratic loss, projection supplies the required finite range:
both $\Pi_r^W(y)$ and the true $h$ lie in $K_r$, and hence
$\ell(y,h)=\norm{\Pi_r^W(y)-h}_W^2\le4\norm W_\op r^2$.
The Gaussian risk and this range determine the risk certificate in
Theorem~\ref{thm:collective}. Thus the radius has a direct statistical
role: it bounds the possible error after projection and specifies the
region on which the approximation is used. Shrinkage uses the same
radius to trade variance against bias; confidence regions and tests use
the unprojected outcome with its Gaussian comparison law.

For finite-radius shrinkage, abbreviate
\begin{gather*}
 A_\eta=\Tr W(V_X+\eta^2I_p)=\Phi_W(Z)+\eta^2\Tr W>0,\\
 B_r=\norm W_\op r^2,\quad\text{\normalfont and}\quad
 \varepsilon=\varepsilon_n(r,\eta,R_{\rm out}).
\end{gather*}
For deterministic $c\in[0,1]$, define
$\widehat h_c=\Pi_r^W(cY)$ and set $c_*=B_r/(A_\eta+B_r)$.
The selection rule reports $\widehat h_{c_*}$ if
$A_\eta B_r/(A_\eta+B_r)+4B_r\varepsilon<B_r$, and zero otherwise.
It uses only the known model and tuning constants.

\begin{corollary}[Finite-radius shrinkage]\label{cor:collective-shrinkage}
Under Theorem~\ref{thm:collective}, every $c\in[0,1]$ satisfies
\begin{equation}\label{eq:collective-shrinkage}
 \sup_{h\in K_r}\E_h\norm{\widehat h_c-h}_W^2
 \le c^2A_\eta+(1-c)^2B_r+4B_r\varepsilon.
\end{equation}
The selection rule above has worst-ball risk at most
\begin{equation}\label{eq:collective-selected-risk}
 \min\left\{B_r,\frac{A_\eta B_r}{A_\eta+B_r}+4B_r\varepsilon\right\}.
\end{equation}
For $r>0$, the displayed certificate is strictly below $B_r$ whenever
$\varepsilon<B_r/[4(A_\eta+B_r)]$. For $r=0$, both the selected risk and $c_*$ are zero.
\end{corollary}
The proof is given in Section~\ref{app:collective-shrinkage-proof}.
This procedure uses the available radius to allow bias. It does not identify the finite-ball minimax risk, and it need not improve on zero at each individual $h$. Confidence regions and tests below continue to use the unprojected output $Y$.
Algorithm~\ref{alg:collective-inference} in the introduction summarizes
the measurement construction and this selection rule.

\begin{corollary}[One fixed-dimensional schedule]\label{cor:collective-rate}
Fix the model, $p,r,W$ and the chosen observables. For sufficiently large $n$, choose
\[
 \eta_n=n^{-1/[2(p+5)]}(\log n)^{p/[2(p+5)]}\le1\quad 
 \text{\normalfont and}\quad
 R_{{\rm out},n}=r+\sqrt{8\norm{V_X+I_p}_\op\log n}.
\]
Then
\begin{equation}\label{eq:collective-rate}
 \varepsilon_n(r,\eta_n,R_{{\rm out},n})+\eta_n^2
 =O\left(n^{-1/(p+5)}(\log n)^{p/(p+5)}\right).
\end{equation}
The same order bounds the excess on the right of~\eqref{eq:collective-risk} above $C_W+\epsilon_{\mathrm{opt}}$.
\end{corollary}
The proof is given in Section~\ref{app:collective-rate-proof}.
The explicit finite-$n$ budget is the substantive bound; the schedule is a conservative existence rate and can be uninformative at moderate $n$. It is not a growing-dimensional rate and does not establish a matching minimax lower bound on $K_r$.
The smoothing term $\eta^2$ controls both the added Gaussian risk
$\eta^2\Tr W$ and the change in the comparison law. In fact, since
$V_X\succ0$, uniformly in $h\in\R^p$,
\begin{equation}\label{eq:gaussian-smoothing-TV}
 \TV\left\{N(h,V_X+\eta^2I_p),N(h,V_X)\right\}
 \le\frac{\eta^2}{2\sqrt2}\norm{V_X^{-1}}_{\mathrm F}.
\end{equation}
To verify this bound directly, let $q_t$ be the density of
$N(h,\Sigma_t)$, where $\Sigma_t=V_X+tI_p$, and let
$G_0\sim N(0,I_p)$. Differentiating with $y$ fixed gives
\[
 \left.\partial_t\log q_t(y)\right|_{y=h+\Sigma_t^{1/2}G_0}
 =\tfrac12\bigl\{G_0^T\Sigma_t^{-1}G_0-\Tr\Sigma_t^{-1}\bigr\}.
\]
This centered quantity has variance
$\tfrac12\Tr\Sigma_t^{-2}$, obtained by diagonalizing
$\Sigma_t^{-1}$ and using $\operatorname{Var}(G_{0,i}^2)=2$.
Hence Cauchy--Schwarz gives
$$\int|\partial_tq_t(y)|\,dy\le\norm{\Sigma_t^{-1}}_{\mathrm F}/\sqrt2.$$
Integrating from $t=0$ to $t=\eta^2$ and using
$\norm{\Sigma_t^{-1}}_{\mathrm F}\le\norm{V_X^{-1}}_{\mathrm F}$
proves~\eqref{eq:gaussian-smoothing-TV}. The triangle inequality therefore yields
\begin{equation}\label{eq:collective-unsmoothed-TV}
 \sup_{h\in K_r}\TV\left\{P_{n,h}^{\eta},N(h,V_X)\right\}
 \le\min\left\{1,\varepsilon_n(r,\eta,R_{\rm out})
       +\frac{\eta^2}{2\sqrt2}\norm{V_X^{-1}}_{\mathrm F}\right\}.
\end{equation}
For the fixed model and observable list of
Corollary~\ref{cor:collective-rate}, the inverse covariance norm is
constant, so its schedule gives the same rate for this comparison
with the unsmoothed Gaussian law.

For confidence regions and testing, put $V_\eta=V_X+\eta^2I_p$ and
$\varepsilon=\varepsilon_n(r,\eta,R_{\rm out})$. For $0<\alpha<1$,
write $\chi_p^2$ for a chi-square random variable with $p$ degrees of
freedom and $q_{p,1-\alpha}$ for its $(1-\alpha)$-quantile. Using the
unprojected outcome $Y$, define
\[
 \mathcal C_\alpha(Y)=\left\{\theta_0+h/\sqrt n:h\in K_r,
 (Y-h)^TV_\eta^{-1}(Y-h)\le q_{p,1-\alpha}\right\}.
\]
For a fixed null $h_0\in K_r$, reject when
$(Y-h_0)^TV_\eta^{-1}(Y-h_0)>q_{p,1-\alpha}$.
At $h\in K_r$, write $\lambda=(h-h_0)^TV_\eta^{-1}(h-h_0)$ and let
$\chi_p^2(\lambda)$ denote a noncentral chi-square random variable
with $p$ degrees of freedom and noncentrality $\lambda$.

\begin{corollary}[Local confidence regions and testing]\label{cor:collective-inference}
Under Theorem~\ref{thm:collective}, the region above satisfies
\begin{equation}\label{eq:collective-coverage}
 \sup_{h\in K_r}\left|
 \Prb_{\theta_0+h/\sqrt n}\left\{\theta_0+h/\sqrt n\in\mathcal C_\alpha(Y)\right\}
 -(1-\alpha)\right|\le\varepsilon.
\end{equation}
In particular it is at least $1-\alpha-\varepsilon$. If
$\varepsilon<\alpha$, the quantile $q_{p,1-\alpha+\varepsilon}$ gives
coverage at least $1-\alpha$.

The test above has size at most $\alpha+\varepsilon$. At every
$h\in K_r$, its power differs by at most $\varepsilon$ from
\[
 \Prb\bigl\{\chi_p^2(\lambda)>q_{p,1-\alpha}\bigr\},
\]
with the probability evaluated under that noncentral law.
\end{corollary}
The proof is given in Section~\ref{app:collective-inference-proof}. These are known-center local guarantees. Projection is used for bounded quadratic loss, but need not preserve the Gaussian statistic and is not used here. The comparison remains with the smoothed covariance $V_\eta$.
This known-center restriction concerns the measurement itself: the observables, ancilla and calibration depend on $\theta_0$. Replacing $\theta_0$ by a pilot estimate requires uniform control of these choices and their conditional error budgets. Appendix~\ref{sec:adaptive} gives pilot-centered inference for specified measured models, and Section~\ref{sec:exact-adaptive} gives an exact scalar implementation. 

For a finite readout, let $\Delta>0$, $R_{\rm out}>r$ and
$V_\eta=V_X+\eta^2I_p$, and abbreviate
$\varepsilon_n=\varepsilon_n(r,\eta,R_{\rm out})$.
Integrate the POVM effects over a finite grid of cells of side at most
$\Delta$ in $[-R_{\rm out},R_{\rm out}]^p$, with one failure outcome
outside. Report a cell midpoint projected by $\Pi_r^W$, and zero on failure.

\begin{corollary}[Finite readout]\label{cor:finite-readout}
Under Theorem~\ref{thm:collective}, the finite readout above has TV
distance at most $\varepsilon_n$ from the same discretization of
$N(h,V_\eta)$, uniformly for $h\in K_r$. Relative to the projected
continuous procedure, its worst risk increases by at most
\[
 2r\norm W_\op\sqrt p\,\Delta+
 4\norm W_\op r^2\left\{2p e^{-(R_{\rm out}-r)^2/(2v_\eta)}+\varepsilon_n\right\}.
\]
\end{corollary}
The proof is given in Section~\ref{app:finite-readout-proof}.
The target must be discretized by the same map: a finite discrete law and a continuous Gaussian have TV distance one.

\section{A noncommuting qubit illustration}\label{sec:examples}
We illustrate the collective construction of Section~\ref{sec:collective}
in a two-parameter qubit model, obtaining an explicit risk bound and
showing how measurement order affects the output means. Numerical
integration of the constructed output law illustrates its Gaussian
approximation and estimation risks at small sample sizes.

\subsection{A noncommuting qubit model}\label{sec:transverse-qubit}
The transverse model has known
longitudinal coordinate $0<s<1$ and unknown transverse parameter $\theta$.
The Pauli matrices used in this model are
\[
 \sigma_x=\begin{pmatrix}0&1\\1&0\end{pmatrix},\quad
 \sigma_y=\begin{pmatrix}0&-i\\i&0\end{pmatrix}
 \quad\text{and}\quad \sigma_z=\begin{pmatrix}1&0\\0&-1\end{pmatrix}.
\]
The model's density on $\norm\theta<\sqrt{1-s^2}$, with nominal
center $\theta_0=0$, is
\begin{equation}\label{eq:qubit}
 \rho_\theta=\tfrac12\bigl\{\Id+s\sigma_z+\theta_1\sigma_x+\theta_2\sigma_y\bigr\}.
\end{equation}
The Pauli anticommutation identities give
$(s\sigma_z+\theta_1\sigma_x+\theta_2\sigma_y)^2=(s^2+\norm\theta^2)\Id$.
Thus the two eigenvalues of the density are
\[
 \lambda_\pm(\theta)=\frac{1\pm\sqrt{s^2+\norm\theta^2}}2.
\]
They are positive precisely on the stated parameter domain. On a closed
physical disk $\norm\theta\le b<\sqrt{1-s^2}$, the lower eigenvalue
is at least $(1-\sqrt{s^2+b^2})/2>0$, giving the uniform faithfulness
needed for local calculations. The trace identities
$\Tr\sigma_a=0$ and $\Tr(\sigma_a\sigma_b)=2\delta_{ab}$ also give
\[
 \Tr\rho_\theta\sigma_x=\theta_1
 \quad\text{and}\quad \Tr\rho_\theta\sigma_y=\theta_2.
\]
Consequently these observables estimate the transverse displacement in
mean throughout the model, and their mean derivatives at zero form
$I_2$. To identify the SLDs, note that
$\partial_j\rho_\theta=\sigma_j/2$, where $\sigma_1=\sigma_x$ and
$\sigma_2=\sigma_y$. Since each anticommutes with $\sigma_z$,
$(\rho_0\sigma_j+\sigma_j\rho_0)/2=\sigma_j/2$.
At the reference point $\theta=0$, the SLDs are therefore $L_1=\sigma_x$, $L_2=\sigma_y$, so
\[
 \mathbb{I}=I_2\quad\text{\normalfont and}\quad J=\begin{pmatrix}1&is\\-is&1\end{pmatrix}.
\]
The imaginary part records the expected commutator. Faithfulness holds on each closed ball strictly inside the parameter domain.

An implementable measurement chooses the $x$ or $y$ axis with equal probability and records the axis and sign. For $a\in\{x,y\}$, its effects are
\[
 M_{a,\pm}=\tfrac14(\Id\pm\sigma_a),
\]
with $p_\theta(a,\pm)=(1\pm\theta_a)/4$, where $\theta_x=\theta_1$
and $\theta_y=\theta_2$. For an outcome on axis $a$, the score has
only one nonzero coordinate, $\pm(1\pm\theta_a)^{-1}$. Summing its
squared value against the two sign probabilities gives
$\{(1+\theta_a)^{-1}+(1-\theta_a)^{-1}\}/4
=1/\{2(1-\theta_a^2)\}$. Its Fisher information is
\[
 F(\theta;M)=\tfrac12\diag\bigl((1-\theta_1^2)^{-1},(1-\theta_2^2)^{-1}\bigr)
 \quad\text{\normalfont and}\quad F(0;M)=I_2/2.
\]
The factor one half reflects allocating each copy to one of two axes,
so this classical outcome carries half the SLD information in each
coordinate at the reference point. The information loss at the reference point illustrates why the statistical
guarantee must be tied to the specified measurement. Appendix~\ref{sec:measured}
provides the corresponding likelihood and coverage analysis for this POVM.

\subsection{The constructed collective procedure in the transverse model}\label{sec:collective-qubit}
For~\eqref{eq:qubit}, $D_1=1,D_2=0$: the directional derivative
$D\rho_\theta[v]=(v_1\sigma_x+v_2\sigma_y)/2$ has eigenvalues
$\pm\norm v/2$, and the density is affine. The choice of calibrated
observables can also be optimized explicitly. Expand each Hermitian
$X_i$ in the real Pauli basis. The derivative constraints fix its
$\sigma_x$ and $\sigma_y$ coefficients, while centering at
$\rho_0=(\Id+s\sigma_z)/2$ fixes the identity coefficient. Every feasible
pair therefore has the form
\[
 X_1=\sigma_x+t_1(\sigma_z-s\Id)
 \quad\text{and}\quad
 X_2=\sigma_y+t_2(\sigma_z-s\Id),\quad t_1,t_2\in\R.
\]
Writing $t=(t_1,t_2)^T$, Pauli multiplication gives
\[
 \Rea Z=I_2+(1-s^2)tt^T
 \quad\text{and}\quad
 \Phi_{I_2}(Z)=2+2s+(1-s^2)\norm t^2.
\]
Indeed $\Ima Z=\left(\begin{smallmatrix}0&s\\-s&0\end{smallmatrix}\right)$
is independent of $t$ and has two singular values equal to $s$.
Because $1-s^2>0$, the minimum occurs at $t=0$. At $W=I_2$, the
feasible observables $X_1=\sigma_x,X_2=\sigma_y$ therefore attain the
known value $C_W=2+2s$. Substituting this
minimizing pair into~\eqref{eq:Vstar} gives
\[
 V_X=(1+s)I_2,\quad C=\begin{pmatrix}s&-is\\is&s\end{pmatrix},
 \quad\text{\normalfont and}\quad \operatorname{rank}C=1.
\]
A two-dimensional pure ancilla suffices: take $\omega=|0\rangle\langle0|$,
where $|0\rangle$ is the $+1$ eigenstate of $\sigma_z$, and set
$A_1=\sqrt s\sigma_x$ and $A_2=-\sqrt s\sigma_y$.
The ancillary means vanish, both ancillary variances equal $s$, and
$\Tr\omega A_1A_2=-is$. Thus the product-state expansion of the
enlarged observables $U_i=X_i\otimes\Id+\Id\otimes A_i$ gives
\[
 \Tr\omega A_iA_j=C_{ij}
 \quad\text{and}\quad
 \Tr(\rho_0\otimes\omega)U_iU_j=Z_{ij}+C_{ij}=(1+s)\delta_{ij}.
\]
The eigenvalues of $C$ are zero and $2s$, so a single ancillary
direction orthogonal to its reference vector realizes the required
Gram matrix. The ancillary covariance cancels the imaginary ordered
covariance and adds the real variance needed to produce $V_X$, while
preserving the parameter response.
This is cancellation in expectation: the operator commutator remains
$[U_1,U_2]=2i(\sigma_z\otimes\Id-s\Id\otimes\sigma_z)$.
The enlarged one-copy dimension is four, $\ell_U=1+\sqrt s$ and
$v_\eta=1+s+\eta^2$. The evaluable projected-risk bound is
\[
 2+2s+2\eta^2+4r^2\varepsilon_n(r,\eta,R_{\rm out}),
 \quad r/\sqrt n\le a<\sqrt{1-s^2}.
\]
At $s=1/2$, $\norm{[U_1,U_2]}_\op=3$ although its reference expectation vanishes, so normalization still uses the ordered kernel argument.
The same construction applies at $s=0$: $\rho_0=\Id/2$ is faithful with repeated eigenvalues, $\mathbb{I}=I_2$, $D_1=1,D_2=0$, $Z=V_X=I_2$ and $C=0$. No ancillary degree of freedom is needed. The Pauli observables still fail to commute, with commutator norm two. These values verify the density bounds~\eqref{eq:trace-regularity} and local unbiasedness constraints~\eqref{eq:local-unbiased-observables}, with faithful signal state and positive SLD information, at a repeated positive spectrum.

The exact output means exhibit the finite-sample effect of the measurement
order. Let $S_{\alpha,n}$ and $T_{\alpha,n}$ be the
$n^{-1/2}$-normalized sums of $\sigma_\alpha$, for $\alpha=x,y$, on the
signal and ancillary factors, respectively; identities act on all
other factors. Then $B_{n,1}=S_{x,n}+\sqrt s\,T_{x,n}$ and
$B_{n,2}=S_{y,n}-\sqrt s\,T_{y,n}$. For the ordered kernel
$K_y=g_{\eta,y_2}(B_{n,2})g_{\eta,y_1}(B_{n,1})$ in the product state
$\bar\rho_{n,h}=(\rho_{h/\sqrt n}\otimes\omega)^{\otimes n}$, integration
of the second outcome first gives
\[
 \E_hY_1=\Tr\bar\rho_{n,h}B_{n,1}
 \quad\text{and}\quad
 \E_hY_2=\Tr\bar\rho_{n,h}\mathcal D_\eta(B_{n,2}),
\]
where $\mathcal D_\eta(A)=\int g_{\eta,y}(B_{n,1})A
g_{\eta,y}(B_{n,1})\,dy$ describes the disturbance from the first
measurement. These identities use the scalar Gaussian moments:
spectral calculus gives $\int g_{\eta,y}(B)^2\,dy=\Id$ and
$\int y g_{\eta,y}(B)^2\,dy=B$. In an eigenbasis of $B_{n,1}$ with
eigenvalues $\lambda_a$, completing the square gives
\[
 \int_{\R}g_{\eta,y}(\lambda_a)g_{\eta,y}(\lambda_b)\,dy
 =e^{-(\lambda_a-\lambda_b)^2/(8\eta^2)}.
\]
The integrand is this factor times the density of
$N((\lambda_a+\lambda_b)/2,\eta^2)$, so the map multiplies entry
$(a,b)$ of $A$ by the displayed factor. Define the commutator map by
$\operatorname{ad}_B(A)=BA-AB$. Since its square multiplies that entry
by $(\lambda_a-\lambda_b)^2$, the dephasing map has the representation
$\mathcal D_\eta=\exp\{-\operatorname{ad}_{B_{n,1}}^2/(8\eta^2)\}$.
The Pauli commutators give
\[
 \operatorname{ad}_{B_{n,1}}^2(S_{y,n})=\frac4nS_{y,n}
 \quad\text{and}\quad \operatorname{ad}_{B_{n,1}}^2(T_{y,n})=\frac{4s}{n}T_{y,n}.
\]
At $\theta=h/\sqrt n$, the signal means are
$\Tr\bar\rho_{n,h}S_{x,n}=\sqrt n\,\theta_1=h_1$ and
$\Tr\bar\rho_{n,h}S_{y,n}=\sqrt n\,\theta_2=h_2$.
Both ancillary transverse means vanish, so the first-moment identities
above give
\begin{equation}\label{eq:qubit-output-means}
 \E_hY_1=h_1
 \quad\text{and}\quad \E_hY_2=e^{-1/(2n\eta^2)}h_2.
\end{equation}
The first mean is unchanged because the later measurement integrates
to the identity. The second mean reflects the first measurement through
$\mathcal D_\eta$, even though both input observables were calibrated.
Thus calibration of the input observables does not make the ordered
readout exactly unbiased at finite $n$. The output-law comparison in
Theorem~\ref{thm:collective} includes this discrepancy. Second moments
and tail bounds for the same instrument are given in
Lemma~\ref{lem:instrument-moments} of Appendix~\ref{app:instrument-moments}.

Table~\ref{tab:collective-output-main} illustrates the statistical benefit
of combining the explicit collective measurement of
Theorem~\ref{thm:collective} with the radius-dependent shrinkage estimator
of Corollary~\ref{cor:collective-shrinkage}: shrinkage reduces the estimated
risk relative to projection alone at every displayed point.
For example, at $n=3$ and $h=(.35,-.2)$, shrinkage lowers the estimated
risk from $.335$ to $.1582$, a reduction of about $53\%$ using the
same measurement output.
The calculations use $s=.5$, $r=.5$, $\eta=.7$ and $W=I_2$, with local
parameters at the center, in the interior and on the boundary of the disk.
The covariance calculation above makes the tuning explicit:
\[
 V_X=1.5I_2
 \quad\text{and}\quad V_X+\eta^2I_2=1.99I_2.
\]
The Gaussian comparison law is therefore $N(h,1.99I_2)$. Its variance
in each coordinate consists of one unit from the signal, $.5$ from the
ancilla and $.49$ from smoothing. These contributions serve distinct
roles: the ancillary moments produce the real covariance target, while
the Gaussian resolution controls the measurement smoothing. For the
shrinkage calculation in Corollary~\ref{cor:collective-shrinkage}, this gives
\[
 A_\eta=\Tr(1.99I_2)=3.98,\quad B_r=r^2=.25
 \quad\text{and}\quad c_*=\frac{.25}{3.98+.25}=\frac{.25}{4.23}.
\]
Here $A_\eta$ measures the total Gaussian fluctuation and $B_r$ bounds
the squared local displacement. Since the radius is small relative to
the fluctuations, the coefficient places substantial weight on the
nominal center. Under the Gaussian comparison law, multiplication by
$c$ changes the covariance to $c^2(1.99I_2)$ and the bias to $(c-1)h$.
Bounding the squared bias by $(1-c)^2B_r$ gives the quadratic
risk bound minimized by $c_*$.

\begin{table}[t]
\caption{Gaussian approximation and the effect of shrinkage for the
constructed collective measurement at the center, an interior point and
a boundary point, with $s=.5$, $r=.5$, $\eta=.7$ and $W=I_2$.
TV and the projected and shrunk squared-error risks are pointwise
estimates from the 300-node quadrature grid. The zero risk is $\norm h^2$.
Risks are in local coordinates; physical-parameter risks are divided by $n$.
Smaller TV errors and risks are better; bold marks the smallest reported risk in each row, including the zero baseline.}
\label{tab:collective-output-main}
\centering\small\setlength{\tabcolsep}{4pt}
\begin{tabular}{rcrrrr}
$n$&$h$&TV estimate&Projected risk&Shrunk risk&Zero risk\\\hline
1 & $(0,0)$ & 0.1004 & 0.240 & 0.0128 & \textbf{0.0000}\\
1 & $(0.35,-0.2)$ & 0.1069 & 0.335 & \textbf{0.1591} & 0.1625\\
1 & $(0.5,0)$ & 0.1060 & 0.368 & \textbf{0.2333} & 0.2500\\
2 & $(0,0)$ & 0.0391 & 0.242 & 0.0130 & \textbf{0.0000}\\
2 & $(0.35,-0.2)$ & 0.0465 & 0.336 & \textbf{0.1585} & 0.1625\\
2 & $(0.5,0)$ & 0.0452 & 0.376 & \textbf{0.2339} & 0.2500\\
3 & $(0,0)$ & 0.0278 & 0.242 & 0.0132 & \textbf{0.0000}\\
3 & $(0.35,-0.2)$ & 0.0333 & 0.335 & \textbf{0.1582} & 0.1625\\
3 & $(0.5,0)$ & 0.0316 & 0.378 & \textbf{0.2342} & 0.2500\\
\end{tabular}
\end{table}

Write $\Pi_r=\Pi_r^{I_2}$ for Euclidean projection onto the local disk.
The projected and shrunk risk columns evaluate $\Pi_r(Y)$ and
$\Pi_r(c_*Y)$, respectively, before the deterministic selection rule;
the zero column gives $\norm h^2$.
Projection uses the known radius to keep the reported displacement in
the parameter set, and shrinkage first reduces the size of the noisy
readout. The reported risks are in local coordinates: the corresponding
physical estimates are $\widehat\theta=\Pi_r(Y)/\sqrt n$ and
$\widehat\theta=\Pi_r(c_*Y)/\sqrt n$, whose squared-error risks equal
the displayed values divided by $n$.

As $n$ increases from one to three, the estimated TV discrepancy decreases
at all three parameter points, from about $.10$ to about $.03$.
At the interior point $h=(.35,-.2)$, shrinkage more than halves the
projected risk, from about $.335$ to about $.159$; at the boundary
$h=(.5,0)$, it reduces the risk from $.368$--$.378$ to about $.234$.
The zero column supplies a pointwise comparison: shrinking improves
on zero at the displayed interior and boundary points, while zero has
zero risk at the center. These quadrature evaluations connect the
constructed output law to its Gaussian approximation and show how
radius-dependent shrinkage improves estimation from that output.
Appendix~\ref{app:instrument-moments} gives the integration method.

Table~\ref{tab:collective-budget-main} makes the paper's finite-sample
contribution quantitative: the constructed measurement admits computable
bounds on both its output approximation and its worst-ball estimation
risk. For example, at $n=10^{16}$ and $r=1$, the selected rule has a
uniform risk bound of $.7954$, about $20.5\%$ below the zero estimator's
worst-ball risk $1$.
Each TV bound is uniform over $\norm h\le r$ by
Theorem~\ref{thm:collective}; the unshrunk and selected risk bounds follow
from~\eqref{eq:collective-risk} and~\eqref{eq:collective-selected-risk}.
The same TV column bounds the coverage and power approximation errors
in Corollary~\ref{cor:collective-inference}.
Here $A_\eta=3+2\eta^2$ and $B_r=r^2$, with row-specific tuning given in
the caption. At the two smaller sample sizes the
selection rule chooses zero. At $n=10^{20},r=4$, the selected risk bound
is $2.613$, compared with the zero baseline $16$.
Thus the bounds give explicit conditions under which using the measured
output improves the guaranteed risk over reporting the nominal center.
These large sample sizes are inputs to deterministic evaluations of the
analytic inequalities, and show the conservativeness of the sufficient
bounds. Table~\ref{tab:collective-output-main} directly evaluates the
output at small $n$, while Table~\ref{tab:collective-budget-main}
quantifies guarantees uniform over the local disk.

\begin{table}[t]
\caption{Uniform analytic bounds for the collective procedure, with
$s=.5$, $W=I_2$, $D_1=1$, $D_2=0$ and $a=.8$.
The bounds for TV, unshrunk risk and selected risk use~\eqref{eq:collective-TV-budget},
\eqref{eq:collective-risk} and~\eqref{eq:collective-selected-risk}, respectively.
Risks use local coordinates; the zero baseline is $r^2$.
Bounds are rounded upward to four significant figures. The tuning pairs
$(\eta,R_{\rm out})$ are $(1.420892,1.05)$, $(1.095466,5.749596)$,
$(.209779,6.090968)$ and $(.192499,10.068808)$ in row order.
Smaller TV and risk bounds are better; bold marks the smallest risk bound or baseline in each row, including ties.}
\label{tab:collective-budget-main}
\centering\small\setlength{\tabcolsep}{5pt}
\begin{tabular}{rrrrrr}
&&TV&Unshrunk&Selected&Zero\\
$n$&$r$&bound&risk bound&risk bound&baseline\\\hline
100 & 1 & 1 & 11.04 & \textbf{1} & \textbf{1}\\
$10^{6}$ & 1 & 0.3716 & 6.887 & \textbf{1} & \textbf{1}\\
$10^{16}$ & 1 & 0.01001 & 3.129 & \textbf{0.7954} & 1\\
$10^{20}$ & 4 & $5.231\,10^{-4}$ & 3.108 & \textbf{2.613} & 16\\
\end{tabular}
\end{table}
\FloatBarrier

\section{Conclusion}\label{sec:conclusion}
This paper gives a constructive route from calibrated quantum observables
to finite-sample statistical guarantees. A finite ancillary system makes the
comparison covariance real, an ordered sequence of Gaussian instruments
defines an exactly normalized collective measurement, and a local
fluctuation estimate controls its output law. The resulting total variation
bound applies uniformly on the prescribed local parameter ball and supports
confidence coverage, testing power, projected quadratic risk, and
radius-dependent shrinkage.

The construction retains the quantities needed to evaluate a procedure:
density derivative bounds, observable norms, smoothing, local radius, and
the gap of the chosen feasible Holevo solution. These quantities connect
measurement design to statistical performance at a specified sample size.
The qubit illustration further separates covariance cancellation from
operator commutation and exhibits the finite-sample effect of measurement
order. Exact instrument moments and numerical output calculations make
these features directly assessable.

The main argument requires two density derivatives and applies to faithful
finite-dimensional models near a known nominal center, including repeated
positive spectra. The accompanying developments provide detailed tools for
specified measured models, pilot-based inference, finite-radius risk
comparison under separate channels, and likelihood-surrogate transfer.
Together they make the assumptions and intermediate calculations available
within one self-contained treatment and establish a general framework for
connecting collective quantum observation to classical inference with
quantitative finite-sample error bounds.

\begin{appendix}
\clearpage
\section{Background for quantum and classical inference}\label{app:background}
This appendix supplements the background in Section~\ref{sec:setup}
and the covariance and measurement construction in
Section~\ref{sec:collective}. State definitions, faithfulness, spectral
calculus and measurement means are developed in
Section~\ref{sec:distances}; their local regularity and score
interpretations are given in Section~\ref{sec:likelihood-setup}.
Sections~\ref{app:bg-unbiased} and~\ref{app:bg-moments} provide
additional calibration examples, componentwise calculations and
product-state moment identities. Section~\ref{app:bg-holevo} gives
the full expansion proving the covariance inequality.
Sections~\ref{app:bg-local} and~\ref{app:bg-qlan} develop quantum
likelihoods and Gaussian shifts. Section~\ref{app:bg-operations}
proves the distance-transfer inequalities and explains channel
comparison. Sections~\ref{app:bg-measured-primer} and~\ref{app:bg-risk-primer}
introduce measured likelihood, pilot localization and finite-radius
risk for the supporting results. The underlying estimation framework
is standard; see \citet{Holevo2011} and Appendix~B.1 of \citet{YFGSupp}.

\subsection{What local unbiasedness calibrates}\label{app:bg-unbiased}
Use the faithful reference state and mean functions
$\mu_i(\theta)=\Tr\rho_\theta X_i$ from Section~\ref{sec:distances},
with the observables held fixed as $\theta$ varies. The calibration
conditions~\eqref{eq:intro-calibration} can be written as
\begin{equation}\label{eq:bg-local-unbiased}
 \mu_i(\theta_0)=0
 \quad\text{and}\quad
 \left.\partial_j\mu_i(\theta)\right|_{\theta=\theta_0}
 =\delta_{ij},\quad 1\le i,j\le p.
\end{equation}
The following calculations give componentwise examples of the local-mean
and estimator interpretations established in Section~\ref{sec:distances}.

For a twice continuously differentiable ($C^2$) density map, the
centering condition $\mu_i(\theta_0)=0$ and the derivative calibration
$\partial_j\mu_i(\theta_0)=\delta_{ij}$ in~\eqref{eq:bg-local-unbiased}
make the mean track the displacement $\theta_i-\theta_{0,i}$ locally.
For a scalar or vector remainder, $O(\norm u^2)$ means that its
absolute value or Euclidean norm is bounded by $C\norm u^2$ near
zero, with a constant $C$ independent of $u$. Taylor expansion at
$\theta_0$ gives
\begin{align*}
 \mu_i(\theta)
 &=\mu_i(\theta_0)+\sum_{j=1}^p\partial_j\mu_i(\theta_0)
       (\theta_j-\theta_{0,j})+O\left(\norm{\theta-\theta_0}^2\right)\\
 &=\sum_{j=1}^p\delta_{ij}(\theta_j-\theta_{0,j})
       +O\left(\norm{\theta-\theta_0}^2\right)
 =\theta_i-\theta_{0,i}+O\left(\norm{\theta-\theta_0}^2\right).
\end{align*}
Centering removes the constant term, and the Kronecker delta retains
only the displacement in coordinate $i$ in the linear term.

For the two-parameter example, let $\theta=(\theta_1,\theta_2)^T$
and suppose the spectral outcome of $X_1$ estimates
$\theta_1-\theta_{0,1}$. Its mean
$\mu_1(\theta)=\Tr\rho_\theta X_1$ should therefore be approximately
$\theta_1-\theta_{0,1}$ near $\theta_0$. In addition to
$\mu_1(\theta_0)=0$, the derivative conditions are
\[
 \left.\frac{\partial\mu_1}{\partial\theta_1}\right|_{\theta=\theta_0}=1
 \quad\text{\normalfont and}\quad
 \left.\frac{\partial\mu_1}{\partial\theta_2}\right|_{\theta=\theta_0}=0.
\]
Substituting these two slopes and the zero mean into the
Taylor expansion shows their separate roles:
\begin{align*}
 \mu_1(\theta)
 &=0+1\cdot(\theta_1-\theta_{0,1})
       +0\cdot(\theta_2-\theta_{0,2})
       +O\left(\norm{\theta-\theta_0}^2\right)\\
 &=\theta_1-\theta_{0,1}+O\left(\norm{\theta-\theta_0}^2\right).
\end{align*}
For a small real perturbation $\varepsilon$, first hold
$\theta_2=\theta_{0,2}$ fixed and set
$\theta_1=\theta_{0,1}+\varepsilon$. Then, as $\varepsilon\to0$,
\[
 \mu_1\bigl(\theta_0+(\varepsilon,0)^T\bigr)
 =\varepsilon+O(\varepsilon^2).
\]
Thus increasing the first parameter by $\varepsilon$ changes the
expected measurement outcome by approximately $\varepsilon$.
Now hold $\theta_1=\theta_{0,1}$ fixed and set
$\theta_2=\theta_{0,2}+\varepsilon$. The displacement being estimated
is still zero, and
\[
 \mu_1\bigl(\theta_0+(0,\varepsilon)^T\bigr)=O(\varepsilon^2).
\]
The mean therefore remains approximately zero. The off-diagonal
condition $\partial_2\mu_1(\theta_0)=0$ prevents a change in the
second parameter from producing a first-order mean response that
would be interpreted as a displacement in the first parameter.

The vector $\mu(\theta)$ collects the separate means of
$X_1,\ldots,X_p$.
To connect the observable calibration to the usual definition of a
locally unbiased estimator, first fix one coordinate $i$. Measure this same
$X_i$ on each of $n$ independently prepared copies of $\rho_\theta$,
obtaining real outcomes $Y_{i,1},\ldots,Y_{i,n}$ with
$\E_\theta Y_{i,k}=\Tr\rho_\theta X_i=\mu_i(\theta)$ for every $k$.
Their average estimates the displacement $\theta_i-\theta_{0,i}$,
so adding the known nominal coordinate gives the estimator
\[
 \widehat\theta_i=\theta_{0,i}+\frac1n\sum_{k=1}^nY_{i,k}.
\]
Linearity of expectation gives the link to the mean function above:
\[
 \E_\theta\widehat\theta_i
 =\theta_{0,i}+\frac1n\sum_{k=1}^n\E_\theta Y_{i,k}
 =\theta_{0,i}+\mu_i(\theta).
\]
Applying the centering and derivative conditions
in~\eqref{eq:bg-local-unbiased} therefore yields
\[
 \E_{\theta_0}\widehat\theta_i
 =\theta_{0,i}+\mu_i(\theta_0)=\theta_{0,i}\quad
 \text{\normalfont and}\quad
 \left.\partial_j\E_\theta\widehat\theta_i\right|_{\theta=\theta_0}
 =\partial_j\mu_i(\theta_0)=\delta_{ij}.
\]
Here $j$ indexes the physical parameter being varied, while the
measured observable $X_i$ remains fixed. Thus the derivative identity
for this one estimator is exactly row $i$ of $D\mu(\theta_0)=I_p$.
The Taylor expansion above now gives the local bias directly:
\[
 \E_\theta\widehat\theta_i-\theta_i
 =\mu_i(\theta)-(\theta_i-\theta_{0,i})
 =O\left(\norm{\theta-\theta_0}^2\right).
\]
This is a local statement as $\theta\to\theta_0$; the estimator can
still be biased away from that point.

Applying this scalar calculation for every $i$ does not, by itself,
specify a measurement that produces all coordinates from the same
$n$ copies. For example,
when $n\ge p$, separate measurements can use fixed disjoint groups
of sizes $n_i\ge1$, with $\sum_i n_i=n$, measuring $X_i$ on group
$i$ and using $n_i$ in its sample average. The resulting vector
estimator has mean $\theta_0+\mu(\theta)$ and hence is locally
unbiased in all coordinates; its covariance depends on the
allocation.

For a specified finite POVM $M=(M_y)$ and fixed reports
$x_i(y)=\widehat\theta_i(y)-\theta_{0,i}$, recall the first-moment
identity from Section~\ref{sec:distances}:
\begin{equation}\label{eq:bg-povm-moments}
 X_i=\sum_y x_i(y)M_y
 \quad\text{and}\quad
 \E_\theta\left[\widehat\theta_i(Y)-\theta_{0,i}\right]
 =\Tr\rho_\theta X_i.
\end{equation}
Keeping the measurement and reporting rule fixed, evaluating and
differentiating~\eqref{eq:bg-povm-moments} gives
\[
 \E_{\theta_0}\widehat\theta_i(Y)-\theta_{0,i}
 =\Tr\rho X_i\quad
 \text{\normalfont and}\quad
 \left.\partial_j\E_\theta\widehat\theta_i(Y)\right|_{\theta=\theta_0}
 =\Tr(\partial_j\rho_{\theta_0})X_i.
\]
Local unbiasedness at $\theta_0$ means that the bias of each reported
coordinate is zero at $\theta_0$ and has zero first derivatives there.
The zero-bias requirement at $\theta_0$ says that the average reported coordinate
equals the true coordinate at the nominal point:
\[
 \E_{\theta_0}\widehat\theta_i(Y)=\theta_{0,i}
 \quad\Longleftrightarrow\quad
 \E_{\theta_0}\widehat\theta_i(Y)-\theta_{0,i}=0.
\]
For the derivative requirement, the target is the coordinate function
$g_i(\theta)=\theta_i$. Varying $\theta_j$ while holding the other
coordinates fixed changes $g_i$ with slope one if $j=i$ and zero
otherwise, so $\partial_jg_i(\theta)=\delta_{ij}$. In contrast,
$\theta_{0,i}$ is a fixed nominal value. Differentiating
the bias therefore gives
\[
 \left.\partial_j\bigl\{\E_\theta\widehat\theta_i(Y)-\theta_i\bigr\}
 \right|_{\theta=\theta_0}
 =\left.\partial_j\E_\theta\widehat\theta_i(Y)\right|_{\theta=\theta_0}
 -\delta_{ij}.
\]
Requiring this derivative of the bias to be zero makes the derivative
of the mean equal to $\delta_{ij}$. Thus the mean of coordinate $i$
must respond with slope one when $\theta_i$ changes and slope zero
when another coordinate $\theta_j$, $j\ne i$, changes. The zero-bias and derivative requirements apply to the expected report.
Substituting $\E_{\theta_0}[\widehat\theta_i(Y)-\theta_{0,i}]=0$ into~\eqref{eq:bg-povm-moments} and using $\left.\partial_j\E_\theta\widehat\theta_i(Y)\right|_{\theta=\theta_0}=\delta_{ij}$ in its derivative gives
\[
 \Tr\rho X_i=0
 \quad\text{\normalfont and}\quad
 \Tr(\partial_j\rho_{\theta_0})X_i=\delta_{ij},
 \quad 1\le i,j\le p.
\]
With $\mu_i(\theta)=\Tr\rho_\theta X_i$, these are exactly the
centering condition $\mu_i(\theta_0)=0$ and the derivative calibration
$\partial_j\mu_i(\theta_0)=\delta_{ij}$ in~\eqref{eq:bg-local-unbiased}.
Conversely, a \emph{feasible} list, satisfying~\eqref{eq:bg-local-unbiased},
only specifies the required means and their local
derivatives. It does not specify a measurement producing all
coordinates with a desired covariance. Constructing such a measurement
is a separate part of
Section~\ref{sec:collective}.

Write $\dot\ell(y)=s(y)$ for the reference score of
Section~\ref{sec:likelihood-setup}, evaluated at $\theta_0$ and held
fixed when taking expectations under nearby parameters.
Since the probabilities sum to one, the
score has mean zero:
\[
 \E_{\theta_0}\dot\ell_j(Y)
 =\sum_y p_{\theta_0}(y)\dot\ell_j(y)
 =\left.\partial_j\sum_y p_\theta(y)\right|_{\theta=\theta_0}=0,
\]
where the final equality follows because $\sum_y p_\theta(y)=1$ for every $\theta$.
Consequently, the classical Fisher information
$F=F(\theta_0;M)$ is the covariance matrix of the score:
\[
 F=\operatorname{Cov}_{\theta_0}\left(\dot\ell(Y)\right)
  =\E_{\theta_0}\left[\dot\ell(Y)\dot\ell(Y)^T\right]
 \quad\text{\normalfont and}\quad
 F_{jk}=\sum_y p_{\theta_0}(y)\dot\ell_j(y)\dot\ell_k(y).
\]
In particular, $F_{jj}$ is the variance of score coordinate $j$,
and $F_{jk}$ is the covariance of coordinates $j$ and $k$.

The score also converts derivatives of expectations into ordinary
expectations. For any fixed real statistic $g(y)$, differentiating
the finite sum gives
\[
 \left.\partial_j\E_\theta g(Y)\right|_{\theta=\theta_0}
 =\sum_y g(y)\left.\partial_jp_\theta(y)\right|_{\theta=\theta_0}=\E_{\theta_0}\left[g(Y)\dot\ell_j(Y)\right].
\]
For densities on a fixed support, the same calculation uses
integrals, provided differentiation under the integral is justified
and the displayed expectations are finite. In the finite-outcome
setting here, no interchange with an infinite sum is needed.
If $F$ is positive definite, choose the reported displacement
\[
 x_i(y)=\sum_{k=1}^p(F^{-1})_{ik}\dot\ell_k(y).
\]
The zero score mean gives
\[
 \E_{\theta_0}x_i(Y)
 =\sum_k(F^{-1})_{ik}\E_{\theta_0}\dot\ell_k(Y)=0.
\]
To verify the derivative condition explicitly, hold $F$ and the
reference score values fixed as $\theta$ varies:
\begin{align*}
 \left.\partial_j\E_\theta x_i(Y)\right|_{\theta=\theta_0}
 &=\sum_y x_i(y)
   \left.\partial_jp_\theta(y)\right|_{\theta=\theta_0}=\E_{\theta_0}\left[x_i(Y)\dot\ell_j(Y)\right]\\
 &=\sum_k(F^{-1})_{ik}
   \E_{\theta_0}\left[\dot\ell_k(Y)\dot\ell_j(Y)\right]=\sum_k(F^{-1})_{ik}F_{kj}\\
 &=(F^{-1}F)_{ij}=(I_p)_{ij}=\delta_{ij}.
\end{align*}
Thus the reporting rule $\widehat\theta_i(y)=\theta_{0,i}+x_i(y)$
is locally unbiased at $\theta_0$. Multiplication by $F^{-1}$ changes
the mean derivative matrix of the fixed score from $F$ to $I_p$.

The SLD equation~\eqref{eq:sld} gives the corresponding matrix
construction. Section~\ref{sec:likelihood-setup} explains its analogy
with the classical score; the following adjoint and trace identities
make the symmetrization explicit.
There is an additional issue for matrices. Since every $\rho_\theta$
is Hermitian, its derivative $\partial_j\rho_{\theta_0}$ is Hermitian
as well. The product $\rho L_j$ need not be Hermitian, because
\[
 (\rho L_j)^*=L_j\rho,
\]
which equals $\rho L_j$ only when the two matrices commute.
The SLD uses their average, whose adjoint is itself:
\[
 \left[\tfrac12(L_j\rho+\rho L_j)\right]^*
 =\tfrac12(\rho L_j+L_j\rho)
 =\tfrac12(L_j\rho+\rho L_j).
\]
This is the meaning of symmetrization here. If $\rho$ and $L_j$
commute, the average reduces to $\rho L_j$, recovering the same
product form as the classical score identity. The connection to a
quantum log likelihood is derived in Section~\ref{app:bg-local}.
Taking the trace of the defining SLD equation gives the zero-mean
identity
\[
 \Tr\rho L_j=\Tr(\partial_j\rho_{\theta_0})
 =\left.\partial_j\Tr\rho_\theta\right|_{\theta=\theta_0}=0,
\]
where the final equality follows because $\Tr\rho_\theta=1$ for every $\theta$, so its derivative with respect to $\theta_j$ is zero.
The derivative identity and the feasibility of
$X_i=\sum_k(\mathbb{I}^{-1})_{ik}L_k$ are verified in
Section~\ref{sec:likelihood-setup}. The measured information $F$ depends
on the chosen POVM, whereas $\mathbb{I}$ describes the quantum state
model before a measurement is specified.

\subsection{Gram representation and product-state moments}\label{app:bg-moments}
For the centered observables of Section~\ref{sec:holevo-ancilla},
recall $Z_{ij}=\Tr\rho X_iX_j$ and write $G=\Rea Z$, $\Gamma=\Ima Z$.
The decomposition and expected-commutator interpretation are given
in~\eqref{eq:main-moment-parts}. If the observables commute, their
common spectral measurement has ordinary real covariance $Z$.
An explicit HS representation of the operator moments is
\[
 \langle X_i\sqrt\rho,X_j\sqrt\rho\rangle_{\HS}
 =\Tr\bigl(\sqrt\rho X_iX_j\sqrt\rho\bigr)=Z_{ij}.
\]
Thus $Z$ records the pairwise inner products of the matrices
$X_i\sqrt\rho$. For observables with nonzero means, the covariance
Gram matrix instead uses $X_i-(\Tr\rho X_i)\Id$ in place of $X_i$.

The normalized sums $X_i^{[n]}=n^{-1/2}\sum_{k=1}^nX_i^{(k)}$
from Section~\ref{sec:distances} preserve the full ordered covariance.
At the reference state,
centering makes all cross-copy terms vanish, giving
\[
 \Tr\rho^{\otimes n}X_i^{[n]}X_j^{[n]}=Z_{ij}.
\]
Recall that at the local parameter $h=\sqrt n(\theta-\theta_0)$ the state is
$\rho_{n,h}=\rho_{\theta_0+h/\sqrt n}^{\otimes n}$.
Each factor of this product state has trace one. Hence the definition
$\mu_i(\theta)=\Tr\rho_\theta X_i$ gives the exact identity
\[
 \Tr\rho_{n,h}X_i^{[n]}
 =\frac1{\sqrt n}\sum_{k=1}^n\Tr\rho_{\theta_0+h/\sqrt n}X_i
 =\sqrt n\,\mu_i(\theta_0+h/\sqrt n).
\]
The centering condition $\mu_i(\theta_0)=0$ and derivative calibration
$\partial_j\mu_i(\theta_0)=\delta_{ij}$ in~\eqref{eq:bg-local-unbiased}
give, by Taylor expansion,
\[
 \mu_i(\theta_0+h/\sqrt n)
 =\mu_i(\theta_0)
   +\frac1{\sqrt n}\sum_{j=1}^p\partial_j\mu_i(\theta_0)h_j
   +O\left(\norm h^2/n\right)=\frac{h_i}{\sqrt n}+O\left(\norm h^2/n\right).
\]
Multiplying by $\sqrt n$ therefore yields
\[
 \Tr\rho_{n,h}X_i^{[n]}
 =\sqrt n\left\{\frac{h_i}{\sqrt n}+O\left(\norm h^2/n\right)\right\}
 =h_i+O\left(\norm h^2/\sqrt n\right).
\]
For fixed observables and a $C^2$ model, the remainder is uniform for
$h$ in a fixed local ball, provided $\theta_0+h/\sqrt n$ stays in the
physical neighborhood where the second derivatives are bounded. This
explains the order-one mean $h$ and covariance scale used for local
inference. The list $X^{[n]}$ still need not be jointly measurable.
Section~\ref{app:bg-local} develops the product-state likelihood
and its quadratic approximation on this local scale.

\subsection{Proof of the covariance inequality for reported estimates}\label{app:bg-holevo}
Section~\ref{sec:holevo-ancilla} relates weighted risk to covariance
and bias. Here let $V=\operatorname{Cov}_{\theta_0}(\widehat\theta)$
for a reported estimator with finite second moments that is locally
unbiased at $\theta_0$. We give the full operator expansion underlying
the covariance constraint used there.

We next compare the covariance $V$ of the reported estimator with
the operator moment matrix $Z$. Recall that
$x_i(y)=\widehat\theta_i(y)-\theta_{0,i}$,
$X_i=\sum_y x_i(y)M_y$ and $Z_{ij}=\Tr\rho X_iX_j$, where
$\rho=\rho_{\theta_0}$. Local unbiasedness implies
$\E_{\theta_0}x_i(Y)=0$, so the covariance is the expectation of
the product of these centered reports. Using the outcome probability
$\Prb_{\theta_0}\{Y=y\}=\Tr\rho M_y$ gives
\[
 V_{ij}
 =\E_{\theta_0}[x_i(Y)x_j(Y)]=\sum_y x_i(y)x_j(y)\Prb_{\theta_0}\{Y=y\}
 =\sum_y x_i(y)x_j(y)\Tr\rho M_y.
\]
The matrix $Z$ is a lower bound for $V$ in the positive semidefinite
order:
\begin{equation}\label{eq:bg-covariance-constraint}
 V-Z\succeq0.
\end{equation}
This means $c^*Vc\ge c^*Zc$ for every $c\in\mathbb C^p$; it is
an inequality between quadratic forms, not between individual
matrix entries. Both matrices are Hermitian, although $Z$ can have
complex entries. To prove $V-Z\succeq0$ in~\eqref{eq:bg-covariance-constraint}, or equivalently $c^*(V-Z)c\ge0$ for every $c$, fix $c\in\mathbb C^p$ and put
$b_y=\sum_i c_ix_i(y)$ and $B_c=\sum_i c_iX_i=\sum_y b_yM_y$.
For each outcome $y$, $b_y\in\mathbb C$ is a scalar: it is a
linear combination of the real reported displacements $x_i(y)$
with complex coefficients $c_i$. In contrast, $B_c$ is a
$d\times d$ matrix, which need not be Hermitian because the
coefficients $c_i$ may be complex.
The numbers $x_i(y)$ are real and the operators $X_i$ are Hermitian,
so the two quadratic forms are
\[
 c^*Vc=\sum_y|b_y|^2\Tr\rho M_y
 \quad\text{\normalfont and}\quad
 c^*Zc=\Tr\rho B_c^*B_c.
\]
Define the operator
$S_c=\sum_y(b_y\Id-B_c)^*M_y(b_y\Id-B_c)$.
Here $b_y\Id$ is the scalar $b_y$ times the $d\times d$ identity
matrix, so each difference $b_y\Id-B_c$ is a matrix.
Expanding its products and using
$\sum_y b_yM_y=B_c$, $\sum_y\overline{b_y}M_y=B_c^*$ and
$\sum_yM_y=\Id$ gives
\begin{align*}
 S_c
 &=\sum_y|b_y|^2M_y
   -\left(\sum_y\overline{b_y}M_y\right)B_c
   -B_c^*\left(\sum_y b_yM_y\right)
   +B_c^*\left(\sum_yM_y\right)B_c\\*
 &=\sum_y|b_y|^2M_y-B_c^*B_c.
\end{align*}
Each summand in $S_c$ is positive semidefinite: for any vector $v$,
its quadratic form is
$((b_y\Id-B_c)v)^*M_y((b_y\Id-B_c)v)\ge0$, since $M_y\succeq0$.
The resulting operator inequality is
\[
 \sum_y|b_y|^2M_y-B_c^*B_c=S_c\succeq0.
\]
It uses only positivity and completeness of the POVM effects. Taking
its expectation in any positive semidefinite state $\rho$ proves
\[
 c^*(V-Z)c
 =\Tr\rho S_c
 =\Tr(\rho^{1/2}S_c\rho^{1/2})\ge0.
\]
This trace step requires neither faithfulness nor commutation of
$\rho$ and $S_c$. Centering the reported values is what identifies
their second-moment matrix with the covariance $V$.
For a real vector $a\in\R^p$, this implies
\[
 \operatorname{Var}_{\theta_0}\!\left(\sum_i a_i x_i(Y)\right)
 =a^TVa\ge a^TGa
 \quad\text{and}\quad G=\Rea Z.
\]
Here $a^TGa$ is the spectral outcome variance of the centered
observable $\sum_i a_iX_i$, as shown in Section~\ref{sec:holevo-ancilla}.
In particular, $V_{ii}\ge Z_{ii}=\Tr\rho X_i^2$.
This explains the lower-bound interpretation: reporting all coordinates
through the specified POVM can add variance beyond their operator
moments. Only unbiasedness at $\theta_0$ is needed for this covariance
comparison; the derivative conditions additionally make the operators
a feasible list. The same argument extends to
integrals when a POVM has continuous outcomes and finite second moments.

The minimization over real covariances is established
in~\eqref{eq:main-holevo-variational}, with optimizer~\eqref{eq:Vstar}.
For a locally unbiased single-copy estimator, its first-moment operators
satisfy~\eqref{eq:local-unbiased-observables}. Combining the covariance
constraint~\eqref{eq:bg-covariance-constraint} with that minimization
and the definition~\eqref{eq:holevo-finite} therefore gives
\[
 \Tr WV\ge\Phi_W(Z(X))\ge C_W.
\]
This is a covariance benchmark for locally unbiased reports. Its
asymptotic attainability with collective measurements is treated under
the regularity and asymptotic unbiasedness conditions of
\citet{YFG2013,FY2023}. The finite-radius procedures and their bias
and approximation costs are analyzed in
Section~\ref{sec:output-inference}.

\subsection{Independent copies, local parameters and likelihoods}
\label{app:bg-local}
A \emph{statistical experiment} specifies the information available
under each possible parameter value. Classically it is a family of
probability laws $\{P_\theta\}$ on an observation space. In the
quantum setting it is a family of states $\{\rho_\theta\}$ on a
specified system; choosing a measurement then produces classical
outcomes. Comparing experiments asks whether procedures based on
one family can be reproduced, exactly or approximately, from the other.

Independent preparations of the same state are represented by the
tensor product $\rho_\theta^{\otimes n}$ on
$\mathcal H^{\otimes n}$, a space of dimension $d^n$.
The operator $A^{(k)}$ from Section~\ref{sec:setup} acts as $A$ on copy $k$ and as the identity
on the other copies. Observables on distinct copies commute, and
their expectations factor under a product state. Thus repeated
measurements of a fixed observable have the usual independent
classical outcomes. A collective measurement instead has effects
on the entire tensor product and need not be a product of separate
measurements.
Section~\ref{sec:distances} derives the local scale
$\theta=\theta_0+h/\sqrt n$ and the corresponding factor $n^{-1}$
for physical quadratic risk. The direct collective construction uses a
known center; estimating it introduces the localization problem of
Appendix~\ref{sec:adaptive}.

A pilot estimate uses an initial group of copies to choose the
measurement for fresh copies. Conditional on the pilot record,
the selected measurement is fixed and the fresh copies still have
their product state at the true parameter. One can therefore analyze
that fixed procedure, then average over the pilot record and account
for the probability of poor localization. Derivatives of the
conditional likelihood hold the selected measurement fixed.

Before a measurement is chosen, a quantum likelihood must compare
states rather than probabilities of observed outcomes. For faithful
states $\rho$ and $\rho_{\theta_0+u}$, the positive square-root
likelihood ratio is the matrix
\[
 R=R(\rho_{\theta_0+u}\mid\rho)
 =\rho^{-1/2}\bigl(\rho^{1/2}\rho_{\theta_0+u}\rho^{1/2}\bigr)^{1/2}
   \rho^{-1/2}.
\]
It is the unique positive solution of $\rho_{\theta_0+u}=R\rho R$:
conjugating $\rho_{\theta_0+u}=R\rho R$ by $\rho^{1/2}$ identifies
$\rho^{1/2}R\rho^{1/2}$ with the unique positive square root in
the formula. Define the Hermitian log-likelihood ratio
$\mathcal L(u)=2\log R$, so that
\[
 \rho_{\theta_0+u}=e^{\mathcal L(u)/2}\rho e^{\mathcal L(u)/2}
 \quad\text{\normalfont and}\quad
 \Tr\rho e^{\mathcal L(u)}=1.
\]
The normalization identity $\Tr\rho e^{\mathcal L(u)}=1$ follows by taking the trace and is the analogue
of a classical likelihood ratio having mean one. If the states are
diagonal in a common basis, with probabilities $p_\theta(a)>0$,
the diagonal entries of $\mathcal L(u)$ are
$\log\{p_{\theta_0+u}(a)/p_{\theta_0}(a)\}$. For noncommuting
states, $\mathcal L(u)$ generally differs from
$\log\rho_{\theta_0+u}-\log\rho$. It is an operator defined before
measurement; a specified measurement instead yields the scalar
log-likelihood ratio of its observed outcome.

At $u=0$, $R=\Id$ and $\mathcal L(0)=0$. Differentiating the
sandwich identity with respect to $u_j$ at zero shows that
$\partial_{u_j}\mathcal L(0)$ solves the SLD equation~\eqref{eq:sld}. Its uniqueness therefore gives
$\partial_{u_j}\mathcal L(0)=L_j$. If
$H_{ij}=\partial_{u_i}\partial_{u_j}\mathcal L(0)$, differentiating
$\Tr\rho e^{\mathcal L(u)}=1$ twice gives
\[
 0=\Tr\rho H_{ij}
   +\tfrac12\Tr\rho(L_iL_j+L_jL_i)
   =\Tr\rho H_{ij}+\mathbb{I}_{ij}.
\]
This explains the curvature term in the quadratic approximation.
The tensorized log likelihood is the sum of the one-copy log-likelihood ratios,
as in the classical independent case:
\[
 \Lambda_n(h)=2\log R(\rho_{n,h}\mid\rho^{\otimes n})
 =\sum_{k=1}^n\mathcal L(h/\sqrt n)^{(k)}.
\]
The normalized SLD sums from Section~\ref{sec:likelihood-setup},
$\Delta_{n,i}=n^{-1/2}\sum_{k=1}^n L_i^{(k)}$, collected in the
observable list $\Delta_n=(\Delta_{n,1},\ldots,\Delta_{n,p})$, give its
first-order term $h^T\Delta_n$. Likelihood normalization determines
the mean second-order term, giving the score quadratic
\[
 Q_n(h)=h^T\Delta_n-\tfrac12h^T\mathbb{I}h\,\Id.
\]
Here $h^T\Delta_n=\sum_i h_i\Delta_{n,i}$, $\Id$ acts on
$\mathcal H^{\otimes n}$, and $\mathbb{I}_{ij}=\Rea\Tr\rho L_iL_j$ is the SLD information.
Using the state-weighted norm of Section~\ref{sec:distances}, the remainder $\Lambda_n(h)-Q_n(h)$ can be measured
using $\norm A_{2,\rho^{\otimes n}}^2=\Tr\rho^{\otimes n}A^*A$;
Lemma~\ref{lem:quadratic-state-support} in Appendix~\ref{app:polynomial-support}
instead uses an explicit operator Taylor remainder to compare normalized states.
For Hermitian $A$, this squared norm is its reference-state mean
square. It weights directions with small reference-state probabilities
lightly. The operator norm $\norm A_\op=\sup_{\norm v=1}\norm{Av}$
controls every unit vector, so a uniform operator-norm bound is stronger.
State-weighted mean-square control alone does not justify replacing exponential likelihood sandwiches by their approximations. The proof of Lemma~\ref{lem:quadratic-state-support} uses operator remainder bounds and normalized amplitude comparisons, which are stronger inputs.

\subsection{qLAN and Gaussian shifts}
\label{app:bg-qlan}
Classical local asymptotic normality (LAN) combines a quadratic expansion
of the local log likelihood with a Gaussian limit for its normalized
score; see \citet{BickelKlaassenRitovWellner1993} for its role in efficient
parametric and semiparametric estimation. It explains why many large-sample local estimation problems
have a Gaussian shift approximation: the local parameter changes
the Gaussian mean while the covariance stays fixed. For example,
for independent outcomes $Y_1,\ldots,Y_n$ from the finite classical
model of Section~\ref{sec:likelihood-setup}, assume positive probabilities
with three continuous derivatives near $\theta_0$ and $F\succ0$.
Write $S_n^{\rm phys}=n^{-1/2}\sum_k\dot\ell(Y_k)$ for the normalized score
in physical parameter coordinates
and $P_{\theta_0}^{\otimes n}$ for their joint reference law. For fixed
$h\in\R^p$, classical LAN has the form
\[
 \sum_{k=1}^n\log\frac{p_{\theta_0+h/\sqrt n}(Y_k)}{p_{\theta_0}(Y_k)}
 =h^TS_n^{\rm phys}-\tfrac12h^TFh+o_{P_{\theta_0}^{\otimes n}}(1)\quad 
 \text{\normalfont and}\quad
 S_n^{\rm phys}\ \Longrightarrow\ N(0,F).
\]
The symbol $o_P(1)$ denotes convergence to zero in probability under
$P$, and $\Longrightarrow$ denotes convergence in distribution.
The quadratic term follows from the identity that the expected
log-probability Hessian at $\theta_0$ equals $-F$; this identity is
obtained by differentiating $\sum_y p_\theta(y)=1$ twice. The normal
limit follows from the classical central limit theorem.

A related development in classical likelihood theory is rescaled local
asymptotic normality (RLAN), introduced by \citet{NingIonidesRitov2021}. RLAN supplements the usual
$n^{-1/2}$-scale LAN expansion with a cubic log-likelihood approximation
on enlarged $n^{-1/4}$-scale neighborhoods. Under suitable regularity
and Monte Carlo error conditions, this framework supports statistically
efficient estimation from noisy likelihood evaluations. It illustrates
the broader role of local approximation methods in connecting the
structure of a statistical experiment with inferential guarantees.

The qLAN framework extends this principle while
retaining the possible noncommutativity of the limiting observables.
A likelihood formulation therefore requires joint noncommutative fluctuation
limits and a remainder negligible in the specified joint tests,
in addition to a quadratic likelihood expansion
\citep{YFG2013,FY2020,FY2023}. A state-weighted remainder bound is
one quantitative ingredient, not by itself a definition of qLAN.
Here joint tests compare expectations of specified operator expressions,
retaining their multiplication order.

To understand the Gaussian limit, first consider only the SLD
fluctuations $\Delta_n$. Their ordered covariance at the reference
state is the complex matrix from~\eqref{eq:derivatives}, with entries
$J_{ij}=\Tr\rho L_iL_j$. Recall its decomposition $J=\mathbb{I}+i\Omega$, where
$\mathbb{I}=\Rea J$ is the SLD information and
$\Omega_{ij}=\Tr\rho[L_i,L_j]/(2i)$. The matrix $\mathbb{I}$ records
symmetrized variances, whereas $\Omega$ records the expected
commutators. A quantum Gaussian state is the analogue of a
multivariate Gaussian law for a possibly noncommuting list of
observables $Y=(Y_1,\ldots,Y_p)$. Write $\phi_h(A)$ for the
expectation of an observable $A$ in this state. The Gaussian limit
of the SLD fluctuations has mean $\phi_h(Y)=\mathbb{I}h$ and centered
ordered covariance $J$, meaning that
\[
 \phi_h\bigl((Y_i-(\mathbb{I}h)_i\Id)(Y_j-(\mathbb{I}h)_j\Id)\bigr)=J_{ij}.
\]
The notation $\phi_h(Y)$ lists the $p$ expectations $\phi_h(Y_i)$;
these Gaussian $Y_i$ are limiting observables, whereas the $Y_k$
in the preceding classical calculation were recorded data.
Its bounded exponential observables,
called Weyl operators, are
$\mathcal W(v)=e^{iv^TY}$ for $v\in\R^p$ and satisfy
\[
 \phi_h(\mathcal W(v))
 =\exp\bigl\{iv^T\mathbb{I}h-\tfrac12v^T\mathbb{I}v\bigr\}.
\]
Thus each single linear combination $v^TY$ has an ordinary normal
outcome distribution with mean $v^T\mathbb{I}h$ and variance $v^T\mathbb{I}v$.

The distinction from a classical Gaussian vector lies in the product
rule: for real $v,w\in\R^p$,
\[
 \mathcal W(v)\mathcal W(w)
 =e^{-iv^T\Omega w}\mathcal W(v+w).
\]
This phase retains the order of the observables. When $\Omega\ne0$,
the list $Y$ does not represent jointly observed classical Gaussian
coordinates. Directions in the kernel $\ker\Omega=\{v\in\R^p:\Omega v=0\}$ commute with
all the limiting coordinates and form a classical part. Because
$\Omega$ is real antisymmetric, its rank is even; write
$\operatorname{rank}\Omega=2m$. A real change
of coordinates separates $p-2m$ commuting coordinates and $m$
noncommuting pairs. The latter can be represented by oscillator
observables, such as position and momentum, on an infinite-dimensional
Hilbert space. This commutation decomposition does not by itself
assert statistical independence of the two parts. When $\Omega=0$,
the entire limit is an ordinary classical Gaussian family; when
both parts are present it is called a classical--quantum, or hybrid,
Gaussian experiment. The limiting system has its own state space,
separate from the original $\mathbb C^d$.
These conventions specify the means, covariances and multiplication order needed for the Gaussian comparisons in this paper.

The mean $\mathbb{I}h$ follows from the SLD response identity
$\Tr(\partial_j\rho_{\theta_0})L_i=\mathbb{I}_{ij}$.
Under the local state $\rho_{n,h}$, Taylor expansion consequently
gives $\Tr\rho_{n,h}\Delta_{n,i}=(\mathbb{I}h)_i+O(\norm h^2/\sqrt n)$
on a fixed local ball. In contrast, the feasible observables $X_i$
in~\eqref{eq:local-unbiased-observables} have response $\delta_{ij}$,
so their normalized sums have limiting mean $h$. The constructed
measurement produces a classical output approximating
$N(h,V_X+\eta^2I_p)$, with covariance enlarged by measurement
noise and smoothing. These are distinct objects: the Gaussian SLD
observables describe score fluctuations, whereas the classical
normal law describes the recorded estimator output.

The $p$ SLD coordinates need not give a sufficient Gaussian
representation of the entire local experiment. In a general model,
the observable family may have to be enlarged. For faithful
independent-copy models with three continuous density derivatives,
a uniform positive lower eigenvalue bound near $\theta_0$, and
positive definite SLD information, one construction uses the following
real linear map on Hermitian matrices, recalled from
Section~\ref{sec:holevo-ancilla}: $D_\rho(A)$ is the unique Hermitian solution of
\[
 \rho D_\rho(A)+D_\rho(A)\rho=i[\rho,A],\quad A=A^*.
\]
In an orthonormal eigenbasis of $\rho$, with eigenvalues
$\lambda_a>0$, its entries are
$[D_\rho(A)]_{ab}=i(\lambda_a-\lambda_b)A_{ab}/(\lambda_a+\lambda_b)$,
so faithfulness again gives existence and uniqueness. A real span
$\mathcal V$ is invariant under this map if
$D_\rho(\mathcal V)\subseteq\mathcal V$. Starting with the SLDs,
one can take
\[
 \mathcal V=\operatorname{span}_{\R}
 \bigl\{D_\rho^k(L_j):k\ge0,\ 1\le j\le p\bigr\},
\]
where $D_\rho^0$ is the identity map and higher powers denote
repeated application. This span is finite dimensional because the
space of Hermitian $d\times d$ matrices has real dimension $d^2$.
The normalized sums of a basis of this enlarged family give a
$D$-extension in the sense of \citet{FY2023}; see Theorem~C.1 of \citet{FY2023Supp}.
The general definition in \citet{FY2023} concerns sequences of
observables and requires joint Gaussian limits and additional
doubled-test limits, involving operators on both sides of a state
square root. A representative doubled test is
$\Tr\sqrt\sigma\, e^{iA}\sqrt\sigma\,e^{iB}$ for a state $\sigma$
and Hermitian $A,B$; it retains information about both sides of
$\sqrt\sigma$, unlike a single expectation $\Tr\sigma e^{iA}$.
Finite $D_\rho$-invariant closure is a way to obtain the joint Gaussian and doubled-test limits in the independent-copy setting, not the definition
for arbitrary sequences. The score example above therefore explains
Gaussian fluctuations without asserting that
every asymptotic measurement can be represented using those $p$
coordinates alone.

Gaussian approximation also has a nonparametric pure-state theory.
\citet{ButuceaGutaNussbaum2018}
establish local equivalence to a quantum Gaussian white-noise experiment,
and \citet{LahiryNussbaum2022} derive sharp asymptotic
minimax and adaptive estimation results. Their infinite-dimensional
pure-state setting differs from the faithful finite-dimensional signal
models used for the direct finite-sample output bound here.
Theorem~\ref{thm:local-char} controls expectations of ordered products
of exponentials, with multiplication order fixed.
Theorem~\ref{thm:collective} supplies the further
integration and tail estimates for the chosen measurement's output.
Comparison of all measurements requires a separate experiment-level
argument, described next.

\subsection{Measurements, channels and operational comparison}
\label{app:bg-operations}
Sections~\ref{sec:holevo-ancilla} and~\ref{sec:collective-output}
give the ancillary construction and ordered measurement operation.
This subsection supplies the distance-transfer proof and explains
comparison of whole experiments through channels.

For the quantum channels defined in Section~\ref{sec:distances},
complete positivity means
that positivity is preserved even when the map is applied to one
part of a larger system. Appending a known ancilla,
performing a unitary operation, discarding a subsystem and recording
a measurement outcome are examples. A channel used to compare
statistical experiments must not depend on the unknown parameter.
Otherwise it could encode information about that parameter instead
of obtaining it from the input state.
The following lemma records the standard distance and bounded-loss
inequalities used throughout the paper.
\begin{lemma}[Distance and bounded-loss transfer]\label{lem:decision-transfer}
Channels contract $d_{\rm tr}$, and every POVM satisfies
$\TV(P_\rho^M,P_\sigma^M)\le d_{\rm tr}(\rho,\sigma)$.
If $0\le\ell\le B$, then $|\E_P\ell-\E_Q\ell|\le B\TV(P,Q)$.
For probability densities $p,q$ and a measurable set $A$,
\begin{equation}\label{eq:compact-TV}
 \TV(P,Q)\le\int_A|p-q|+Q(A^c).
\end{equation}
\end{lemma}
\begin{proof}
For traceless Hermitian $H=H_+-H_-$, its positive and negative parts
satisfy $\Tr H_+=\Tr H_-=\norm H_1/2$. For every effect $0\preceq E\preceq\Id$,
$|\Tr HE|\le\norm H_1/2$, with equality for the support projection of
$H_+$. Thus $\norm H_1/2=\sup_{0\preceq E\preceq\Id}|\Tr HE|$.
The channel adjoint is positive and unital, so it sends effects to effects.
For $H=\rho-\sigma$ this gives
\[
 d_{\rm tr}(\mathcal T(\rho),\mathcal T(\sigma))
 =\sup_{0\preceq E\preceq\Id}\bigl|\Tr H\mathcal T^*(E)\bigr|
 \le d_{\rm tr}(\rho,\sigma).
\]
Likewise, each $M(B)$ is an effect, so taking the supremum of
$|\Tr HM(B)|$ over measurable $B$ gives the measurement inequality.
For $0\le\ell\le B$, the layer-cake representation yields
\[
 |\E_P\ell-\E_Q\ell|
 =\left|\int_0^B\bigl[P(\ell>t)-Q(\ell>t)\bigr]\,dt\right|
 \le B\TV(P,Q).
\]
Finally, if $a_A=\int_A|p-q|$, normalization gives
\[
 P(A^c)-Q(A^c)=-\int_A(p-q)\le a_A.
\]
Using $|p-q|\le p+q$ on $A^c$, we obtain
\[
 2\TV(P,Q)\le a_A+P(A^c)+Q(A^c)
             \le2a_A+2Q(A^c),
\]
which proves~\eqref{eq:compact-TV}. The integral term in~\eqref{eq:compact-TV} has no factor $1/2$: normalization is also used to control the unknown $P$-tail.
\end{proof}

For hybrid Gaussian states
$\phi,\psi$, the same operational definition is
\[
 d_{\rm tr}(\phi,\psi)
 =\sup_{0\preceq E\preceq\Id}|\phi(E)-\psi(E)|,
\]
where $E$ ranges over measurement effects of the hybrid system.
Here $\phi$ and $\psi$ assign expectations to observables.
These definitions apply directly to comparisons between hybrid
Gaussian experiments. A channel approximation
transfers measurement procedures: a measurement
on the target experiment can be applied to the input experiment by
first passing its state through the channel. In a strong qLAN
comparison, let $\{\gamma_h:h\in K_r\}$ denote the full limiting
Gaussian experiment, which may include more coordinates than the
SLD limit $\phi_h$ in Section~\ref{app:bg-qlan}. One constructs channels $\mathcal T_n$
from the original system to this experiment and $\mathcal R_n$ in
the reverse direction, neither depending on $h$, such that
\[
 \sup_{h\in K_r}d_{\rm tr}(\mathcal T_n(\rho_{n,h}),\gamma_h)\to0\quad 
 \text{\normalfont and}\quad
 \sup_{h\in K_r}d_{\rm tr}(\mathcal R_n(\gamma_h),\rho_{n,h})\to0.
\]
Such channel results are proved for suitable models and local
parameter sets in \citet{GK2006,KG2009}. The relevant model, spectrum and neighborhood
restrictions are part of those theorems. Appendix~\ref{sec:decision}
states such a comparison separately in Assumption~\ref{ass:channels}
and proves its decision consequences.

By comparison, our direct output theorem controls the TV
distance for one constructed measurement, uniformly for $h\in K_r$.
It therefore controls all events and bounded losses computed from
that output, but does not assert a reverse channel or a simultaneous
representation of arbitrary measurements. Section~\ref{sec:output-inference} derives the resulting event and
bounded-loss guarantees, with the projection bound
in~\eqref{eq:projection-facts} controlling quadratic risk.

\subsection{Classical likelihood, localization and pilot inference}\label{app:bg-measured-primer}
Once a fixed measurement has outcome probabilities $p_\theta(y)$, inference
uses an ordinary classical experiment. Assume finitely many outcomes,
$p_\theta(y)>0$ near an interior reference point $\theta_0$, and three
continuous derivatives. For independent recorded outcomes $Y_1,\ldots,Y_n$,
the log likelihood is $\mathsf l_n(\theta)=\sum_{k=1}^n\log p_\theta(Y_k)$.
The score of one observation is the column vector
$s_\theta(y)=\nabla_\theta\log p_\theta(y)$, where the gradient takes
one derivative in each real parameter coordinate. Normalization gives
\[
 \E_\theta s_\theta(Y)=\sum_y\nabla_\theta p_\theta(y)=0.
\]
The Fisher matrix already introduced in Section~\ref{sec:distances} is
$F(\theta)=\E_\theta[s_\theta(Y)s_\theta(Y)^T]$, suppressing the fixed measurement in its notation.
Differentiating $\sum_y p_\theta(y)=1$ twice gives the curvature identity
\[
 \E_\theta\nabla_\theta^2\log p_\theta(Y)=-F(\theta).
\]
The matrix $F(\theta)$ depends on the chosen measurement. Positivity of $F=F(\theta_0)$ means that every local
parameter direction changes the measured distribution to first order.

Assume $F=F(\theta_0)\succ0$. The transformed coordinate $z=\sqrt nF^{1/2}(\theta-\theta_0)$ puts the
reference information at the identity. Define the standardized score
$\xi(y)=F^{-1/2}s_{\theta_0}(y)$ and normalized sum
$S_n=n^{-1/2}\sum_k\xi(Y_k)$. They satisfy
$\E_{\theta_0}S_n=0$ and $\operatorname{Cov}_{\theta_0}S_n=I_p$.
Taylor expansion suggests the local log likelihood
\[
 \ell_n(z)=\mathsf l_n(\theta_0+F^{-1/2}z/\sqrt n)-\mathsf l_n(\theta_0)
 \approx z^TS_n-\tfrac12\norm z^2.
\]
The quadratic on the right is maximized at $z=S_n$. A rigorous Fisher
expansion bounds $\widehat z-S_n$, where $\widehat z$ maximizes the actual
likelihood. A Wilks expansion instead bounds
$2\ell_n(\widehat z)-\norm{S_n}^2$. These are different approximations.
A Gaussian approximation $S_n\approx N(0,I_p)$ then gives the chi-square
reference law for squared norms and likelihood-ratio statistics. The physical estimate is $\widehat\theta=\theta_0+F^{-1/2}\widehat z/\sqrt n$. Recall that a
$\chi_p^2$ variable is the sum of squares of $p$ independent standard
normal variables. For $0<\alpha<1$, its quantile $q_{p,1-\alpha}$ is chosen so that its
probability below that value is $1-\alpha$.

Localization is needed before using a Taylor approximation at the maximizing
point. For example, if the Hessian $\nabla_z^2\ell_n(z)$ throughout a ball differs from $-I_p$
in operator norm by at most $\delta<1$, the likelihood is strictly concave
there. An inward-pointing gradient at its boundary then puts the maximum in
the interior, where the score equation applies. When that ball is centered
at an unknown true parameter, the constrained estimator is an oracle
construction. A known center, a global concavity argument, or a separately
proved pilot localization condition is required to implement it.

At the likelihood maximizer $\widehat\theta$, the observed information substitutes the data-dependent negative Hessian,
\[
 \widehat F=-\frac1n\sum_{k=1}^n
       \nabla_\theta^2\log p_{\widehat\theta}(Y_k),
\]
for the reference matrix. The resulting confidence region is
$\{\theta:n(\widehat\theta-\theta)^T\widehat F
(\widehat\theta-\theta)\le q_{p,1-\alpha}\}$ when $\widehat F\succ0$.
An observed-information error bound controls the additional substitution. A prescribed positive fallback
matrix defines the procedure on data sets where the Hessian is unsuitable.
Coverage means the probability, at the true parameter, that this random set
contains that parameter.

For adaptation, use $n_0$ copies for a pilot estimate $\vartheta$ and the
remaining $N=n-n_0$ copies for a measurement $M_\vartheta$. The pilot record
generates a sigma-field $\mathcal F_0$, the collection of events determined
by that record. Conditional on $\mathcal F_0$, both $\vartheta$ and the
chosen measurement are fixed, and the fresh outcomes are independent with
probabilities $p_{\theta;\vartheta}(y)=\Tr\rho_\theta M_{\vartheta,y}$.
Derivatives in $\theta$ hold $\vartheta$ and $M_\vartheta$ fixed.
For a pilot value $\vartheta\in\Theta$, suppose the selected nonzero-effect
probabilities are positive and differentiable near $\vartheta$ and its
Fisher matrix $F_\vartheta$ is positive definite. A one-step estimator
then takes one information-weighted score correction:
\[
 \widehat\theta_{\rm os}=\vartheta+
 F_\vartheta^{-1}\frac1N\sum_{k=1}^N
 \left.\nabla_\theta\log p_{\theta;\vartheta}(Y_k)\right|_{\theta=\vartheta},
\]
where $F_\vartheta$ is the Fisher information of that fixed measurement
evaluated at $\vartheta$. To define a procedure on every pilot record,
an inadmissible pilot or an undefined score or inverse must trigger a
specified fallback measurement and report; Appendix~\ref{sec:adaptive}
gives these rules. The correction removes the first-order pilot
error when the model is regular. In nonlinear models, the remaining bias
is generally quadratic in $\vartheta-\theta$; affine examples may cancel
it exactly. If a conditional guarantee has failure probability at most
$\alpha_1$ whenever the pilot lies in a stated good set, and that good set
fails with probability at most $\alpha_0$, the unconditional failure is at
most $\alpha_0+\alpha_1$. Uniform constants on the good set are essential.

Sample allocation depends on the loss being controlled. A larger pilot can
reduce measurement mismatch while leaving fewer copies for the final mean.
For quadratic risk, conditioning gives the exact decomposition
\[
 \E_\theta\norm{\widehat\theta-\theta}^2
 =\E_\theta\operatorname{Tr}\operatorname{Cov}_\theta
       (\widehat\theta\mid\mathcal F_0)
  +\E_\theta\norm{\E_\theta(\widehat\theta\mid\mathcal F_0)-\theta}^2.
\]
Thus conditional unbiasedness removes the second term, but leaves the pilot
choice in the first. Sections~\ref{app:adaptive-risk} and~\ref{sec:exact-adaptive} track these
terms explicitly, rather than identifying a distributional error rate with
a risk-optimal pilot allocation.

\subsection{Finite-radius risk and external Gaussian comparison}\label{app:bg-risk-primer}
For an experiment $\mathcal E=\{P_h:h\in K_r\}$, a loss
$\ell(a,h)$ and admissible estimators, minimax risk is
$\mathfrak R_{\mathcal E}(\ell,r)=\inf_{\widehat h}\sup_{h\in K_r}
\E_h\ell(\widehat h,h)$. In a quantum experiment the infimum includes
both the measurement and the rule for reporting its outcome. Restricting
measurements to individual copies, disjoint blocks or arbitrary collective
measurements changes the admissible class and can change the value.
A lower bound proved for locally unbiased estimators need not hold for
arbitrary biased estimators. Integrating risk against a probability
distribution on the parameter ball gives a Bayes risk, which cannot exceed
the supremum risk of the same procedure. A lower bound valid for every
procedure under that prior therefore also bounds minimax risk without an
unbiasedness assumption. Theorem~\ref{thm:radius} implements this argument
for the classical Gaussian shift using an explicit smooth prior.

For the classical Gaussian shift $Y\sim N(h,V)$ and weight $W\succ0$, define $\ell_W(a,h)=(a-h)^TW(a-h)$ and let $\mathfrak R(\ell_W,r)$ denote its minimax risk on $K_r$. The
unrestricted rule $Y$ has constant quadratic risk $C=\operatorname{Tr}(WV)$.
On the finite ball $K_r$, the rule zero has worst risk at most
$\norm W_\op r^2$. Hence the finite-ball minimax risk can be smaller than
$C$. The radius correction $\beta_W(r)=C-\mathfrak R(\ell_W,r)$ accounts
for that difference when the full-shift minimax value is $C$.
Projection onto $K_r$ bounds the loss by $4\norm W_\op r^2$, so a
channel approximation error $\epsilon$ yields a risk-transfer error at most
$4\norm W_\op r^2\epsilon$. Two-sided risk comparison requires channels
in both directions. Appendix~\ref{sec:decision} states the channel assumptions,
proves their consequences and includes a classical Gaussian radius bound.

\section{Inference after a specified measurement}\label{sec:measured}
Once a POVM has been fixed, the resulting model is classical. The
nonasymptotic likelihood analysis below uses the local quadratic-bracketing
viewpoint developed by \citet{Spokoiny2012}, specialized to finite
outcomes and uniformly bounded derivatives. We include elementary proofs
and explicit constants for the measurement-dependent and pilot-adaptive
applications.
We first derive measured likelihood bounds and localize the maximizer in Section~\ref{app:measured-local}. Section~\ref{app:measured-distribution} gives distribution and coverage bounds; Sections~\ref{app:global-localization} and~\ref{app:block-measurements} treat global localization and block measurements. Section~\ref{app:information-loss} relates measured information to the intrinsic SLD information.

\subsection{Local likelihood bounds and localization}\label{app:measured-local}
Fix a finite-outcome POVM $M=(M_y)_{y\in\mathcal Y}$ and measure each copy separately. The outcomes $Y_1,\ldots,Y_n$ are independent and identically distributed (i.i.d.). They have the probabilities $p_\theta(y)=\Tr\rho_\theta M_y$ of Section~\ref{app:bg-measured-primer}. Throughout this appendix, $\Prb$ and $\E$ without a parameter subscript refer to the reference point $\theta_0$. Remove zero effects; under local faithfulness the remaining probabilities are positive. Assume the reference measured information $F\succ0$. In information coordinates $t=F^{1/2}(\theta-\theta_0)$, introduce the log-probability functions $f_y$ and their Hessians $G_y$, and abbreviate the standardized scores $\xi(y)$ of Section~\ref{app:bg-measured-primer} by $s_y$:
\[
 f_y(t)=\log p_{\theta_0+F^{-1/2}t}(y),\quad
 s_y=\nabla f_y(0),\quad\text{\normalfont and}\quad G_y=\nabla^2f_y(0).
\]
Then $\E_0s_Y=0$, $\E_0s_Ys_Y^T=I_p$, and $\E_0G_Y=-I_p$. Here $I_p$ is the identity matrix, distinct from the SLD information $\mathbb{I}$.

\begin{assumption}[Measured derivative bounds]\label{ass:meas}
For some $a_M>0$, each $f_y$ is $C^3$ on a neighborhood of the closed ball $\norm t\le a_M$, whose image in physical coordinates is contained in $\Theta$. At the reference point, assume for every $y$ that
\[
 \norm{s_y}\le A\quad\text{\normalfont and}\quad \norm{G_y+I_p}_\op\le B.
\]
On the whole ball, require the uniform third derivative bound
\[
 \sup_{\norm t\le a_M}\sup_{\norm u=\norm v=\norm w=1}
 |D^3f_y(t)[u,v,w]|\le K.
\]
\end{assumption}
For a fixed $C^3$ faithful density model and finite POVM with $F\succ0$,
the bounds in Assumption~\ref{ass:meas} hold on a sufficiently small
compact neighborhood. Faithfulness makes the nonzero effects' outcome
probabilities positive; the $C^3$ assumption and compactness then bound
the derivatives of their logarithms. The local log likelihood and normalized score from Section~\ref{app:bg-measured-primer} then take the forms
\[
 \ell_n(z)=\sum_{k=1}^n\bigl\{f_{Y_k}(z/\sqrt n)-f_{Y_k}(0)\bigr\}\quad\text{\normalfont and}\quad
 S_n=n^{-1/2}\sum_{k=1}^n s_{Y_k}.
\]
Here $z=\sqrt nF^{1/2}(\theta-\theta_0)$ and $r_{\rm M}$ denotes a radius in $z$ coordinates, whereas $h=\sqrt n(\theta-\theta_0)$ and $r$ use the intrinsic coordinates. For each $z$, the likelihood $\ell_n(z)$ and score vector $S_n$ are classical random quantities on one common observation space. For $x>0$, the score and Hessian net thresholds are
\[
 t_s=x+p\log5+\log4\quad\text{\normalfont and}\quad
 t_H=x+p\log9+\log4.
\]
The thresholds $t_s,t_H$ define the score and Hessian bounds
\begin{gather}
 z_n(x)=2\left\{\sqrt{2t_s}+\frac{2At_s}{3\sqrt n}\right\}\quad
 \text{\normalfont and}\quad
 b_n^H(x)=2B\left\{\sqrt{\frac{2t_H}{n}}+\frac{2t_H}{3n}\right\}.\label{eq:bn}
\end{gather}
Adding the deterministic Hessian variation over the ball gives
\begin{equation}\label{eq:delta}
 \delta_n(r_{\rm M},x)=b_n^H(x)+Kr_{\rm M}/\sqrt n.
\end{equation}

A sharper alternative uses the matrix variance
\[
 \sigma_G^2=\norm{\E(G_Y+I_p)^2}_\op\le B^2
\]
and replaces the net bound $b_n^H(x)$ by
\begin{equation}\label{eq:matrix-bern}
 \widetilde b_n^H(x)=\sqrt{2\sigma_G^2\{x+\log(4p)\}/n}
 +\frac{2B\{x+\log(4p)\}}{3n}.
\end{equation}
The self-adjoint matrix Bernstein inequality of \citet{Tropp2012}, applied to $G_Y+I_p$ and its negative, gives failure probability at most $e^{-x}/2$. Either deterministic choice can be used throughout, with $\delta$ defined accordingly. The net formula~\eqref{eq:bn} remains the baseline when no choice is stated.

\begin{theorem}[Measured quadratic bracketing]\label{thm:measured}
Under Assumption~\ref{ass:meas}, if $r_{\rm M}/\sqrt n\le a_M$, there is an event $\mathcal A_x$ of probability at least $1-e^{-x}$ under $\theta_0$ on which $\norm{S_n}\le z_n(x)$ and
\begin{equation}\label{eq:hessian}
 \sup_{\norm z\le r_{\rm M}}\norm{\nabla^2\ell_n(z)+I_p}_\op\le\delta\quad\text{\normalfont and}\quad\delta=\delta_n(r_{\rm M},x).
\end{equation}
Consequently, simultaneously for $\norm z\le r_{\rm M}$,
\begin{equation}\label{eq:meas-bracket}
 z^TS_n-\tfrac12(1+\delta)\norm z^2\le\ell_n(z)\le
 z^TS_n-\tfrac12(1-\delta)\norm z^2.
\end{equation}
\end{theorem}
\begin{proof}
For unit $v$, $v^Ts_Y$ is centered with variance one and absolute value at most $A$. Bernstein's inequality on a $1/2$-net of at most $5^p$ points gives $\norm{S_n}\le z_n(x)$ except with probability $e^{-x}/2$. For $D_n=n^{-1}\sum_k(G_{Y_k}+I_p)$, the scalar $v^T(G_Y+I_p)v$ is centered, bounded by $B$, and has variance at most $B^2$. A $1/4$-net of at most $9^p$ points gives $\norm{D_n}_\op\le b_n^H(x)$ except with probability $e^{-x}/2$. The net inequalities are in Appendix~\ref{app:nets}. For the Hessian away from zero, the chain rule and the bound $K$ in Assumption~\ref{ass:meas} give
\[
 \nabla^2\ell_n(z)=\frac1n\sum_k\nabla^2f_{Y_k}(z/\sqrt n)\quad\text{\normalfont and}\quad
 \norm{\nabla^2\ell_n(z)-\nabla^2\ell_n(0)}_\op
 \le K\norm z/\sqrt n.
\]
Together with $\nabla^2\ell_n(0)+I_p=D_n$, this proves~\eqref{eq:hessian}. Since $\ell_n(0)=0$ and $\nabla\ell_n(0)=S_n$, integration along the segment gives
\[
 \ell_n(z)-z^TS_n
 =\int_0^1(1-t)z^T\nabla^2\ell_n(tz)z\,dt.
\]
Bounding the integrand between $-(1+\delta)\norm z^2$ and $-(1-\delta)\norm z^2$, and using $\int_0^1(1-t)dt=1/2$, proves~\eqref{eq:meas-bracket}.
\end{proof}

\begin{corollary}[Interior localization and Fisher expansion]\label{cor:fisher}
In Theorem~\ref{thm:measured}, suppose
\begin{equation}\label{eq:localize}
 \delta<1\quad\text{\normalfont and}\quad z_n(x)<(1-\delta)r_{\rm M}.
\end{equation}
On $\mathcal A_x$, the maximizer $\widehat z$ of $\ell_n$ over $\{z:\norm z\le r_{\rm M}\}$ is unique and interior. Its physical-parameter version is $\widehat\theta=\theta_0+F^{-1/2}\widehat z/\sqrt n$, and
\begin{equation}\label{eq:fisher}
 \norm{\widehat z-S_n}\le\frac{\delta}{1-\delta}\norm{S_n}\quad\text{\normalfont and}\quad
 \norm{\widehat z}\le\frac{\norm{S_n}}{1-\delta}.
\end{equation}
\end{corollary}
\begin{proof}
Work on $\mathcal A_x$. Compactness and continuity give a maximizer,
and~\eqref{eq:hessian} gives
$\nabla^2\ell_n(z)\preceq-(1-\delta)I_p$ throughout the ball.
For a boundary point $\norm z=r_{\rm M}$, integration of the Hessian gives
\[
 z^T\nabla\ell_n(z)
 =z^TS_n+\int_0^1z^T\nabla^2\ell_n(tz)z\,dt
 \le r_{\rm M}\bigl\{z_n(x)-(1-\delta)r_{\rm M}\bigr\}<0.
\]
The derivative in the feasible inward direction $-z$ is therefore
positive, excluding a boundary maximizer. Strict concavity gives
uniqueness. At the interior maximizer, the score equation reads
\[
 0=S_n-\widehat z+
 \int_0^1\bigl\{\nabla^2\ell_n(t\widehat z)+I_p\bigr\}\widehat z\,dt.
\]
The integral has norm at most $\delta\norm{\widehat z}$, so
\[
 \norm{\widehat z}\le\norm{S_n}+\delta\norm{\widehat z}
 \quad\text{and}\quad
 \norm{\widehat z-S_n}\le\delta\norm{\widehat z}.
\]
Rearranging the first inequality and substituting it into the second
proves~\eqref{eq:fisher}.
\end{proof}

\begin{corollary}[Wilks expansion]\label{cor:wilks}
Under the conditions of Corollary~\ref{cor:fisher}, on $\mathcal A_x$,
\begin{equation}\label{eq:wilks}
 \abs{2\ell_n(\widehat z)-\norm{S_n}^2}\le\frac{\delta}{1-\delta}\norm{S_n}^2.
\end{equation}
\end{corollary}
\begin{proof}
For $a>0$, completing the square gives
\[
 2z^TS_n-a\norm z^2
 =\frac{\norm{S_n}^2}{a}
   -a\left\|z-\frac{S_n}{a}\right\|^2.
\]
Apply this identity to the upper quadratic bracket with $a=1-\delta$.
For the lower bracket, the point $z_-=S_n/(1+\delta)$ is feasible because
\[
 \norm{z_-}\le\frac{z_n(x)}{1+\delta}
 <\frac{1-\delta}{1+\delta}r_{\rm M}\le r_{\rm M}.
\]
Maximality of $\widehat z$ and the lower bracket at $z_-$ therefore give
\[
 \frac{\norm{S_n}^2}{1+\delta}
 \le2\ell_n(\widehat z)
 \le\frac{\norm{S_n}^2}{1-\delta}.
\]
Subtract $\norm{S_n}^2$ and use
$\delta/(1+\delta)\le\delta/(1-\delta)$ to obtain~\eqref{eq:wilks}.
\end{proof}

The probability bounds in this appendix are pointwise under
$P_{\theta_0}$. When $\theta_0$ is a known nominal center, the restricted
maximizer in Corollary~\ref{cor:fisher} is computable and its coverage
is calibrated at that center. When $\theta_0$ denotes the unknown
true parameter, this restriction defines an oracle estimator;
implementation then requires global localization or the pilot-centered
construction in Appendix~\ref{sec:adaptive}.

\subsection{Distribution and coverage bounds}\label{app:measured-distribution}
Let $\Phi$ and $\phi$ be the distribution function and density of $N(0,1)$, and write $$d_{\rm K}(U,V)=\sup_t|\Prb(U\le t)-\Prb(V\le t)|$$ for Kolmogorov distance between scalar laws. Let $Z\sim N(0,I_p)$, $\beta_3=\E_0\norm{s_Y}^3\le Ap$, and
\begin{equation}\label{eq:BE}
 b_n^{\rm BE}=\min\bigl\{1,400p^{1/4}\beta_3/\sqrt n\bigr\}.
\end{equation}
We use the convex-set Berry--Esseen inequality of \citet{Bentkus2003}: for i.i.d. centered identity-covariance vectors, the convex-set distance between their normalized sum and $Z$ is at most~\eqref{eq:BE}.
For halfspaces one can use the scalar Berry--Esseen inequality directly. Define
\[
 \beta_{3,\rm dir}=\sup_{\norm v=1}\E|v^Ts_Y|^3\le A\quad\text{\normalfont and}\quad
 b_n^{\rm dir}=\min\bigl\{1,\beta_{3,\rm dir}/\sqrt n\bigr\}.
\]
The constant one is conservative; the scalar result of \citet{Shevtsova2011} is stronger. Each projection has variance one. 

The maximum chi-square density is $c_2=1/2$ when $p=2$ and, when $p>2$,
\[
 c_p=\frac{(p-2)^{p/2-1}e^{-(p-2)/2}}{2^{p/2}\Gamma(p/2)}.
\]
For an interval of length $a\ge0$, the chi-square concentration function is bounded by
\begin{equation}\label{eq:omega}
 \omega_p(a)=\begin{cases}
 \min\bigl\{1,\sqrt{2a/\pi}\bigr\},&p=1,\\
 \min\{1,c_pa\},&p\ge2.
 \end{cases}
\end{equation}
Indeed the $\chi_1^2$ density is decreasing and its probability on $[0,a]$ is at most $\sqrt{2a/\pi}$. For $p\ge2$, maximize the density at $p-2$. In particular, a linear bound in $a$ is not valid uniformly near zero when $p=1$.

For the distributional bounds, choose a fixed measurable maximizing rule
outside $\mathcal A_x$ for the restricted estimator of
Corollary~\ref{cor:fisher}. Write $z_*=z_n(x)$ and introduce the
deterministic error budgets
\[
 e_F=\frac{\delta z_*}{1-\delta},\quad
 w_W=\frac{\delta z_*^2}{1-\delta}
 \quad\text{and}\quad w_F=e_F(2z_*+e_F).
\]

\begin{theorem}[Distributional Fisher and Wilks bounds]\label{thm:distribution}
Under the conditions of Corollary~\ref{cor:fisher},
\begin{gather}
 \sup_{\norm v=1,t\in\R}\left|\Prb(v^T\widehat z\le t)-\Phi(t)\right|
 \le b_n^{\rm dir}+e^{-x}+e_F/\sqrt{2\pi},\label{eq:normal-dist}\\
 \sup_t\left|\Prb(2\ell_n(\widehat z)\le t)-\Prb(\chi_p^2\le t)\right|
 \le b_n^{\rm BE}+e^{-x}+\omega_p(w_W),\label{eq:wilks-dist}\\
 \text{\normalfont and}\quad \sup_t\left|\Prb\left(\norm{\widehat z}^2\le t\right)-\Prb(\chi_p^2\le t)\right|
 \le b_n^{\rm BE}+e^{-x}+\omega_p(w_F).\label{eq:wald-dist}
\end{gather}
Bounds may be truncated at one. If $q_{p,1-\alpha}$ is a chi-square quantile, the ellipsoid
$\{\theta:n(\widehat\theta-\theta)^TF(\widehat\theta-\theta)\le q_{p,1-\alpha}\}$
has coverage error at $\theta_0$ at most the right side of~\eqref{eq:wald-dist}, when $F=F(\theta_0;M)$ is known. Likelihood-ratio test coverage is bounded by~\eqref{eq:wilks-dist}.
\end{theorem}
\begin{proof}
On $\mathcal A_x$, $\norm{\widehat z-S_n}\le e_F$, and
$\Prb(\mathcal A_x^c)\le e^{-x}$. For every unit vector $v$ and real $t$,
\[
 \Prb(v^TS_n\le t-e_F)-e^{-x}
 \le\Prb(v^T\widehat z\le t)
 \le\Prb(v^TS_n\le t+e_F)+e^{-x}.
\]
The scalar Berry--Esseen bound controls each score probability by
$\Phi$ with error $b_n^{\rm dir}$. Since
$|\Phi(t\pm e_F)-\Phi(t)|\le e_F/\sqrt{2\pi}$,
this proves~\eqref{eq:normal-dist} uniformly in $v,t$.

Put $V_n=\norm{S_n}^2$ and $F_p(t)=\Prb(\chi_p^2\le t)$, with
$F_p(t)=0$ for $t<0$. Euclidean balls are convex, so
\[
 \sup_t|\Prb(V_n\le t)-F_p(t)|\le b_n^{\rm BE}.
\]
On $\mathcal A_x$, Corollary~\ref{cor:wilks} gives
$|2\ell_n(\widehat z)-V_n|\le w_W$, while
\[
 \left|\norm{\widehat z}^2-V_n\right|
 \le\norm{\widehat z-S_n}\left(\norm{\widehat z}+\norm{S_n}\right)
 \le e_F(2z_*+e_F)=w_F.
\]
For either statistic $T_n$ and its corresponding bound $w$, the same
 event inclusion yields
\[
 F_p(t-w)-b_n^{\rm BE}-e^{-x}
 \le\Prb(T_n\le t)
 \le F_p(t+w)+b_n^{\rm BE}+e^{-x}.
\]
Each interval between $t$ and $t\pm w$ has length $w$, so
\eqref{eq:omega} bounds its chi-square probability by $\omega_p(w)$.
This proves~\eqref{eq:wilks-dist} and~\eqref{eq:wald-dist}.
At $t=q_{p,1-\alpha}$, $F_p(t)=1-\alpha$, giving the stated
coverage bounds.
\end{proof}
For observed-information inference, recall $\widehat F$ from
Section~\ref{app:bg-measured-primer} and use the error budgets above to set
\[
 \widehat F=-\frac1n\sum_{k=1}^n\nabla_\theta^2\log p_{\widehat\theta}(Y_k)\quad\text{\normalfont and}\quad
 w_{\rm obs}=w_F+\frac{\delta z_*^2}{(1-\delta)^2}.
\]
Use $I_p$ if the displayed matrix is nonfinite or has a nonpositive eigenvalue. Choose a measurable maximizing rule; for the disk-constrained Pauli procedure we choose the minimizer of Euclidean norm among all maximizers. Its uniqueness follows because the concave likelihood has a compact convex maximizer set and the squared Euclidean norm is strictly convex. These rules define the statistic on every data set.

\begin{proposition}[Observed-information confidence ellipsoid]\label{prop:observed}
Under the conditions of Corollary~\ref{cor:fisher}, the observed-information
statistic defined above satisfies
\begin{equation}\label{eq:observed}
 \sup_t\left|\Prb\bigl\{n(\widehat\theta-\theta_0)^T\widehat F(\widehat\theta-\theta_0)\le t\bigr\}
 -\Prb(\chi_p^2\le t)\right|
 \le b_n^{\rm BE}+e^{-x}+\omega_p(w_{\rm obs}).
\end{equation}
Consequently the ellipse with center $\widehat\theta$, matrix $n\widehat F$, and squared radius $q_{p,1-\alpha}$ has coverage error bounded by~\eqref{eq:observed}. It is computable whenever $\widehat\theta$ is implementable, as with the global or pilot-localized estimators.
\end{proposition}
\begin{proof}
Work on $\mathcal A_x$ and set
$H_n=F^{-1/2}\widehat F F^{-1/2}$. The change to information coordinates
and~\eqref{eq:hessian} give
\[
 H_n=-\nabla_z^2\ell_n(\widehat z)
 \quad\text{and}\quad \norm{H_n-I_p}_\op\le\delta<1.
\]
Thus $H_n\succ0$, so the information fallback is not used on this event.
Write $T_{\rm obs}=n(\widehat\theta-\theta_0)^T\widehat F
(\widehat\theta-\theta_0)=\widehat z^TH_n\widehat z$ and
$V_n=\norm{S_n}^2$. The Fisher expansion and the definition of $w_F$ imply
\[
 \begin{aligned}
 |T_{\rm obs}-V_n|
 &\le\bigl|\widehat z^T(H_n-I_p)\widehat z\bigr|
       +\left|\norm{\widehat z}^2-\norm{S_n}^2\right|\\
 &\le\delta\norm{\widehat z}^2+w_F
 \le\frac{\delta z_*^2}{(1-\delta)^2}+w_F=w_{\rm obs}.
 \end{aligned}
\]
Since $\Prb(\mathcal A_x^c)\le e^{-x}$, for every real $t$,
\[
 \Prb(V_n\le t-w_{\rm obs})-e^{-x}
 \le\Prb(T_{\rm obs}\le t)
 \le\Prb(V_n\le t+w_{\rm obs})+e^{-x}.
\]
The convex-set Berry--Esseen bound used in
Theorem~\ref{thm:distribution} bounds the distributional error of $V_n$
by $b_n^{\rm BE}$. If $F_p(t)=\Prb(\chi_p^2\le t)$, the definition
of $\omega_p$ bounds both
$F_p(t+w_{\rm obs})-F_p(t)$ and
$F_p(t)-F_p(t-w_{\rm obs})$ by $\omega_p(w_{\rm obs})$.
The preceding probability bounds prove~\eqref{eq:observed}.
At $t=q_{p,1-\alpha}$, the event $T_{\rm obs}\le t$ is precisely
coverage of $\theta_0$ by the reported ellipsoid, proving the final claim.
\end{proof}
More generally, if an information estimator has relative operator error at most $\eta_I$ except with probability $\alpha_I$, the bound holds with $w_F+\eta_I(z_*+e_F)^2$ and an added $\alpha_I$. Independence is unnecessary. Conditional application to Theorem~\ref{thm:adaptive}, with $N$ observations, adds the pilot failure $\alpha_0$ and uses the corresponding uniform constants.

\subsection{An assumption-based global localization result}\label{app:global-localization}
Write $T$ for the parameter set in information coordinates and
$\widehat\theta_{\rm global}$ for a global maximum likelihood estimator
(MLE), when one exists.
\begin{proposition}[Global concavity localization]\label{prop:global}
In addition to the measured score assumptions, suppose $T$ is compact
and convex, contains $\{t:\norm t\le r_{\rm M}/\sqrt n\}$, and
$\nabla^2f_y(t)\preceq-cI_p$ for every $y$ and $t\in T$, with $c>0$.
Then the global MLE exists, is unique, and
\[
 \Prb_0\left\{\norm{\sqrt nF^{1/2}(\widehat\theta_{\rm global}-\theta_0)}>z_n(x)/c\right\}
 \le e^{-x}/2.
\]
If $z_n(x)<cr_{\rm M}$, the global MLE lies in the local ball on the event $\{\norm{S_n}\le z_n(x)\}$; the local expansions then apply to the actual global MLE.
\end{proposition}
\begin{proof}
Put $Z_n=\sqrt n\,T$. On this set the chain rule gives
\[
 \nabla^2\ell_n(z)=\frac1n\sum_{k=1}^n
 \nabla^2f_{Y_k}(z/\sqrt n)\preceq-cI_p.
\]
Continuity and compactness give a maximizer $\widehat z$, and strict
concavity makes it unique. Since $0\in Z_n$, constrained optimality
gives $\nabla\ell_n(\widehat z)^T(0-\widehat z)\le0$, including when
$\widehat z$ lies on the boundary. The segment from zero to $\widehat z$
lies in $Z_n$, and $\nabla\ell_n(0)=S_n$, so
\[
 \begin{aligned}
 0\le\widehat z^T\nabla\ell_n(\widehat z)
 =\widehat z^TS_n+
   \int_0^1\widehat z^T\nabla^2\ell_n(t\widehat z)\widehat z\,dt\le\norm{\widehat z}\norm{S_n}-c\norm{\widehat z}^2.
 \end{aligned}
\]
Thus $\norm{\widehat z}\le\norm{S_n}/c$, also when $\widehat z=0$.
The score concentration step in the proof of
Theorem~\ref{thm:measured} gives
$\Prb_0\{\norm{S_n}>z_n(x)\}\le e^{-x}/2$. Consequently,
\[
 \Prb_0\left\{\norm{\widehat z}>z_n(x)/c\right\}
 \le\Prb_0\left\{\norm{S_n}>z_n(x)\right\}\le e^{-x}/2.
\]
Since $\widehat z=\sqrt nF^{1/2}(\widehat\theta_{\rm global}-\theta_0)$,
this is the asserted bound. If $z_n(x)<cr_{\rm M}$, the same score event
gives $\norm{\widehat z}<r_{\rm M}$, placing the global MLE inside the
local ball. 
\end{proof}

\subsection{Fixed-block collective measurements}\label{app:block-measurements}
For an integer block size $1\le m\le n$, let $N=\lfloor n/m\rfloor$ and measure a finite POVM $M^{(m)}$ independently on each of $N$ blocks. Put
\[
 p_\theta^{(m)}(y)=\Tr\rho_\theta^{\otimes m}M_y^{(m)}
 \quad\text{and}\quad F_m=\text{\normalfont its one-block Fisher information}.
\]
\begin{corollary}[Block measurements]\label{cor:block}
For the block experiment above, assume $F_m\succ0$ at $\theta_0$ and the
block version of Assumption~\ref{ass:meas} with constants $A_m,B_m,K_m$.
Then the measured likelihood, Fisher, Wilks and confidence bounds in
Sections~\ref{app:measured-local}--\ref{app:measured-distribution} hold
with sample size $N$ and information $F_m$, subject to their stated
localization conditions. The estimator normalization is
$\sqrt N F_m^{1/2}(\widehat\theta-\theta_0)$, equivalently
$\sqrt{mN}(F_m/m)^{1/2}(\widehat\theta-\theta_0)$.
\end{corollary}
\begin{proof}
For block outcomes $y_1,\ldots,y_N$, the tensor-product trace factors as
\[
 \Prb_\theta(Y_1=y_1,\ldots,Y_N=y_N)
 =\Tr\left\{(\rho_\theta^{\otimes m})^{\otimes N}
       \bigotimes_{j=1}^N M_{y_j}^{(m)}\right\}
 =\prod_{j=1}^N p_\theta^{(m)}(y_j).
\]
Hence the block outcomes are i.i.d. Their physical-parameter scores
are centered at $\theta_0$ and have covariance $F_m$, so independence
gives total Fisher information $NF_m$. Define
\[
 f_y^{(m)}(t)=\log p_{\theta_0+F_m^{-1/2}t}^{(m)}(y)
 \quad\text{and}\quad z=\sqrt N F_m^{1/2}(\theta-\theta_0).
\]
The local log likelihood and its score are then
\[
 \ell_N^{(m)}(z)=\sum_{j=1}^N
 \left\{f_{Y_j}^{(m)}(z/\sqrt N)-f_{Y_j}^{(m)}(0)\right\}
 \quad\text{and}\quad
 S_N^{(m)}=N^{-1/2}\sum_{j=1}^N\nabla f_{Y_j}^{(m)}(0).
\]
The normalized score has mean zero and covariance $I_p$. The block
derivative assumptions therefore give exactly the hypotheses of
Sections~\ref{app:measured-local}--\ref{app:measured-distribution}, with
$(n,F,A,B,K)$ replaced by $(N,F_m,A_m,B_m,K_m)$ and the corresponding
neighborhood and localization conditions. Applying those results proves
the asserted bounds. Finally,
$\sqrt N F_m^{1/2}=\sqrt{mN}(F_m/m)^{1/2}$, which proves the equivalent
normalization in physical-copy units.
\end{proof}

\subsection{Information loss under measurement}\label{app:information-loss}
\begin{proposition}\label{prop:info-loss}
The Braunstein--Caves information inequality~\citep{BraunsteinCaves1994} states that measured Fisher information satisfies $F\preceq \mathbb{I}$.
\end{proposition}
\begin{proof}
Discard zero effects; faithfulness implies $p_y=\Tr\rho M_y>0$ for
each remaining outcome. For real $v$, put $L=L[v]$ and use the SLD equation:
\[
 \dot p_y=\Tr(\partial_v\rho)M_y
 =\tfrac12\Tr(\rho L+L\rho)M_y
 =\Rea\Tr\rho LM_y.
\]
For $A_y=\sqrt\rho\sqrt{M_y}$ and
$B_y=\sqrt\rho L\sqrt{M_y}$, cyclicity of the trace gives
$\Tr A_y^*B_y=\Tr\rho LM_y$,
$\norm{A_y}_{\HS}^2=p_y$ and
$\norm{B_y}_{\HS}^2=\Tr M_yL\rho L$. The Cauchy--Schwarz inequality
for the HS inner product therefore yields
\[
 (\dot p_y)^2
 =\bigl(\Rea\Tr A_y^*B_y\bigr)^2
 \le\bigl|\Tr A_y^*B_y\bigr|^2
 \le p_y\Tr M_yL\rho L.
\]
Divide by $p_y$, sum over outcomes and use $\sum_yM_y=\Id$:
\[
 v^TFv=\sum_y\frac{(\dot p_y)^2}{p_y}
 \le\Tr\left(\sum_yM_y\right)L\rho L
 =\Tr\rho L^2=v^T\mathbb I v.
\]
Since this holds for every real $v$, $F\preceq\mathbb I$.
\end{proof}
The score in~\eqref{eq:fisher} is a measured score. Identifying it with a simultaneous observation of every entry of $\Delta_n$ requires an additional compatibility argument.

\begin{proposition}[Equality and score compatibility]\label{prop:equality}
For a faithful model, a finite POVM attains $F=\mathbb{I}$ in every parameter direction only if the SLDs commute. Conversely, a measurement in a common eigenbasis of the SLDs attains $F=\mathbb{I}$ at the reference point. In particular, a measurement in the eigenbasis of the SLD for a one-parameter model attains scalar SLD information locally.
\end{proposition}
\begin{proof}
For each coordinate $i$, the proof of Proposition~\ref{prop:info-loss}
expresses $\mathbb I_{ii}-F_{ii}$ as a sum of nonnegative terms over
outcomes. If $F=\mathbb I$, each term vanishes. Equality in both the
real-part bound and the Cauchy--Schwarz inequality for the HS inner product therefore gives,
for every nonzero effect,
\[
 \sqrt\rho L_i\sqrt{M_y}
 =\lambda_{iy}\sqrt\rho\sqrt{M_y}
 \quad\text{with}\quad \lambda_{iy}\in\R.
\]
Faithfulness permits multiplication by $\rho^{-1/2}$, yielding
$L_i\sqrt{M_y}=\lambda_{iy}\sqrt{M_y}$. Thus the range of
$\sqrt{M_y}$ consists of common eigenvectors, and
\[
 [L_i,L_j]\sqrt{M_y}=0
 \quad\text{and}\quad
 [L_i,L_j]=\sum_y[L_i,L_j]M_y=0.
\]
The last equality uses completeness of the POVM.
Conversely, take an orthonormal common eigenbasis $e_y$ with
$L_i e_y=\lambda_{iy}e_y$ and effects $M_y=|e_y\rangle\langle e_y|$.
The SLD equation gives $\partial_i p_y=\lambda_{iy}p_y$, and hence
\[
 F_{ij}=\sum_y p_y\lambda_{iy}\lambda_{jy}
 =\Rea\Tr\rho L_iL_j=\mathbb I_{ij}.
\]
For one parameter, a Hermitian SLD always has such an eigenbasis.
\end{proof}

\section{Adaptive local quantum inference}\label{sec:adaptive}
Adaptive quantum estimation has an established asymptotic theory; see
\citet{Fujiwara2006} and the corrigendum by \citet{Fujiwara2011}.
Here the guarantees are conditional finite-sample bounds for a specified
pilot and measurement rule.
Section~\ref{app:pilot-inference} conditions on a pilot estimate to obtain local likelihood and one-step inference bounds. Section~\ref{app:adaptive-risk} derives conditional quadratic risks and compares sample allocations for that loss. Section~\ref{sec:exact-adaptive} gives the exact affine scalar calculation.

\subsection{Pilot localization and one-step inference}\label{app:pilot-inference}
Continue with the pilot split of Section~\ref{app:bg-measured-primer}: for $1\le n_0<n$, the pilot $\vartheta=\widetilde\theta$ uses $n_0$ copies and the POVM $M_\vartheta$ uses the remaining $N=n-n_0$ copies. Conditional on the pilot sigma-field $\mathcal F_0$, the latter observations have probabilities
$p_\theta(y;\vartheta)=\Tr\rho_\theta M_{\vartheta,y}$, with the pilot index now placed after the outcome. Abbreviate $F(\theta;M_\vartheta)$ by $F(\theta;\vartheta)$; the pilot information already introduced in Section~\ref{app:bg-measured-primer} is $F_\vartheta=F(\vartheta;\vartheta)$. All derivatives in $\theta$ hold the selected POVM $M_\vartheta$ fixed. For pilot values with $F(\theta_0;\vartheta)\succ0$, the information coordinate and local likelihood specialize to $z=\sqrt N F(\theta_0;\vartheta)^{1/2}(\theta-\theta_0)$ and
\[
 \ell_{N,\vartheta}(z)=\sum_{k=1}^N
 \log\frac{p_{\theta_0+F(\theta_0;\vartheta)^{-1/2}z/\sqrt N}(Y_k;\vartheta)}
 {p_{\theta_0}(Y_k;\vartheta)}
 \quad\text{\normalfont and}\quad S_N=\nabla_z\ell_{N,\vartheta}(0).
\]
The same selected POVM is applied separately to each of the $N$ fresh
copies, so their outcomes are i.i.d.
conditional on $\mathcal F_0$. The pilot and remaining copies are
independent before measurement. All
measurement selection and reported estimators below are measurable functions
of available data. To define a procedure on every pilot record, fix a known
parameter $\theta_{\rm safe}\in\Theta$ and the one-outcome POVM
$M_{\rm safe}=(\Id)$. If the proposed $M_\vartheta$ is undefined or is not a
valid finite POVM, use $M_{\rm safe}$ for the remaining copies and report
$\theta_{\rm safe}$. These checks involve only the pilot and the known model.
Each theorem below assumes that this fallback is unnecessary on its
analytical good-pilot event $\mathcal A_0$; the unknown event
$\mathcal A_0$ is never used to select the measurement or the reported value.

For pilot-centered maximization, choose a deterministic radius $R>0$
and form the candidate ball
\[
 \mathcal D_\vartheta=\left\{\theta:\norm{\theta-\vartheta}\le R/\sqrt N\right\}.
\]
Before the second-stage measurement, admit the record only if
$M_\vartheta$ is valid, $\mathcal D_\vartheta\subset\Theta$, and its
nonzero-effect probabilities are positive and continuous throughout
$\mathcal D_\vartheta$. Require this admission rule to be measurable.
On an admitted record, define $\widehat\theta$ by maximizing the
conditional second-stage likelihood over the compact ball
$\mathcal D_\vartheta$, using a fixed measurable choice of maximizer.
Continuity guarantees existence. Otherwise use $M_{\rm safe}$ and
report $\theta_{\rm safe}$ as specified above. Thus domain inclusion is
required on admitted records, not on every possible pilot value; no
likelihood is evaluated on a rejected domain.

\begin{lemma}[Geometry of the pilot-centered domain]\label{lem:pilot-domain}
Fix a pilot value satisfying $\norm{\vartheta-\theta_0}\le e_0$ and
$mI_p\preceq F_0\preceq MI_p$, where $F_0=F(\theta_0;\vartheta)$,
$e_0\ge0$ and $0<m\le M<\infty$. If $R>\sqrt N e_0$, set
\[
 r_-=\sqrt m(R-\sqrt N e_0)
 \quad\text{and}\quad
 r_+=\sqrt M(R+\sqrt N e_0).
\]
Then $\mathcal Z_\vartheta=\sqrt N F_0^{1/2}(\mathcal D_\vartheta-\theta_0)$
is compact and convex, and
\[
 \left\{z:\norm z\le r_-\right\}\subseteq\mathcal Z_\vartheta
 \subseteq\left\{z:\norm z\le r_+\right\}.
\]
In particular, zero is an interior point and every boundary point of
$\mathcal Z_\vartheta$ has norm at least $r_-$.
\end{lemma}
\begin{proof}
For $\norm z\le r_-$, the inverse image satisfies
\[
 \norm{\theta_0+N^{-1/2}F_0^{-1/2}z-\vartheta}
 \le e_0+\frac{\norm z}{\sqrt{Nm}}\le\frac R{\sqrt N},
\]
which proves the first inclusion. Conversely, for $\theta\in\mathcal D_\vartheta$,
\[
 \norm{\sqrt N F_0^{1/2}(\theta-\theta_0)}
 \le\sqrt{NM}\left(\norm{\theta-\vartheta}+e_0\right)
 \le\sqrt M(R+\sqrt N e_0)=r_+.
\]
The image of the compact convex ball under this invertible affine map
is compact and convex. Since $r_->0$, the first inclusion also gives
the assertions about zero and the boundary.
\end{proof}

\begin{theorem}[Pilot-centered maximization]\label{thm:adaptive}
For the procedure above, let $e_0\ge0$ be a deterministic pilot-error
bound, with failure tolerance $0<\alpha_0<1$, and choose $R>\sqrt N e_0$. Suppose
$\mathcal A_0=\{\norm{\vartheta-\theta_0}\le e_0\}$ has probability
at least $1-\alpha_0$ and every pilot value in $\mathcal A_0$ is admitted.
Uniformly over $\mathcal A_0$, assume
$mI_p\preceq F(\theta_0;\vartheta)\preceq MI_p$, with finite constants
$0<m\le M$, and Assumption~\ref{ass:meas} holds at the true parameter
with constants $A,B,K$ on $\norm t\le a_M$, where
$t=F(\theta_0;\vartheta)^{1/2}(\theta-\theta_0)$.
With $r_-$ and $r_+$ from Lemma~\ref{lem:pilot-domain}, assume
$r_+/\sqrt N\le a_M$, $\delta=\delta_N(r_+,x)<1$ and
$z_N(x)<(1-\delta)r_-$.
With probability at least $1-\alpha_0-e^{-x}$, the maximizer is interior
and unique, and for
\[
 \widehat z_N=\sqrt N F(\theta_0;\vartheta)^{1/2}(\widehat\theta-\theta_0),
\]
the Fisher and Wilks bounds~\eqref{eq:fisher}--\eqref{eq:wilks} hold with $\widehat z_N$, the true-centered conditional score $S_N$, and $N$. The likelihood ratio compares $\widehat\theta$ with $\theta_0$ within the conditional likelihood.
\end{theorem}
\begin{proof}
Fix a pilot in $\mathcal A_0$. Lemma~\ref{lem:pilot-domain} gives the
compact convex domain $\mathcal Z_\vartheta$ and its inner and outer
radius bounds. Conditional on $\mathcal F_0$,
Theorem~\ref{thm:measured} supplies an
event $\mathcal E_\vartheta$ with failure probability at most $e^{-x}$.
On this event, throughout $\mathcal Z_\vartheta$,
$\nabla^2\ell_{N,\vartheta}\preceq-(1-\delta)I_p$ and
$\norm{S_N}\le z_N(x)$. Every boundary point has $\norm z\ge r_-$, so
\[
 z^T\nabla\ell_{N,\vartheta}(z)
 \le\norm z\left\{z_N(x)-(1-\delta)\norm z\right\}
 \le\norm z\left\{z_N(x)-(1-\delta)r_-\right\}<0.
\]
Moving toward the feasible point zero increases the likelihood.
The maximizer is therefore interior and, by strict concavity, unique.
Its score equation gives the Fisher bound exactly as in
Corollary~\ref{cor:fisher}. For the lower Wilks bracket,
$\norm{S_N/(1+\delta)}\le z_N(x)/(1+\delta)<r_-$,
so the evaluation point belongs to $\mathcal Z_\vartheta$;
the upper quadratic bracket can be maximized over all of $\R^p$.

Finally, the good-pilot event is measurable with respect to
$\mathcal F_0$, and
\[
 \Prb(\mathcal A_0\cap\mathcal E_\vartheta^c)
 =\E\left[\Id_{\mathcal A_0}\Prb(\mathcal E_\vartheta^c\mid\mathcal F_0)\right]
 \le e^{-x}.
\]
Adding $\Prb(\mathcal A_0^c)\le\alpha_0$ gives the stated probability.
Every good-pilot record is admitted, so rejected records add no
further failure probability.
\end{proof}
The likelihood is analyzed at the true point, where the conditional score is
centered, even though its maximization domain is centered at the pilot.
Expanding at the pilot as if it were the truth would omit a score bias.
For a normalization computable from the pilot, let
$\overline F_\vartheta=F_\vartheta$ on admitted records for which
$F_\vartheta$ is defined and positive definite, and let
$\overline F_\vartheta=I_p$ otherwise. This identity-matrix convention
makes the statistic below defined on every record. In addition to the assumptions of
Theorem~\ref{thm:adaptive}, suppose $F_\vartheta\succ0$ on $\mathcal A_0$
and, uniformly there,
\[
 \norm{F_\vartheta^{1/2}F(\theta_0;\vartheta)^{-1/2}-I_p}_\op
 \le L_F e_0
\]
for a finite constant $L_F\ge0$. Define
\[
 T_{\rm pil}=\sqrt N\,\overline F_\vartheta^{1/2}
                    (\widehat\theta-\theta_0),\quad
 z_*=z_N(x)
 \quad\text{and}\quad E_{\rm pil}=\frac{(\delta+L_Fe_0)z_*}{1-\delta}.
\]
On the good-pilot and conditional likelihood event of
Theorem~\ref{thm:adaptive},
\[
 \norm{T_{\rm pil}-S_N}
 \le\norm{\widehat z_N-S_N}
   +\norm{F_\vartheta^{1/2}F(\theta_0;\vartheta)^{-1/2}-I_p}_\op
       \norm{\widehat z_N}
 \le E_{\rm pil}.
\]
The uniform score bound $A$ in Theorem~\ref{thm:adaptive} bounds the conditional
directional third moment by $A$ and the conditional norm third moment
by $Ap$. Put
\[
 b_N^{\rm dir}=\min\bigl\{1,A/\sqrt N\bigr\}
 \quad\text{and}\quad b_N^{\rm BE}=\min\bigl\{1,400p^{1/4}Ap/\sqrt N\bigr\}.
\]
Conditional application of the upper and lower probability bounds in
Theorem~\ref{thm:distribution}, followed by integration over the pilot,
therefore gives the explicit unconditional bounds
\begin{align}
 \sup_{\norm v=1,t\in\R}
 \left|\Prb_{\theta_0}\bigl\{v^TT_{\rm pil}\le t\bigr\}-\Phi(t)\right|
 &\le\alpha_0+e^{-x}+b_N^{\rm dir}
       +E_{\rm pil}/\sqrt{2\pi},\label{eq:pilot-normalized-halfspace}\\
 \sup_t\left|\Prb_{\theta_0}\left\{\norm{T_{\rm pil}}^2\le t\right\}
                 -\Prb\bigl\{\chi_p^2\le t\bigr\}\right|
 &\le\alpha_0+e^{-x}+b_N^{\rm BE}
       +\omega_p\!\left(E_{\rm pil}(2z_*+E_{\rm pil})\right).
       \label{eq:pilot-normalized-wald}
\end{align}
Indeed, on the same event,
$|\norm{T_{\rm pil}}^2-\norm{S_N}^2|
 \le E_{\rm pil}(2z_*+E_{\rm pil})$.
Both bounds may be truncated at one. The likelihood-ratio statistic is
unchanged by information normalization: its bound remains
$\alpha_0+e^{-x}+b_N^{\rm BE}+\omega_p(\delta z_*^2/(1-\delta))$,
with the statistic set to zero on rejected records. The enlarged
$E_{\rm pil}$ is needed for the pilot-normalized estimator and its
squared norm, whereas the Wilks remainder still concerns the original
conditional likelihood.

Write the one-step rule of Section~\ref{app:bg-measured-primer} as $\widehat\theta_1$. Whenever the
score below is defined and $F_\vartheta\succ0$, introduce its standardized form $\psi_\vartheta$; the estimator then has the representation
\[
 \psi_\vartheta(y)=F_\vartheta^{-1/2}\nabla_\theta\log p_\theta(y;\vartheta)|_{\theta=\vartheta}\quad\text{\normalfont and}\quad \widehat\theta_1=\vartheta+F_\vartheta^{-1/2}\frac1N\sum_{k=1}^N\psi_\vartheta(Y_k).
\]
Use the measurement-validity fallback preceding
Theorem~\ref{thm:adaptive}; the maximization-domain test in its preceding setup
is not required for this one-step rule. If the selected score formula
is undefined or $F_\vartheta$ is not positive definite, report the known
parameter $\theta_{\rm safe}$. Define an everywhere-defined normalization
matrix $\widetilde F_\vartheta$ to equal $F_\vartheta$ when the one-step
formula is defined and $I_p$ otherwise. On fallback records this identity
matrix is a reporting convention, not the selected measurement's Fisher
information.
For a pilot where the score formula is defined, write
$\mu_\vartheta=\E_{\theta_0}[\psi_\vartheta(Y)\mid\mathcal F_0]$ and
$\Sigma_\vartheta=\operatorname{Cov}_{\theta_0}(\psi_\vartheta(Y)\mid\mathcal F_0)$
for the one-observation conditional moments. The moment conditions used
below, with finite constants $C_0,C_1,A_\psi\ge0$, are
\begin{equation}\label{eq:pilot-moments}
 \norm{\mu_\vartheta-F_\vartheta^{1/2}(\theta_0-\vartheta)}\le C_0e_0^2,\quad
 \norm{\Sigma_\vartheta-I_p}_\op\le C_1e_0\le1/2
 \quad\text{and}\quad \norm{\psi_\vartheta(y)}\le A_\psi.
\end{equation}
Let $z_N^*(x)$ be~\eqref{eq:bn} with $n=N$ and $A=2\sqrt2A_\psi$, and set
\[
 E_1=C_1e_0\,z_N^*(x)+C_0\sqrt N e_0^2
 \quad\text{and}\quad b_* =\min\{1,2\sqrt2A_\psi/\sqrt N\}.
\]

\begin{lemma}[Conditional standardization]\label{lem:pilot-standardization}
Fix a pilot for which~\eqref{eq:pilot-moments} holds. Then
$\Sigma_\vartheta\succeq I_p/2$, and the standardized sum
\[
 Z_N=N^{-1/2}\sum_{k=1}^N\Sigma_\vartheta^{-1/2}
       (\psi_\vartheta(Y_k)-\mu_\vartheta)
\]
satisfies $\Prb(\norm{Z_N}>z_N^*(x)\mid\mathcal F_0)\le e^{-x}$.
For every unit vector $v$, its conditional normal approximation satisfies
\[
 \sup_t|\Prb(v^TZ_N\le t\mid\mathcal F_0)-\Phi(t)|\le b_*.
\]
\end{lemma}
\begin{proof}
The covariance bound in~\eqref{eq:pilot-moments} gives
$\Sigma_\vartheta\succeq I_p/2$. Write
$\xi_\vartheta=\Sigma_\vartheta^{-1/2}(\psi_\vartheta(Y)-\mu_\vartheta)$.
Conditional on $\mathcal F_0$, its mean is zero and its covariance is
$I_p$. Since $\norm{\mu_\vartheta}\le A_\psi$ and
$\Sigma_\vartheta\succeq I_p/2$,
\[
 \norm{\xi_\vartheta}
 \le\norm{\Sigma_\vartheta^{-1/2}}_\op
       \left(\norm{\psi_\vartheta(Y)}+\norm{\mu_\vartheta}\right)
 \le2\sqrt2A_\psi.
\]
Thus the score concentration argument gives
$\Prb(\norm{Z_N}>z_N^*(x)\mid\mathcal F_0)\le e^{-x}$.
For each unit $v$, the standardized scalar summand has variance one and
\[
 \E\left[|v^T\xi_\vartheta|^3\mid\mathcal F_0\right]
 \le2\sqrt2A_\psi\E\left[(v^T\xi_\vartheta)^2\mid\mathcal F_0\right]
 =2\sqrt2A_\psi.
\]
Scalar Berry--Esseen gives the asserted conditional normal approximation.
\end{proof}

\begin{theorem}[Adaptive one-step estimator]\label{thm:onestep}
Suppose the pilot event $\mathcal A_0$ of Theorem~\ref{thm:adaptive}
has probability at least $1-\alpha_0$. Assume the measurement and
one-step formulas above are defined on $\mathcal A_0$ and the moment
conditions~\eqref{eq:pilot-moments} hold uniformly there.
Use $Z_N$ from Lemma~\ref{lem:pilot-standardization} on $\mathcal A_0$
and set it to zero otherwise. With probability at least $1-\alpha_0-e^{-x}$,
\begin{equation}\label{eq:onestep}
 \norm{\sqrt N \widetilde F_\vartheta^{1/2}(\widehat\theta_1-\theta_0)-Z_N}
 \le C_1e_0\,z_N^*(x)+C_0\sqrt N e_0^2=E_1.
\end{equation}
The halfspace normal-approximation error for
$\sqrt N\,\widetilde F_\vartheta^{1/2}(\widehat\theta_1-\theta_0)$
is at most $\alpha_0+e^{-x}+b_*+E_1/\sqrt{2\pi}$.
\end{theorem}
\begin{proof}
Fix a pilot in $\mathcal A_0$, where
$\widetilde F_\vartheta=F_\vartheta$, and apply
Lemma~\ref{lem:pilot-standardization}.
Put $r_\vartheta=\mu_\vartheta-F_\vartheta^{1/2}(\theta_0-\vartheta)$.
The estimator representation gives the exact identity
\[
 T_N:=\sqrt N F_\vartheta^{1/2}(\widehat\theta_1-\theta_0)
 =\Sigma_\vartheta^{1/2}Z_N+\sqrt N r_\vartheta.
\]
For every eigenvalue $\lambda$ of $\Sigma_\vartheta$,
$|\sqrt\lambda-1|=|\lambda-1|/(\sqrt\lambda+1)\le|\lambda-1|$.
Consequently, on the score event,
\[
 \norm{T_N-Z_N}
 \le C_1e_0\norm{Z_N}+C_0\sqrt N e_0^2
 \le C_1e_0z_N^*(x)+C_0\sqrt N e_0^2=E_1.
\]
For every unit $v$, comparison at thresholds $t-E_1$ and $t+E_1$,
followed by the normal density bound, gives
\[
 \sup_t|\Prb(v^TT_N\le t\mid\mathcal F_0)-\Phi(t)|
 \le b_*+e^{-x}+E_1/\sqrt{2\pi}.
\]
These bounds are uniform over good pilots. Integrating over
$\mathcal A_0$ and adding its complement's probability $\alpha_0$
proves both claims, including the prescribed fallback records.
\end{proof}
The conditional mean and covariance bounds in~\eqref{eq:pilot-moments} can be checked by Taylor expansion at the pilot, without assuming that the pilot equals the true parameter. At $\theta=\vartheta$, normalization gives mean zero and derivative $F_\vartheta^{1/2}$ for $\theta\mapsto\sum_y p_\theta(y;\vartheta)\psi_\vartheta(y)$. The line segment joining $\vartheta$ to $\theta_0$ must lie in the
neighborhood on which the second-derivative bound is uniform; this is
immediate if both points lie in a fixed convex compact neighborhood.
A uniform bound $2C_0$ on the directional second derivatives of this mean
map along that segment gives
$\norm{\mu_\vartheta-F_\vartheta^{1/2}(\theta_0-\vartheta)}\le C_0e_0^2$
by Taylor's formula. A covariance Lipschitz constant $C_1$ gives $\norm{\Sigma_\vartheta-I_p}_\op\le C_1e_0$; requiring $C_1e_0\le1/2$ yields the covariance condition in~\eqref{eq:pilot-moments}. Smooth positive finite-outcome probabilities and uniformly positive information on a compact neighborhood supply finite constants. When a smooth SLD eigenbasis can be chosen in a scalar model, Proposition~\ref{prop:equality} links this implementable adaptive construction directly to intrinsic SLD information.

For scalar information rescaling with a constant $L_{\mathbb I}\ge0$,
define the enlarged error budget
\[
 E_{\rm SLD}=(1+L_{\mathbb{I}}e_0)E_1+L_{\mathbb{I}}e_0\,z_N^*(x).
\]
\begin{corollary}[Pilot-adaptive scalar SLD inference]\label{cor:adaptive-sld}
Under Theorem~\ref{thm:onestep}, consider a one-parameter faithful model
and measure in the SLD eigenbasis at the pilot, with a measurable locally
smooth choice satisfying~\eqref{eq:pilot-moments}. Then
$F_\vartheta=\mathbb I(\vartheta)$. If
$|\sqrt{\mathbb I(\theta_0)/\mathbb I(\vartheta)}-1|\le L_{\mathbb I}e_0$
on $\mathcal A_0$, the Kolmogorov distance between
$\sqrt{N\mathbb I(\theta_0)}(\widehat\theta_1-\theta_0)$ and $N(0,1)$
is at most
\[
 \alpha_0+e^{-x}+b_*+E_{\rm SLD}/\sqrt{2\pi}.
\]
\end{corollary}
\begin{proof}
By Proposition~\ref{prop:equality}, $F_\vartheta=\mathbb I(\vartheta)$.
On $\mathcal A_0$, put
$q_\vartheta=\sqrt{\mathbb I(\theta_0)/\mathbb I(\vartheta)}$ and
$T_N=\sqrt{N\mathbb I(\vartheta)}(\widehat\theta_1-\theta_0)$.
Theorem~\ref{thm:onestep} gives $|T_N-Z_N|\le E_1$ on its score event.
Since $|q_\vartheta-1|\le L_{\mathbb I}e_0$, on that event
\[
 \begin{aligned}
 |q_\vartheta T_N-Z_N|
 &\le q_\vartheta|T_N-Z_N|+|q_\vartheta-1||Z_N|\\
 &\le(1+L_{\mathbb I}e_0)E_1
       +L_{\mathbb I}e_0z_N^*(x)=E_{\rm SLD}.
 \end{aligned}
\]
Here $q_\vartheta T_N=\sqrt{N\mathbb I(\theta_0)}
(\widehat\theta_1-\theta_0)$. Apply the conditional scalar normal
comparison from Lemma~\ref{lem:pilot-standardization}, shifting thresholds by
$E_{\rm SLD}$, and add the score-event and pilot failure probabilities.
\end{proof}

For allocation comparisons, identify a random vector with its law and write
\[
 d_{\rm H}(U,V)=\sup_{\norm u=1,\,t\in\R}
 \left|\Prb\{u^TU\le t\}-\Prb\{u^TV\le t\}\right|
\]
for halfspace distance. Put $T_N=\sqrt N\,\widetilde F_\vartheta^{1/2}
(\widehat\theta_1-\theta_0)$, $L_0=\log(2/\alpha_0)$, and
\[
 \varepsilon_{\rm alloc}
 =\sqrt{L_0/n_0}\,z_N^*(x)+\sqrt N\,L_0/n_0.
\]

\begin{lemma}[Pilot allocation and total-sample normalization]\label{lem:pilot-allocation}
Under Theorem~\ref{thm:onestep}, suppose $p$ is fixed,
$C_0,C_1,A_\psi$ are uniform in $n$, $n_0=o(n)$, and
$e_0^2\le CL_0/n_0$ for a constant $C$ independent of $n$.
Then the remainder in~\eqref{eq:onestep} is $O(\varepsilon_{\rm alloc})$,
and
\[
 d_{\rm H}\bigl(\sqrt{n/N}\,T_N,N(0,I_p)\bigr)
 \le\alpha_0+e^{-x}+b_*+O(\varepsilon_{\rm alloc}+n_0/n).
\]
\end{lemma}
\begin{proof}
Substitution of the pilot bound into~\eqref{eq:onestep} gives the
remainder estimate. Let $Z\sim N(0,I_p)$ and $a_n=\sqrt{n/N}$.
Positive scalar rescaling preserves halfspaces, so
\begin{equation}\label{eq:adaptive-scale}
 d_{\rm H}(a_nT_N,Z)
 \le d_{\rm H}(T_N,Z)+\sup_t|\Phi(t/a_n)-\Phi(t)|
 \le d_{\rm H}(T_N,Z)+O(|a_n-1|).
\end{equation}
The final estimate follows by differentiating in $\log a_n$ and using
$\sup_t|t\phi(t)|<\infty$. Since $n_0=o(n)$,
$|a_n-1|=O(n_0/n)$. The normal-approximation bound in
Theorem~\ref{thm:onestep} now proves the claim.
\end{proof}

For the power allocation $n_0=\lceil n^\gamma\rceil<n$ with
$1/2<\gamma<1$, Lemma~\ref{lem:pilot-allocation} gives the remainder
\[
 O\left(n^{-\gamma/2}\sqrt{\log(2/\alpha_0)}\,z_N^*(x)
       +n^{1/2-\gamma}\log(2/\alpha_0)\right)
\]
and an additional scale discrepancy $O(n^{\gamma-1})$ under
total-sample normalization. Ignoring logarithms and balancing this
discrepancy with the pilot-bias term gives $\gamma=3/4$ and order
$n^{-1/4}$ for these upper-bound contributions. The full bound also
contains $\alpha_0+e^{-x}$ and the Berry--Esseen term $b_*$.
For an explicit schedule including the failure probabilities, take
\[
 \alpha_{0,n}=n^{-1},\quad x_n=\log n,\quad
 n_0=\left\lceil n^{3/4}\sqrt{\log n}\right\rceil
 \quad\text{and}\quad N=n-n_0.
\]
\begin{corollary}[Pilot allocation]\label{cor:split}
Suppose the conditions of Theorem~\ref{thm:onestep} hold uniformly for
the displayed schedule, with fixed $p$, constants $C_0,C_1,A_\psi$
uniform in $n$, and pilot bounds
$\Prb_{\theta_0}\{\norm{\vartheta-\theta_0}\le e_0\}\ge1-n^{-1}$ and
$e_0^2\le C\log(2n)/n_0$, where $C$ is independent of $n$.
Then $n_0<n$ for all sufficiently large $n$, $N/n\to1$, and
\begin{equation}\label{eq:adaptive-full-rate}
 d_{\rm H}\!\left(
 \sqrt n\,\widetilde F_\vartheta^{1/2}(\widehat\theta_1-\theta_0),
 N(0,I_p)\right)=O\!\left(n^{-1/4}\sqrt{\log n}\right).
\end{equation}
\end{corollary}
\begin{proof}
For the specified sequence,
$n_0/n=O(n^{-1/4}\sqrt{\log n})\to0$, and hence $N/n\to1$.
The individual contributions satisfy
\begin{align*}
 e_0^2&=O(\log n/n_0)=O(n^{-3/4}\sqrt{\log n}),
 &z_N^*(x_n)&=O(\sqrt{\log n}),\\
 e_0z_N^*(x_n)&=O(n^{-3/8}(\log n)^{3/4}),
 &\sqrt N e_0^2&=O(n^{-1/4}\sqrt{\log n}).
\end{align*}
Thus $C_1e_0\le1/2$ eventually, and the covariance-error contribution
$C_1e_0z_N^*(x_n)$ is smaller than the leading rate in
\eqref{eq:adaptive-full-rate}. The Berry--Esseen term is $O(n^{-1/2})$,
while $\alpha_{0,n}+e^{-x_n}=2/n$. Since
$n_0/n=O(n^{-1/4}\sqrt{\log n})$, Lemma~\ref{lem:pilot-allocation}
proves~\eqref{eq:adaptive-full-rate}.
\end{proof}
A concrete pilot for a Bloch vector measures each Pauli coordinate on $k$ copies and projects the three empirical means onto a known closed interior Bloch ball containing the truth. Hoeffding's inequality and nonexpansiveness of projection give
$e_0^2=6\log(6/\alpha_0)/k$ with failure probability at most $\alpha_0$ ($n_0=3k$). Thus the pilot assumption is achievable with explicit constants. 

\subsection{Exact conditional risk and allocation}\label{app:adaptive-risk}
For the one-step estimator, write
$$r_\vartheta=\mu_\vartheta-F_\vartheta^{1/2}(\theta_0-\vartheta).$$
Given $W\succ0$ and a known compact convex set $K$ of squared
$W$-diameter $D_W^2$, let $\widehat\theta_K$ be the $W$-metric projection
of the one-step output onto $K$, using an arbitrary $K$-valued output
on otherwise undefined inputs.
\begin{proposition}[Adaptive one-step risk]\label{prop:adaptive-risk}
Under Theorem~\ref{thm:onestep}, conditional on any pilot in $\mathcal A_0$,
\begin{equation}\label{eq:adaptive-risk-exact}
 \E\left[\norm{\sqrt N F_\vartheta^{1/2}(\widehat\theta_1-\theta_0)}^2\,\middle|\,\mathcal F_0\right]
 =\Tr\Sigma_\vartheta+N\norm{r_\vartheta}^2 
 \le p(1+C_1e_0)+NC_0^2e_0^4.
\end{equation}
If $\theta_0\in K$, $F_\vartheta\succeq mI_p$, and
$\Tr(WF_\vartheta^{-1})\le C_{\rm target}+L_Fe_0$ on $\mathcal A_0$,
then the projected estimator defined above satisfies
\begin{align}\label{eq:adaptive-risk}
 &n\E\norm{W^{1/2}(\widehat\theta_K-\theta_0)}^2\\
 &
 \le\frac nN\bigl\{C_{\rm target}+(L_F+C_1\Tr W/m)e_0\bigr\}
 +\frac{n\norm W_\op C_0^2e_0^4}{m}+nD_W^2\alpha_0.\nonumber
\end{align}
\end{proposition}
\begin{proof}
Fix a pilot in $\mathcal A_0$. In the identity
$T_N=\Sigma_\vartheta^{1/2}Z_N+\sqrt N r_\vartheta$ from the proof of
Theorem~\ref{thm:onestep}, the conditional mean and covariance of $Z_N$
are zero and $I_p$. The cross term therefore vanishes, giving
\[
 \E\left[\norm{T_N}^2\,\middle|\,\mathcal F_0\right]
 =\Tr\Sigma_\vartheta+N\norm{r_\vartheta}^2
 \le p(1+C_1e_0)+NC_0^2e_0^4.
\]
For weighted loss, put $H_\vartheta=F_\vartheta^{-1/2}WF_\vartheta^{-1/2}$.
The same calculation yields
\[
 N\E\left[\norm{W^{1/2}(\widehat\theta_1-\theta_0)}^2\,\middle|\,\mathcal F_0\right]
 =\Tr(H_\vartheta\Sigma_\vartheta)
  +N r_\vartheta^TH_\vartheta r_\vartheta.
\]
Since $H_\vartheta\succeq0$ and
$\Sigma_\vartheta\preceq(1+C_1e_0)I_p$,
\[
 \Tr(H_\vartheta\Sigma_\vartheta)
 \le\Tr(WF_\vartheta^{-1})+C_1e_0\Tr H_\vartheta
 \le C_{\rm target}+(L_F+C_1\Tr W/m)e_0.
\]
Here $F_\vartheta^{-1}\preceq m^{-1}I_p$ gives
$\Tr H_\vartheta\le\Tr W/m$ and
$\norm{H_\vartheta}_\op\le\norm W_\op/m$, so
\[
 r_\vartheta^TH_\vartheta r_\vartheta
 \le\frac{\norm W_\op}{m}\norm{r_\vartheta}^2
 \le\frac{\norm W_\op C_0^2e_0^4}{m}.
\]
Projection cannot increase loss to $\theta_0\in K$, and every reported
value lies in $K$. Thus, writing $L_K=\norm{W^{1/2}(\widehat\theta_K-\theta_0)}^2$,
\[
 \E L_K
 \le\E\left[\Id_{\mathcal A_0}
       \E\left[\norm{W^{1/2}(\widehat\theta_1-\theta_0)}^2
                   \,\middle|\,\mathcal F_0\right]\right]
       +D_W^2\Prb(\mathcal A_0^c).
\]
Insert the uniform conditional bounds above, use
$\Prb(\mathcal A_0^c)\le\alpha_0$, and multiply by $n$ to obtain
\eqref{eq:adaptive-risk}.
\end{proof}
If $e_0^2=O(\log n/n_0)$ and $\alpha_0\le n^{-2}$, the excess upper bound in~\eqref{eq:adaptive-risk} has orders $n_0/n$, $\sqrt{\log n/n_0}$, and $n(\log n)^2/n_0^2$. Ignoring logarithms, $n_0\asymp n^{2/3}$ balances these at $n^{-1/3}$. The target $C_{\rm target}$ is an assumed measurement-information benchmark, not automatically the Holevo value. For example, a scalar pilot-selected SLD eigenbasis has $F_\vartheta=\mathbb{I}(\vartheta)$; a Lipschitz bound for $1/\mathbb{I}$ gives $C_{\rm target}=1/\mathbb{I}(\theta_0)$ with $W=1$. The $n^{2/3}$ allocation therefore concerns this generic quadratic-risk
bound, whereas the $n^{3/4}\sqrt{\log n}$ allocation in Corollary~\ref{cor:split}
concerns its distributional bound with total-sample normalization.
The order $\sqrt n$ pilot in Section~\ref{sec:exact-adaptive} instead
uses the exact affine scalar risk calculation.

\subsection{Exact adaptive scalar-qubit efficiency}\label{sec:exact-adaptive}
Adaptive quantum estimation has an established asymptotic theory; see
\citet{Fujiwara2006,Fujiwara2011}, including the corrigendum.
The following affine model permits an exact finite-sample risk calculation.
\begin{lemma}[Affine conditional unbiasedness]\label{lem:affine-unbiased}
For an affine family $\rho_\theta=\rho_*+\sum_i\theta_iV_i$, fix a pilot center $v$ in the physical model and hold its selected finite-outcome POVM $M_v$ fixed. Write $p_\theta(y;v)=\Tr\rho_\theta M_{v,y}$ for its outcome probabilities. Assume $p_v(y;v)>0$ for every retained outcome and let $F_v=F(v;M_v)\succ0$ be its Fisher matrix at $v$. Then the influence value
\[
 T_v(y)=v+F_v^{-1}\nabla_\theta\log p_\theta(y;v)|_{\theta=v}
\]
satisfies $\E_\theta T_v(Y)=\theta$ for every physical $\theta$. Thus the conditional bias bound in Theorem~\ref{thm:onestep} is zero in this setting.
\end{lemma}
\begin{proof}
Fix the pilot value $v$. Let $g_v(y)$ be the gradient of
$p_\theta(y;v)$ at $\theta=v$, and set
$s_v(y)=g_v(y)/p_v(y;v)$. Affinity of the model and normalization of
the outcome probabilities give
\[
 p_\theta(y;v)=p_v(y;v)+g_v(y)^T(\theta-v)
 \quad\text{and}\quad \sum_y g_v(y)=0.
\]
Since $F_v=\sum_y g_v(y)g_v(y)^T/p_v(y;v)$, the score mean is
\[
 \E_\theta\bigl[s_v(Y)\bigr]
 =\sum_y g_v(y)
  +\left(\sum_y\frac{g_v(y)g_v(y)^T}{p_v(y;v)}\right)(\theta-v)
 =F_v(\theta-v).
\]
Substituting into the definition of $T_v$ yields
\[
 \E_\theta\bigl[T_v(Y)\bigr]
 =v+F_v^{-1}\E_\theta\bigl[s_v(Y)\bigr]
 =v+F_v^{-1}F_v(\theta-v)=\theta.
\]
For the standardized score $\psi_v=F_v^{-1/2}s_v$, this also gives
$\E_\theta\psi_v(Y)=F_v^{1/2}(\theta-v)$. Thus the conditional mean
remainder in~\eqref{eq:pilot-moments} is identically zero, and its bias
bound holds with $C_0=0$.
\end{proof}

For the scalar qubit model below, we also compute the conditional variance
explicitly, yielding an exact finite-sample risk identity. Consider the
scalar submodel $\rho_\theta=(\Id+s\sigma_z+\theta\sigma_x)/2$ of~\eqref{eq:qubit}, with the same known
$0<s<1$ and $a_s=1-s^2$, restricted to $|\theta|<\sqrt{a_s}$. Its scalar SLD is
\begin{equation}\label{eq:scalar-sld}
 L(\theta)=\frac{-\theta\Id+a_s\sigma_x+s\theta\sigma_z}
                  {a_s-\theta^2}.
\end{equation}
Direct multiplication verifies
\[
 \tfrac12\bigl\{\rho_\theta L(\theta)+L(\theta)\rho_\theta\bigr\}
 =\tfrac12\sigma_x=\partial_\theta\rho_\theta
 \quad\text{and}\quad \Tr\rho_\theta L(\theta)^2
 =\Tr(\partial_\theta\rho_\theta)L(\theta)
 =\frac{a_s}{a_s-\theta^2}.
\]
Measure $\sigma_x$ on $n_0$ copies and let $v$ be the sample mean.
As in the general pilot split, $N=n-n_0$. The normalizer and measurement direction for each
remaining copy are
\[
 D_v=(a_s^2+s^2v^2)^{1/2}\quad\text{\normalfont and}\quad
 m(v)=\frac{(a_s,0,sv)}{D_v}.
\]
For the outcome $Y\in\{-1,1\}$, the calibrated estimator is
\[
 T_v(Y)=\frac{D_vY-s^2v}{a_s}.
\]
Report $\widehat\theta_A=N^{-1}\sum_kT_v(Y_k)$. For physical $v$,
$m(v)$ is the normalized Pauli coefficient vector in~\eqref{eq:scalar-sld},
so measuring $m(v)\cdot\sigma$ implements a projective measurement in the eigenbasis of the SLD at $v$.
The displayed direction is defined for every pilot outcome, including values outside the physical interval; there it is simply a specified continuation, not an SLD at a nonexistent state.
Indeed, $D_v\ge a_s>0$, and for every real $v$ and physical $\theta$,
\[
 D_v^2-(a_s\theta+s^2v)^2
 =a_s\bigl\{a_s-\theta^2+s^2(v-\theta)^2\bigr\}>0.
\]
Consequently the conditional outcome probabilities
\[
 \Prb_\theta(Y=\pm1\mid v)
 =\frac12\left(1\pm\frac{a_s\theta+s^2v}{D_v}\right)
\]
are strictly between zero and one and sum to one. Thus the continuation
defines a valid measurement law even when the pilot is outside the
physical parameter interval.

\begin{theorem}[Exact adaptive scalar risk]\label{thm:exact-adaptive}
For every integer $1\le n_0<n$ and every physical $\theta$, the preceding implementable procedure is unbiased and
\begin{equation}\label{eq:exact-adaptive}
 n\E_\theta(\widehat\theta_A-\theta)^2
 =\frac{n}{n-n_0}\left\{1-\frac{\theta^2}{a_s}
 +\frac{s^2}{a_s}\frac{1-\theta^2}{n_0}\right\}.
\end{equation}
The SLD information is $\mathbb{I}(\theta)=a_s/(a_s-\theta^2)$.
For fixed $s\in(0,1)$, choosing $n_0\asymp\sqrt n$ with fixed positive
upper and lower multipliers gives
\begin{equation}\label{eq:adaptive-uniform-risk}
 \sup_{|\theta|<\sqrt{a_s}}
 \left|n\E_\theta(\widehat\theta_A-\theta)^2-\mathbb{I}(\theta)^{-1}\right|
 =O(n^{-1/2}).
\end{equation}
This is uniform additive convergence of normalized risk for the unprojected
estimator $\widehat\theta_A$; the constant may depend on the fixed $s$.
\end{theorem}
\begin{proof}
Conditional on $v$, $\E_\theta Y=(a_s\theta+s^2v)/D_v$. Algebra gives
\begin{equation}\label{eq:adaptive-conditional}
 \E_\theta(T_v(Y)\mid v)=\theta\quad\text{\normalfont and}\quad
 \operatorname{Var}_\theta(T_v(Y)\mid v)
 =1-\theta^2/a_s+(s^2/a_s)(v-\theta)^2.
\end{equation}
The pilot has $\E(v-\theta)^2=(1-\theta^2)/n_0$. Average the conditional variance, divide by $N$, and multiply by $n$; conditional unbiasedness eliminates the between-pilot mean variance. The scalar SLD equation gives the stated $\mathbb{I}(\theta)$.
Subtracting its reciprocal from~\eqref{eq:exact-adaptive} gives
\[
 n\E_\theta(\widehat\theta_A-\theta)^2-\mathbb{I}(\theta)^{-1}
 =\frac{n_0}{n-n_0}\left(1-\frac{\theta^2}{a_s}\right)
 +\frac{ns^2}{a_sn_0(n-n_0)}(1-\theta^2).
\]
Throughout $|\theta|<\sqrt{a_s}$, both parenthesized factors belong to
$(0,1]$. Consequently
\[
 0\le n\E_\theta(\widehat\theta_A-\theta)^2-\mathbb{I}(\theta)^{-1}
 \le\frac{n_0}{n-n_0}+\frac{ns^2}{a_sn_0(n-n_0)}.
\]
For fixed $s$ and $n_0\asymp\sqrt n$, the right side is
$O(n^{-1/2})$, independently of $\theta$, proving
\eqref{eq:adaptive-uniform-risk}. No pilot-good event or localization
probability is required.
\end{proof}
Uniform additive convergence does not imply uniform relative efficiency:
$\mathbb{I}(\theta)^{-1}=1-\theta^2/a_s$ tends to zero at the physical boundary.
The pilot observations can also be retained through
$\widehat\theta_C=(n_0/n)v+(N/n)\widehat\theta_A$. Conditional
unbiasedness implies zero covariance of $v$ and $\widehat\theta_A$,
although they are not generally independent: since both means equal
$\theta$,
\[
 \operatorname{Cov}_\theta(v,\widehat\theta_A)
 =\E_\theta\!\left[(v-\theta)
   \E_\theta(\widehat\theta_A-\theta\mid v)\right]=0.
\]
Consequently
\begin{equation}\label{eq:adaptive-combined}
 n\operatorname{Var}_\theta(\widehat\theta_C)
 =\mathbb{I}(\theta)^{-1}+\frac{s^2}{a_s}\left\{\frac{n_0}{n}\theta^2
 +\frac{N}{nn_0}(1-\theta^2)\right\}.
\end{equation}
This follows by adding their weighted variances using~\eqref{eq:exact-adaptive}. \limitationtext{Projection of either estimator onto a known interval containing the truth decreases squared loss; the corresponding equality becomes an upper bound.} The generic $n^{-1/3}$ excess-risk calculation in Proposition~\ref{prop:adaptive-risk} uses coarser assumptions and is not a universal cost of quantum adaptation.

\section{External Gaussian comparison and quadratic risk}\label{sec:decision}
Gaussian comparison also supports inference in nonparametric pure-state
models: \citet{ButuceaGutaNussbaum2018}
establish local equivalence to quantum Gaussian white noise, and \citet{LahiryNussbaum2022} obtain sharp asymptotic minimax and
adaptive estimation results. Those models differ from the faithful
finite-dimensional experiments considered here; their comparison theorems
are not premises of the results below.
Section~\ref{app:channel-transfer} defines channel deficiency and proves bounded-risk transfer. Section~\ref{app:gaussian-radius} explains the finite-radius correction and proves a classical Gaussian bound. Section~\ref{app:channel-qubit} applies an external qubit channel theorem under its stated restrictions.

\subsection{Channel comparison and risk transfer}\label{app:channel-transfer}
Let $\cG_r=\{\phi_h:\norm h\le r\}$ be another experiment, possibly a hybrid quantum Gaussian experiment. For hybrids, trace norm means the norm of normal states on the associated classical--quantum algebra. For ordinary density matrices it is the Schatten trace norm. We retain the distance conventions of Section~\ref{sec:distances}: $d_{\mathrm{tr}}(\sigma,\tau)=\tfrac12\norm{\sigma-\tau}_1$ and $\TV(P,Q)=\sup_A|P(A)-Q(A)|$.
To define one-sided deficiency and two-sided distance, write $\cQ=\cQ_{n,r}$ and $\cG=\cG_r$. Then
\[
 \delta(\cQ,\cG)=\inf_T\sup_{\norm h\le r}d_{\mathrm{tr}}(T(\rho_{n,h}),\phi_h)\quad\text{\normalfont and}\quad \Delta(\cQ,\cG)=\max\bigl\{\delta(\cQ,\cG),\delta(\cG,\cQ)\bigr\}.
\]
The infima range over parameter-independent channels. Our risk comparisons are the standard bounded-deficiency implications, including their direction and loss-dependent constants.

\begin{assumption}[Separate channel comparison]\label{ass:channels}
There are parameter-independent channels $T:\cQ_{n,r}\to\cG_r$ and $S:\cG_r\to\cQ_{n,r}$ satisfying
\begin{equation}\label{eq:channels}
 \sup_{\norm h\le r}d_{\mathrm{tr}}(T(\rho_{n,h}),\phi_h)\le\epsilon_+\quad 
 \text{\normalfont and}\quad
 \sup_{\norm h\le r}d_{\mathrm{tr}}(S(\phi_h),\rho_{n,h})\le\epsilon_-.
\end{equation}
\end{assumption}
Theorem~\ref{thm:collective} and Lemma~\ref{lem:quadratic-state-support} do not establish the forward and reverse Gaussian channel bounds in Assumption~\ref{ass:channels}. When a model is covered by a strong qLAN theorem, its channels must be imported with that theorem's spectral and neighborhood restrictions. In particular, the result of \citet{KG2009} cannot be treated as a dimension-uniform consequence of a one-copy Taylor bound.

For outcomes in an event $B$, the outcome-law convention of Section~\ref{sec:distances} gives $P_{n,h}^M(B)=\Tr\rho_{n,h}M(B)$ and $P_h^N(B)=\phi_h(N(B))$. For an action-valued POVM $M$ with measurable action space $\mathcal D$, whose outcome $a$ is the reported decision, and nonnegative loss $\ell(a,h)$, let $R_{\mathcal E}(M,h)$ denote its risk in experiment $\mathcal E$. The minimax risk $\mathfrak R_{\mathcal E}(\ell,r)$ from Section~\ref{app:bg-risk-primer} is the infimum over these POVMs of the supremum risk on $\norm h\le r$.

\begin{proposition}[Representation and bounded-risk transfer]\label{thm:transfer}
Under Assumption~\ref{ass:channels}, every POVM $M$ on $\cQ_{n,r}$ has a POVM $N=S^*M$ on $\cG_r$ such that
\[
 \sup_{\norm h\le r}\TV(P_{n,h}^M,P_h^N)\le\epsilon_-.
\]
Every POVM $N$ on $\cG_r$ pulls back to $T^*N$ with error at most $\epsilon_+$. For $0\le\ell\le B_\ell$,
\begin{equation}\label{eq:bounded-risk}
 \mathfrak R_{\cG}(\ell,r)-B_\ell\epsilon_-\le\mathfrak R_{\cQ}(\ell,r)
 \le\mathfrak R_{\cG}(\ell,r)+B_\ell\epsilon_+.
\end{equation}
\end{proposition}
\begin{proof}
The channel adjoint is positive and unital, so $N(A)=S^*(M(A))$ is a
POVM. For every measurable $A$, trace duality gives
\[
 \left|\Tr\rho_{n,h}M(A)-\phi_h\bigl(S^*M(A)\bigr)\right|
 =\left|\Tr\bigl\{\rho_{n,h}-S(\phi_h)\bigr\}M(A)\right|
 \le d_{\mathrm{tr}}\bigl(\rho_{n,h},S(\phi_h)\bigr).
\]
Taking the supremum over events proves the TV bound; the same argument
with $T$ proves the forward bound. For each fixed $h$, integrating the
level sets $A_t(h)=\{a:\ell(a,h)>t\}$ gives
\[
 \left|R_{\cQ}(M,h)-R_{\cG}(S^*M,h)\right|
 \le\int_0^{B_\ell}
       \left|P_{n,h}^M\bigl(A_t(h)\bigr)-P_h^{S^*M}\bigl(A_t(h)\bigr)\right|dt
 \le B_\ell\epsilon_-.
\]
Consequently, for every $M$,
\[
 \mathfrak R_{\cG}(\ell,r)
 \le\sup_{h\in K_r}R_{\cG}(S^*M,h)
 \le\sup_{h\in K_r}R_{\cQ}(M,h)+B_\ell\epsilon_-.
\]
Taking the infimum over $M$ proves the lower bound in
\eqref{eq:bounded-risk}. Conversely, choose a Gaussian POVM $N_\zeta$
whose supremum risk is within $\zeta>0$ of its infimum. Its pullback obeys
\[
 \mathfrak R_{\cQ}(\ell,r)
 \le\sup_{h\in K_r}R_{\cQ}(T^*N_\zeta,h)
 \le\mathfrak R_{\cG}(\ell,r)+\zeta+B_\ell\epsilon_+.
\]
Letting $\zeta\downarrow0$ proves the upper bound.
\end{proof}

For quadratic loss on a compact parameter ball, projection avoids a universal moment assumption on arbitrary estimators. Use the loss $\ell_W(a,h)=(a-h)^TW(a-h)$ from Section~\ref{app:bg-risk-primer}, with $W\succ0$, and the projection $\Pi_r^W$ onto $K_r$ from Section~\ref{sec:distances}.

\begin{proposition}[Compact-ball quadratic-risk transfer]\label{thm:quadratic}
Under Assumption~\ref{ass:channels},
\begin{equation}\label{eq:quad-risk}
 \mathfrak R_{\cG}(\ell_W,r)-4\norm W_\op r^2\epsilon_-
 \le\mathfrak R_{\cQ}(\ell_W,r) 
 \le\mathfrak R_{\cG}(\ell_W,r)+4\norm W_\op r^2\epsilon_+.
\end{equation}
For the upper bound, a near-optimal Gaussian POVM may be projected by $\Pi_r^W$ and pulled back through $T$.
\end{proposition}
\begin{proof}
For $b=\Pi_r^W(a)$ and $h\in K_r$, the projection variational
inequality gives $\ip{a-b}{h-b}_W\le0$. Hence
\[
 \norm{a-h}_W^2
 =\norm{a-b}_W^2+\norm{b-h}_W^2
    +2\ip{a-b}{b-h}_W
 \ge\norm{b-h}_W^2.
\]
Projecting the action of any POVM therefore cannot increase its risk
at any $h\in K_r$. Since projected POVMs are also admissible in the
unrestricted problem, the two infima coincide in each experiment.
On this common restricted action space,
\[
 0\le\ell_W(a,h)
 \le\norm W_\op\norm{a-h}^2
 \le\norm W_\op\left(\norm a+\norm h\right)^2
 \le4\norm W_\op r^2.
\]
Proposition~\ref{thm:transfer} with $B_\ell=4\norm W_\op r^2$
now proves~\eqref{eq:quad-risk}. Applying its upper-bound construction
to a projected near-optimal Gaussian POVM proves the final assertion.
\end{proof}

\subsection{The Gaussian radius correction}\label{app:gaussian-radius}
Suppose the full Gaussian shift experiment has quadratic minimax value $C_W$,
with a procedure of constant risk $C_W$. For the standard quantum Gaussian
shift, the minimax lower inequality is given by \citet[Theorem 5.5]{FY2023}. An unbiased Gaussian measurement
attaining the Holevo value is constructed in \citet[Theorem B.7]{YFGSupp}; Corollary B.6 in the same supplement
identifies that Gaussian value with the original model's Holevo value.
Together these identify the full-shift minimax value and a constant-risk
attaining procedure in that setting. The radius correction from Section~\ref{app:bg-risk-primer}, now for this Gaussian experiment with full-shift value $C_W$, is
\begin{equation}\label{eq:beta}
 \beta_W(r)=C_W-\mathfrak R_{\cG}(\ell_W,r)\ge0.
\end{equation}
\begin{corollary}[Conditional finite-sample Holevo-type comparison]\label{cor:holevo}
Under Assumption~\ref{ass:channels}, if the full Gaussian shift has constant minimax risk $C_W$, then
\begin{equation}\label{eq:holevo-conditional}
 C_W-\beta_W(r)-4\norm W_\op r^2\epsilon_-
 \le\mathfrak R_{\cQ}(\ell_W,r)
 \le C_W+4\norm W_\op r^2\epsilon_+.
\end{equation}
For fixed $W$, let $\epsilon_{n,+},\epsilon_{n,-}$ be the two channel errors at sample size $n$ and radius $r_n$. Convergence to $C_W$ follows if
$\beta_W(r_n)\to0$, $r_n^2\epsilon_{n,+}\to0$, and $r_n^2\epsilon_{n,-}\to0$.
\end{corollary}
\begin{proof}
Write $R_Q(r)=\mathfrak R_{\cQ}(\ell_W,r)$ and
$R_G(r)=\mathfrak R_{\cG}(\ell_W,r)$, and set
$B_r=4\norm W_\op r^2$. The Gaussian procedure with constant risk
$C_W$ remains admissible when the parameter set is restricted to $K_r$.
Consequently,
\[
 R_G(r)\le\sup_{h\in K_r}C_W=C_W
 \quad\text{and}\quad
 \beta_W(r)=C_W-R_G(r)\ge0.
\]
Proposition~\ref{thm:quadratic} compares the two minimax risks on the
same ball and gives
\[
 R_G(r)-B_r\epsilon_-\le R_Q(r)
 \le R_G(r)+B_r\epsilon_+.
\]
Substituting $R_G(r)=C_W-\beta_W(r)$ into both sides yields
\[
 C_W-\beta_W(r)-B_r\epsilon_-
 \le R_Q(r)
 \le C_W-\beta_W(r)+B_r\epsilon_+.
\]
Since $\beta_W(r)\ge0$, the upper endpoint is at most
$C_W+B_r\epsilon_+$. Together with the unchanged lower endpoint,
this proves~\eqref{eq:holevo-conditional}. The radius correction
therefore measures the difference between the Gaussian minimax risks
on the full parameter space and on the restricted ball; the two
channel errors control transfer in the two directions.

For the convergence assertion, allow the experiments and the radius
to depend on $n$, and put
$R_{Q,n}=\mathfrak R_{\cQ_n}(\ell_W,r_n)$ and
$B_n=4\norm W_\op r_n^2$.
Subtracting $C_W$ from the preceding bounds gives
\[
 -\beta_W(r_n)-B_n\epsilon_{n,-}
 \le R_{Q,n}-C_W\le B_n\epsilon_{n,+}.
\]
All error terms are nonnegative. Hence
\[
 |R_{Q,n}-C_W|
 \le\beta_W(r_n)
   +4\norm W_\op r_n^2\max\{\epsilon_{n,+},\epsilon_{n,-}\}.
\]
For fixed $W$, each term on the right tends to zero under the three
stated conditions, proving $R_{Q,n}\to C_W$.
\end{proof}
The radius term is necessary: the constant estimator zero has risk at most $\norm W_\op r^2$ on the ball, below a positive $C_W$ for small $r$. In original coordinates, $\widehat\theta=\theta_0+\widehat h/\sqrt n$ gives
$\ell_W(\widehat h,h)=n(\widehat\theta-\theta)^TW(\widehat\theta-\theta)$.

\begin{theorem}[Explicit classical Gaussian radius correction]\label{thm:radius}
For the classical Gaussian shift $Y\sim N(h,V)$, $V\succ0$, $r>0$, $\norm h\le r$, and $W\succ0$, the full-shift value specializes to $C_W=\Tr WV$. Then
\begin{equation}\label{eq:radius-bound}
 \frac{C_W^2}{C_W+\norm W_\op r^2}
 \le\beta_W(r)\le
 \Tr W\left\{V-(V^{-1}+\pi^2p\,r^{-2}I_p)^{-1}\right\}
 \le\frac{\pi^2p}{r^2}\Tr WV^2.
\end{equation}
In particular, as $r\to\infty$ with $p,V,W$ fixed, the radius correction is of order $r^{-2}$.
\end{theorem}
\begin{proof}
Put $a=r/\sqrt p$ and give $h$ the product density
$\pi(h)=\prod_{j=1}^p a^{-1}\cos^2(\pi h_j/(2a))$ on $[-a,a]^p$. This cube is contained in the radius-$r$ ball. For the one-dimensional marginal
$q_a(u)=a^{-1}\cos^2(\pi u/(2a))$, the score and its second moment are
\[
 (\log q_a)'(u)=-\frac\pi a\tan\!\left(\frac{\pi u}{2a}\right)
 \quad\text{and}\quad
 \int_{-a}^a\frac{q_a'(u)^2}{q_a(u)}\,du
 =\frac{\pi^2}{a^3}\int_{-a}^a\sin^2\!\left(\frac{\pi u}{2a}\right)du
 =\frac{\pi^2}{a^2}.
\]
The prior vanishes on its boundary. Its coordinate scores have mean
zero and are independent, so its information is $jI_p$ with
$j=\pi^2/a^2=\pi^2p/r^2$. For the joint density of $(Y,h)$, its parameter score is
$U=V^{-1}(Y-h)+\nabla\log\pi(h)$, with $\E UU^T=V^{-1}+jI_p$.
For completeness, write $p_h(y)$ for the Gaussian density and
$q(y,h)=p_h(y)\pi(h)$ for the joint density. The parameter score is
$U=\nabla_h\log q$. Project estimators onto $K_r$ in the $W$ norm, which does
not increase loss and makes their actions bounded. Entrywise integration by parts gives
\[
 \begin{aligned}
 \E\left[(\widehat h_i(Y)-h_i)U_j\right]
 &=\int\!\int(\widehat h_i(y)-h_i)\partial_{h_j}q(y,h)\,dh\,dy\\
 &=\delta_{ij}\int\!\int q(y,h)\,dh\,dy=\delta_{ij}.
 \end{aligned}
\]
The boundary term vanishes because the prior is zero on each cube face.
The derivative acts on $-h_i$; no unbiasedness condition on the estimator
is used. This is the van Trees identity~\citep{GillLevit1995}.
Thus $\E[(\widehat h(Y)-h)U^T]=I_p$. Put $e=\widehat h-h$ and
$J=\E UU^T=V^{-1}+jI_p$. Expanding a positive second-moment matrix gives
\[
 0\preceq\E\left[(e-J^{-1}U)(e-J^{-1}U)^T\right]
       =\E ee^T-J^{-1}.
\]
Thus $\E ee^T\succeq(V^{-1}+jI_p)^{-1}$. Taking a weighted trace preserves this order because $\Tr W H=\Tr(W^{1/2}HW^{1/2})\ge0$ for $H\succeq0$.
Minimax risk dominates this Bayes risk. Since
$V-(V^{-1}+jI_p)^{-1}=jV^2(I_p+jV)^{-1}\preceq jV^2$,
the upper bounds follow. For the lower bound, the estimator $cY$ has worst risk at most $c^2C_W+(1-c)^2\norm W_\op r^2$. Writing $B_r=\norm W_\op r^2$, the minimizing coefficient and risk bound are
\[
 c_* =\frac{B_r}{C_W+B_r}
 \quad\text{and}\quad
 \inf_c\{c^2C_W+(1-c)^2B_r\}=\frac{C_WB_r}{C_W+B_r}.
\]
Subtracting this risk upper bound from $C_W$ gives
$\beta_W(r)\ge C_W^2/(C_W+B_r)$, as asserted.
\end{proof}

\subsection{A concrete strong-qLAN risk consequence}\label{app:channel-qubit}
Fix $1/2<\mu<1$. For a physical displacement $v\in\R^3$ near zero, with $-\mu<v_z<1-\mu$, write its components as $v_x,v_y,v_z$ and define
\[
 U(v)=e^{i(v_x\sigma_x+v_y\sigma_y)}.
\]
The physical qubit chart and its local coordinate are
\[
 \rho_v=U(v)\diag(\mu+v_z,1-\mu-v_z)U(v)^*
 \quad\text{\normalfont and}\quad h=\sqrt n\,v.
\]
Theorem~3.1 and equations~(3.8)--(3.9) of \citet{GJK2008}, with the chart defined in their equations~(3.2)--(3.6), give forward and reverse channels on $\norm h\le n^\eta$, $0\le\eta<1/4$, with trace-norm errors $O(n^{-1/4+\eta+\zeta})$ for every $\zeta>0$. Let $|k\rangle$, $k\ge0$, be the oscillator number basis and $D(z)$ the oscillator displacement by a complex amplitude $z$. Put $q=(1-\mu)/\mu$ and define the thermal state $\tau_\mu=(1-q)\sum_{k\ge0}q^k|k\rangle\langle k|$. Their target is
\[
 N(h_z,\mu(1-\mu))\otimes
 D\left(\sqrt{2\mu-1}(-h_y+ih_x)\right)\tau_\mu D\left(\sqrt{2\mu-1}(-h_y+ih_x)\right)^*,
\]
We keep $\mu$ fixed away from the spectral boundaries. Write
\[
 \cQ_n=\left\{\rho_{h/\sqrt n}^{\otimes n}:h\in\R^3,\ \norm h\le n^\eta\right\}
\]
for all sufficiently large $n$ for which this ball lies in the chart, and write $\cG$ for the displayed Gaussian target experiment.

\begin{corollary}[Qubit risk transfer with a rate]\label{cor:qubit-channel}
For this chart, fixed $W\succ0$, and $r_n=n^\eta$,
\begin{equation}\label{eq:qubit-channel-risk}
 |\mathfrak R_{\cQ_n}(\ell_W,r_n)-\mathfrak R_{\cG}(\ell_W,r_n)|
 =O(n^{-1/4+3\eta+\zeta}).
\end{equation}
Hence the risk difference vanishes when $3\eta+\zeta<1/4$.
\end{corollary}
\begin{proof}
The imported trace-norm bounds give
$\epsilon_{n,+},\epsilon_{n,-}=O(n^{-1/4+\eta+\zeta})$ in trace
distance as well, since that distance is half the trace norm.
Proposition~\ref{thm:quadratic}, with $r_n=n^\eta$, yields
\[
 \left|\mathfrak R_{\cQ_n}(\ell_W,r_n)-\mathfrak R_{\cG}(\ell_W,r_n)\right|
 \le4\norm W_\op n^{2\eta}\max\{\epsilon_{n,+},\epsilon_{n,-}\}
 =O(n^{-1/4+3\eta+\zeta}).
\]
The exponent is negative precisely when $3\eta+\zeta<1/4$.
\end{proof}
This is an eventual rate with constants inherited from the channel theorem. The comparison is with the finite-radius Gaussian risk; replacement by $C_W$ still requires its radius correction. Restricting the chart to $h_x=h_y=0$ leaves a parameter-independent oscillator and a scalar classical Gaussian with variance $\mu(1-\mu)$; Theorem~\ref{thm:radius} then gives an explicit $O(r_n^{-2})$ radius correction as well. This yields an unconditional rate consequence for that commuting submodel, while the full noncommuting comparison retains $\beta_W(r_n)$.
In the commuting submodel, combining the two bounds gives
$O(n^{-1/4+3\eta+\zeta}+n^{-2\eta})$. Choosing $\eta=1/20$ balances the powers and yields $O(n^{-1/10+\zeta})$ for sufficiently small $\zeta>0$. This is optimization of the available upper bounds.

\section{An optional likelihood-surrogate construction}\label{app:polynomial-support}
This appendix gives the assumptions and proofs for an optional route to
collective inference through normalized quadratic states. It proves the
degree-two state comparison used in Corollary~\ref{cor:collective-polynomial}.
Its stronger likelihood remainder assumptions are separate from the direct $C^2$ density
argument. Section~\ref{sec:likelihood-support} recalls the positive likelihood
ratio from Section~\ref{app:bg-local} and proves normalization and tensorization;
Section~\ref{sec:polynomial-states} proves the state approximation and its
collective-measurement consequence.
\subsection{Positive likelihood ratios and regularity}\label{sec:likelihood-support}
For positive definite $\sigma$ and $\rho$, the positive square-root likelihood ratio of Section~\ref{app:bg-local} is the unique positive solution of $\sigma=R\rho R$:
\begin{equation}\label{eq:root-ratio}
 R(\sigma\mid\rho)=\rho^{-1/2}\left(\rho^{1/2}\sigma\rho^{1/2}\right)^{1/2}\rho^{-1/2}.
\end{equation}
For a physical increment $u\in\R^p$ with $\theta_0+u\in\Theta$
and both $\rho$ and $\rho_{\theta_0+u}$ positive definite, the one-copy and tensorized log-likelihood ratios recalled from Section~\ref{app:bg-local} are, using the principal Hermitian matrix logarithm,
\begin{equation}\label{eq:log-ratio}
 \mathcal L(u)=2\log R(\rho_{\theta_0+u}\mid\rho)\quad\text{\normalfont and}\quad
 \Lambda_n(h)=2\log R(\rho_{n,h}\mid\rho^{\otimes n}).
\end{equation}
The tensor ratio is defined for $h$ such that $h/\sqrt n$ lies in this faithful domain. The one-copy ratio satisfies $\rho_{\theta_0+u}=e^{\mathcal L(u)/2}\rho e^{\mathcal L(u)/2}$ and $\Tr\rho e^{\mathcal L(u)}=1$. It generally differs from $\log\rho_{\theta_0+u}-\log\rho$.

When $\mathcal L$ is twice differentiable at zero, the SLDs from~\eqref{eq:sld}, the likelihood Hessian from Section~\ref{app:bg-local}, and the score covariance from~\eqref{eq:derivatives} satisfy
\begin{equation}\label{eq:likelihood-derivatives}
 L_i=\partial_i\mathcal L(0),\quad H_{ij}=\partial_i\partial_j\mathcal L(0),\quad
 J_{ij}=\Tr\rho L_iL_j,\quad \mathbb{I}=\Rea J
 \quad\text{and}\quad \Omega=\Ima J.
\end{equation}
Our ordering convention for $J$ is $L_iL_j$; some cited papers use its transpose. Thus $J=\mathbb{I}+i\Omega$ is Hermitian, $\Omega$ is real antisymmetric, and $h^TJh=h^T\mathbb{I}h$ for real $h$. Recall also the normalized SLD list and score quadratic from Section~\ref{app:bg-local}:
\begin{equation}\label{eq:delta-q}
 \Delta_{n,i}=n^{-1/2}\sum_{k=1}^n L_i^{(k)}\quad\text{\normalfont and}\quad
 Q_n(h)=h^T\Delta_n-\tfrac12h^T\mathbb{I}h\,\Id.
\end{equation}
Retain the score list $L=(L_1,\ldots,L_p)$ and linear form $L[u]=\sum_i u_iL_i$ from Section~\ref{sec:likelihood-setup}, and define the bilinear Hessian operator $H[u,v]=\sum_{ij}u_iv_jH_{ij}$. These contractions are distinct from the nonlinear $\mathcal L(u)$. The vector $\Delta_n=(\Delta_{n,1},\ldots,\Delta_{n,p})$ is a list of observables, not generally a jointly observable random vector. Set $M_2=\sup_{\norm v=1}\norm{H[v,v]}_\op$ for the reference Hessian bound.

\begin{assumption}[Local regularity]\label{ass:regular}
There is a closed ball $U=\{u:\norm u\le a\}$ of radius $a>0$ with $\theta_0+U\subset\Theta$ on which $\rho_{\theta_0+u}\succeq\kappa\Id$ for some $\kappa>0$. The map $u\mapsto\rho_{\theta_0+u}$ is $C^3$ on a neighborhood of $U$. The induced map $\mathcal L$ has
\[
 M_3:=\sup_{u\in U}\sup_{\norm v=1}\norm{D^3\mathcal L(u)[v,v,v]}_\op<\infty.
\]
The real SLD information matrix $\mathbb{I}$ in~\eqref{eq:likelihood-derivatives} is positive definite.
\end{assumption}
Smooth functional calculus gives the asserted smoothness and finite $M_3$ on a fixed compact ball. This additional likelihood regularity is used only for the optional surrogate construction and likelihood calculations in this appendix. The direct collective theorem requires the weaker $C^2$ density assumptions stated in the main text.
The following identities give the normalization and tensorization properties
used in the quadratic surrogate construction.
\begin{lemma}[Normalization and tensorization]\label{lem:normal}
Under Assumption~\ref{ass:regular}, for $h/\sqrt n\in U$,
\begin{gather}
 \partial_i\rho_{\theta_0}=\tfrac12(L_i\rho+\rho L_i)
 \quad\text{and}\quad \Tr\rho L_i=0,\label{eq:likelihood-sld}\\
 \Tr\rho H_{ij}=-\mathbb{I}_{ij},\quad\text{\normalfont and}\quad
 \Lambda_n(h)=\sum_{k=1}^n\mathcal L(h/\sqrt n)^{(k)}.\label{eq:identities}
\end{gather}
\end{lemma}
\begin{proof}
Since $\mathcal L(0)=0$, differentiating
$\rho_{\theta_0+u}=e^{\mathcal L(u)/2}\rho e^{\mathcal L(u)/2}$ at zero gives
\[
 \partial_i\rho_{\theta_0}=\tfrac12(L_i\rho+\rho L_i)
 \quad\text{and}\quad
 \Tr\rho L_i=\Tr\partial_i\rho_{\theta_0}=0.
\]
For real $v$, use
$\mathcal L(tv)=tL[v]+t^2H[v,v]/2+o(t^2)$ in the normalization
$\Tr\rho e^{\mathcal L(tv)}=1$. The second-order term gives
\[
 \Tr\rho H[v,v]=-\Tr\rho L[v]^2=-v^T\mathbb I v.
\]
Both sides are real quadratic forms in $v$. Polarization therefore yields
$\Tr\rho H_{ij}=-\mathbb I_{ij}$.
Finally, write $R=R(\rho_{\theta_0+h/\sqrt n}\mid\rho)$.
The positive operator $R^{\otimes n}$ satisfies the defining sandwich
identity for the product states, so uniqueness identifies it as their
positive likelihood ratio. Operators on distinct tensor factors commute,
and spectral calculus gives
\[
 \Lambda_n(h)=2\log(R^{\otimes n})
 =\sum_{k=1}^n(2\log R)^{(k)}
 =\sum_{k=1}^n\mathcal L(h/\sqrt n)^{(k)}.
\]
\end{proof}

\subsection{Quadratic states and the supporting comparison}\label{sec:polynomial-states}
For the polynomial experiments, the channel deficiency $\delta$ and the
two-sided distance $\Delta$ use the definitions in
Appendix~\ref{app:channel-transfer}; identity channels are admissible because
both state families live on the same Hilbert space.
Assume $\rho$ and $\rho_{\theta_0+u}$ are faithful for $u\in U$ and
$\mathcal L$ is twice continuously differentiable near zero. Its quadratic
Taylor polynomial is
\[
 P_2(u)=L[u]+\tfrac12H[u,u].
\]
For real $u$, the polynomial is Hermitian and $P_2(0)=0$. Define the
normalizing factor, state and local experiment by
\begin{gather*}
 Z_2(u)=\Tr\rho e^{P_2(u)},\quad
 \tau_u^{[2]}=\frac{e^{P_2(u)/2}\rho e^{P_2(u)/2}}{Z_2(u)},\\
 \text{and}\quad \mathcal T_{n,r}^{[2]}=\left\{(\tau_{h/\sqrt n}^{[2]})^{\otimes n}:\norm h\le r\right\}.
\end{gather*}
These are normalized states on the original Hilbert space; a quadratic
exponent does not make them Gaussian states. For the comparison below,
consider finite nonnegative constants $C_2,K_a$ satisfying, on $\norm u\le a$,
\[
 \norm{\mathcal L(u)-P_2(u)}_\op\le C_2\norm u^3
 \quad\text{\normalfont and}\quad
 \max\left\{\norm{\mathcal L(u)}_\op,\norm{P_2(u)}_\op\right\}\le K_a.
\]
The next lemma supplies these constants under the likelihood regularity
assumption; the state comparison also applies whenever the displayed
bounds are verified directly.

\begin{lemma}[Quadratic likelihood remainder constants]\label{lem:quadratic-remainder}
Under Assumption~\ref{ass:regular}, the preceding bounds hold with
$C_2=M_3/6$ and
\[
 K_a=a\sup_{\norm v=1}\norm{L[v]}_\op
 +\frac{a^2}{2}\sup_{\norm v=1}\norm{H[v,v]}_\op+M_3a^3/6.
\]
\end{lemma}
\begin{proof}
Taylor's integral remainder gives
\[
 \mathcal L(u)-P_2(u)
 =\frac12\int_0^1(1-t)^2D^3\mathcal L(tu)[u,u,u]\,dt.
\]
Taking operator norms gives
$\norm{\mathcal L(u)-P_2(u)}_\op\le M_3\norm u^3/6$.
The triangle inequality bounds $\norm{P_2(u)}_\op$ by the first two
terms in $K_a$; adding the remainder bounds $\norm{\mathcal L(u)}_\op$.
\end{proof}

\begin{lemma}[Supporting quadratic-state comparison]\label{lem:quadratic-state-support}
For the faithful model and quadratic polynomial above, suppose the bounds
on $\mathcal L-P_2$ and on $\mathcal L,P_2$ hold with constants $C_2,K_a$.
For $r/\sqrt n\le a$, the identity-comparison error
$\sup_{\norm h\le r}d_{\rm tr}(\rho_{n,h},(\tau_{h/\sqrt n}^{[2]})^{\otimes n})$
is at most
\begin{equation}\label{eq:quadratic-state-support}
 \epsilon_2(n,r)=\min\bigl\{1,e^{K_a/2}C_2r^3/n\bigr\}.
\end{equation}
Identity channels therefore give
$\Delta(\mathcal Q_{n,r},\mathcal T_{n,r}^{[2]})\le\epsilon_2(n,r)$.
\end{lemma}
\begin{proof}
For $C_2,K_a$ satisfying the stated bounds, compare normalized one-copy
amplitudes and then tensorize their purification overlap.
Put $A=e^{\mathcal L(u)/2}\sqrt\rho$ and
$C=e^{P_2(u)/2}\sqrt\rho$. Since $P_2(u)$ is Hermitian, its
exponential is invertible; also $\sqrt\rho\ne0$ because $\Tr\rho=1$.
Thus $C\ne0$, and cyclicity of trace gives
\[
 \norm C_\HS^2=\Tr\rho e^{P_2(u)}=Z_2(u)>0.
\]
The normalized amplitude $B=C/\norm C_\HS$ is therefore well-defined.
We have $AA^*=\rho_{\theta_0+u}$, $BB^*=\tau_u^{[2]}$, and both $A$
and $B$ have HS norm one. Writing $H=\mathcal L(u)$ and
$P=P_2(u)$, the Duhamel identity is
\[
 e^{H/2}-e^{P/2}
 =\tfrac12\int_0^1 e^{(1-t)H/2}(H-P)e^{tP/2}\,dt.
\]
It follows by differentiating $e^{(1-t)H/2}e^{tP/2}$, and does not require $H$ and $P$ to commute. Since $\norm{\sqrt\rho}_\HS=1$ and the product of the two exponential norm bounds is at most $e^{K_a/2}$,
\[
 \norm{A-C}_\HS\le\tfrac12e^{K_a/2}C_2\norm u^3.
\]
The reverse triangle inequality yields
\[
 \norm{A-B}_\HS\le\norm{A-C}_\HS+\left|1-\norm C_\HS\right|
                    \le2\norm{A-C}_\HS.
\]
Vectorization is the map
\[
 |A\rangle\!\rangle=\sum_{ij}A_{ij}|i\rangle\otimes|j\rangle.
\]
This vector is a unit purification of $AA^*$; the analogous statement holds for $B$. Their overlap is $z=\langle A,B\rangle_\HS$, with $|z|\le1$, and $d=\norm{A-B}_\HS$ satisfies $d^2=2-2\Rea z$. In particular,
\[
 1-|z|^2\le2-2\Rea z=d^2
 \quad\text{and}\quad 1-|z|^{2n}=(1-|z|^2)\sum_{k=0}^{n-1}|z|^{2k}\le nd^2.
\]
The inequality $1-|z|^2\le2-2\Rea z$ is equivalent to $|1-z|^2\ge0$. The tensor purifications have overlap $z^n$, so their pure-state trace distance is $\sqrt{1-|z|^{2n}}$. Partial trace cannot increase distance, giving the bound $\min\{1,\sqrt n d\}$ for the mixed tensor states. This explains the $\sqrt n$ accumulation rather than a telescoping bound proportional to $n$. For $u=h/\sqrt n$, the amplitude bound gives explicitly
\[
 \sqrt n\,d\le\sqrt n\,e^{K_a/2}C_2\norm u^3
       =e^{K_a/2}C_2\norm h^3/n.
\]
Taking the supremum over $\norm h\le r$ proves~\eqref{eq:quadratic-state-support}.
Both experiments have the same Hilbert space, so the identity channel in each
direction proves the deficiency claim.
\end{proof}
The bound $e^{K_a/2}M_3r^3/(6n)$ under Assumption~\ref{ass:regular}
reflects second-order matching of the one-copy amplitudes. The exponential
constant must be tracked if the model varies. This remains a noncommutative
state comparison on the original space, not a Gaussian approximation.

Every POVM's outcome laws differ in TV by at most
\eqref{eq:quadratic-state-support}; hence tests and bounded losses transfer
directly. For $W\succ0$ and actions projected in the $W$ metric onto
$\norm h\le r$, the squared loss is at most $4\norm W_\op r^2$.
The minimax risks in $\mathcal Q_{n,r}$ and $\mathcal T_{n,r}^{[2]}$
therefore differ by at most $4\norm W_\op r^2\epsilon_2(n,r)$.
With $C_2=M_3/6$, this is bounded by
$(2/3)\norm W_\op e^{K_a/2}M_3r^5/n$, capped by the bounded-loss
estimate. These constants are the ones used in the inference application.
The argument uses full one-copy operators and does not exponentiate a
compressed quadratic-form bound.

\begin{corollary}[Likelihood-surrogate route]\label{cor:collective-polynomial}
Under Lemma~\ref{lem:quadratic-state-support}, apply Theorem~\ref{thm:collective} to the density family $\tau_u^{[2]}$. Use its trace derivative constants $D_1^{\tau},D_2^{\tau}$ on $U$ to compute $\varepsilon_n^{\tau}$ in~\eqref{eq:collective-TV-budget}. On the original experiment the same POVM satisfies~\eqref{eq:collective-TV} with
\begin{equation}\label{eq:collective-modular}
 \varepsilon_n\quad\text{\normalfont replaced by}\quad
 \min\bigl\{1,\epsilon_2(n,r)+\varepsilon_n^{\tau}\bigr\}.
\end{equation}
This optional route can be useful when surrogate derivatives or structured calculations are simpler. The same substitution applies in~\eqref{eq:collective-risk} and Corollary~\ref{cor:collective-shrinkage}, with the original $C_W,V_X$ and optimization gap.
\end{corollary}
\begin{proof}
At zero, $P_2(0)=0$, $Z_2(0)=1$, and
$\partial_iZ_2(0)=\Tr\rho L_i=0$. Differentiating the normalized
surrogate therefore gives
\[
 \tau_0^{[2]}=\rho
 \quad\text{and}\quad
 \partial_i\tau_0^{[2]}=\tfrac12(L_i\rho+\rho L_i)
 =\partial_i\rho_{\theta_0}.
\]
Thus the calibration constraints, feasible observables, covariance and
Holevo objective agree with those of the original model. The surrogate
is faithful, and smooth finite-dimensional calculus makes its trace
derivative bounds $D_j^\tau$ finite on $U$.
Let $P_h$, $P_h^\tau$ be the output laws of the same POVM under the
original and surrogate product states, and put
$Q_h=N(h,V_X+\eta^2I_p)$. Theorem~\ref{thm:collective} gives
$\TV(P_h^\tau,Q_h)\le\varepsilon_n^\tau$ uniformly on $K_r$.
Measurement contraction and Lemma~\ref{lem:quadratic-state-support} give
\[
 \TV(P_h,Q_h)
 \le\TV(P_h,P_h^\tau)+\TV(P_h^\tau,Q_h)
 \le\epsilon_2(n,r)+\varepsilon_n^\tau.
\]
TV is at most one, proving~\eqref{eq:collective-modular}.
For either projected reporting rule, its loss lies in
$[0,4\norm W_\op r^2]$. Lemma~\ref{lem:decision-transfer} therefore
transfers the corresponding Gaussian risk bound with the same
replacement of the TV budget, proving both risk claims.
\end{proof}

\section{Scalar concentration and measured nets}\label{app:bern}
The measured likelihood and pilot-inference bounds use the following standard supporting concentration inequality, proved here for completeness. The subsequent net arguments and growing-dimensional result provide the remaining measured-inference estimates.
\begin{lemma}[Bounded-variable Bernstein inequality]\label{lem:bern}
Let $X_1,\ldots,X_n$ be independent centered real variables with $|X_k|\le b$ and $\sum_k\E X_k^2\le v$ for constants $b,v\ge0$. For $x>0$,
\[
 \Prb\Big\{|\sum_kX_k|>\sqrt{2vx}+2bx/3\Big\}\le2e^{-x}.
\]
\end{lemma}
\begin{proof}
If $v=0$, every $X_k$ is zero almost surely, so the claim is immediate.
Suppose $v>0$. For $m\ge2$, boundedness gives
$|\E X_k^m|\le b^{m-2}\E X_k^2$, and $m!\ge2\cdot3^{m-2}$.
Since $\E X_k=0$, the exponential series and a geometric sum give,
for $|\lambda|b<3$,
\[
 \E e^{\lambda X_k}
 \le1+\frac{\lambda^2\E X_k^2}{2}
          \sum_{j=0}^\infty\left(\frac{|\lambda|b}{3}\right)^j
 =1+\frac{\lambda^2\E X_k^2}{2(1-|\lambda|b/3)}.
\]
Writing $S=\sum_kX_k$, the inequality $\log(1+u)\le u$ and
independence therefore yield
\[
 \log\E e^{\lambda S}
 =\sum_k\log\E e^{\lambda X_k}
 \le\frac{\lambda^2v}{2(1-|\lambda|b/3)}.
\]
For $t>0$, choose $\lambda=t/(v+bt/3)>0$, which satisfies
$\lambda b<3$. Exponential Markov's inequality gives
\[
 \Prb\{S\ge t\}
 \le\exp\left\{-\lambda t+\frac{\lambda^2v}{2(1-\lambda b/3)}\right\}
 =\exp\left\{-\frac{t^2}{2(v+bt/3)}\right\}.
\]
At $t=\sqrt{2vx}+2bx/3$,
\[
 t^2-2x(v+bt/3)=\frac{2bx}{3}\sqrt{2vx}\ge0,
\]
so the upper-tail probability is at most $e^{-x}$.
Applying the same argument to $-X_k$ and adding the two tail
probabilities proves the stated bound.
\end{proof}
\subsection{Nets and measured derivative calculations}\label{app:nets}
For $0<\epsilon<1$, a Euclidean $\epsilon$-net of the unit sphere exists with at most $(1+2/\epsilon)^p$ points. A maximal $\epsilon$-separated set is a net, and disjoint balls of radius $\epsilon/2$ centered at its points fit in the ball of radius $1+\epsilon/2$, proving the volume bound. For such a net $\mathcal N_\epsilon$,
\[
 \norm z\le(1-\epsilon)^{-1}\max_{v\in\mathcal N_\epsilon}|v^Tz|.
\]
Approximate a unit vector maximizing the absolute inner product by a net point. Similarly, for symmetric $D$ and $\epsilon<1/2$,
\[
 \norm D_\op\le(1-2\epsilon)^{-1}\max_{v\in\mathcal N_\epsilon}|v^TDv|,
\]
since $|v^TDv-w^TDw|\le2\epsilon\norm D_\op$ for unit vectors within $\epsilon$. These facts justify the factors 2 and net sizes in~\eqref{eq:bn}--\eqref{eq:bn}.
Differentiate $\sum_y p_{\theta_0+F^{-1/2}t}(y)=1$ once and twice to get $\E_0s_Y=0$ and $\E_0G_Y=-I_p$. The covariance identity follows from the definition of $F$ and $\theta=\theta_0+F^{-1/2}t$. Finite outcomes justify differentiation under the sum. 

\subsection{Measured growing-dimensional regimes}\label{app:dimension}
A subscript $n$ denotes the constant for the $n$th measured model.
Thus $A_n,B_n,K_n$ are the bounds in Assumption~\ref{ass:meas} and
$\beta_{3,n}$ is the third standardized score moment. We take $x_n>0$
and a fixed positive radius multiplier $C$. Write $a_{{\rm M},n}$ for
the neighborhood radius in Assumption~\ref{ass:meas}, and set
$r_{{\rm M},n}=C\sqrt{p_n+x_n}$. The localization conditions are
\[
 \frac{r_{{\rm M},n}}{\sqrt n}\le a_{{\rm M},n},\quad
 \delta_n=\delta_n(r_{{\rm M},n},x_n)<1
 \quad\text{and}\quad z_n(x_n)<(1-\delta_n)r_{{\rm M},n}.
\]
The fixed $C$ is chosen sufficiently large for the last inequality along
the specified model sequence.
\begin{corollary}[Measured growing-dimensional inference]\label{cor:dimension-measured}
Assume the preceding localization conditions and Assumption~\ref{ass:meas}
at each $n$, with $B_n,K_n=O(1)$ and
$A_n\sqrt{(p_n+x_n)/n}=O(1)$.
If $(p_n+x_n)/n\to0$, the Fisher and additive Wilks bounds are respectively
\begin{equation}\label{eq:dim-measured}
 O\left((p_n+x_n)/\sqrt n\right)\quad\text{\normalfont and}\quad O\left((p_n+x_n)^{3/2}/\sqrt n\right).
\end{equation}
They vanish under $(p_n+x_n)^2/n\to0$ and $(p_n+x_n)^3/n\to0$, respectively.
If also $x_n\to\infty$ and $p_n^{1/4}\beta_{3,n}/\sqrt n\to0$, vanishing
Wilks/Wald errors imply the chi-square and ellipsoid coverage conclusions.
For $A_n=O(\sqrt{p_n})$, $p_n^{7/2}/n\to0$ suffices for Berry--Esseen.
\end{corollary}
\begin{proof}
Write $q_n=p_n+x_n$. The thresholds in~\eqref{eq:bn}--\eqref{eq:bn}
are of order $q_n$, and the assumptions give
\[
 \begin{gathered}
 z_n=O\left(\sqrt{q_n}+A_nq_n/\sqrt n\right)=O\left(\sqrt{q_n}\right)
 \\ \text{and}\quad
 \delta_n=O\left(\sqrt{q_n/n}+q_n/n\right)=O\left(\sqrt{q_n/n}\right).
 \end{gathered}
\]
Since $q_n/n\to0$, we have $\delta_n\to0$. Substitution into
Corollaries~\ref{cor:fisher}--\ref{cor:wilks} yields
\[
 e_F=\frac{\delta_nz_n}{1-\delta_n}=O\left(q_n/\sqrt n\right)
 \quad\text{and}\quad
 w_W=\frac{\delta_nz_n^2}{1-\delta_n}=O\left(q_n^{3/2}/\sqrt n\right).
\]
Moreover, $w_F=e_F(2z_n+e_F)=O(q_n^{3/2}/\sqrt n)$, because
$q_n^2/n=o(q_n^{3/2}/\sqrt n)$. The additional observed-information
term $\delta_nz_n^2/(1-\delta_n)^2$ has the same order.
This proves the expansion rates and their stated sufficient conditions
for convergence to zero.

For the distributional bounds, $x_n\to\infty$ removes the exceptional
probability, while $p_n^{1/4}\beta_{3,n}/\sqrt n\to0$ removes the
Berry--Esseen term. The chi-square concentration functions satisfy
$\omega_1(a)\le\sqrt{2a/\pi}$ and $\omega_p(a)\le a/2$ for $p\ge2$:
the latter follows from the $\chi_2^2$ density bound and convolution
with an independent $\chi_{p-2}^2$ variable when $p>2$.
Thus vanishing remainders also remove the concentration terms.
Finally, if $A_n=O(\sqrt{p_n})$, then
\[
 \frac{p_n^{1/4}\beta_{3,n}}{\sqrt n}
 \le\frac{A_np_n^{5/4}}{\sqrt n}
 =O\!\left(\frac{p_n^{7/4}}{\sqrt n}\right)
 =O\!\left(\sqrt{\frac{p_n^{7/2}}n}\right),
\]
which proves the sufficient Berry--Esseen condition.
\end{proof}

\section{Exact moments of the two-coordinate instrument}\label{app:instrument-moments}
Let $B_1,B_2$ be finite-dimensional self-adjoint operators and $\sigma$ a density operator. For $\eta>0$, use the Gaussian kernels $g_{\eta,y_i}$ defined in Section~\ref{sec:collective-output} and set
\[
 K_y=g_{\eta,y_2}(B_2)g_{\eta,y_1}(B_1).
\]
Define the dephasing map $\mathcal D_\eta$ on operators $A$ by its entries in a $B_1$ eigenbasis with eigenvalues $\lambda_a$:
\[
 \bigl[\mathcal D_\eta(A)\bigr]_{ab}
 =e^{-(\lambda_a-\lambda_b)^2/(8\eta^2)}A_{ab}.
\]
Let $Y=(Y_1,Y_2)$ have density $\Tr\sigma K_y^*K_y$.
The braces $\{A,B\}=AB+BA$ denote the anticommutator introduced in
Section~\ref{sec:setup}.

\begin{lemma}[Sequential Gaussian moments]\label{lem:instrument-moments}
For the instrument and state just defined, the first moments are
\[
 \E Y_1=\Tr\sigma B_1\quad\text{\normalfont and}\quad
 \E Y_2=\Tr\sigma\mathcal D_\eta(B_2).
\]
The second moments and the mixed moment satisfy
\begin{gather*}
 \E Y_1^2=\eta^2+\Tr\sigma B_1^2,\quad
 \E Y_2^2=\eta^2+\Tr\sigma\mathcal D_\eta(B_2^2),\\
 \text{and}\quad \E Y_1Y_2=\tfrac12\Tr\sigma\bigl\{B_1,\mathcal D_\eta(B_2)\bigr\}.
\end{gather*}
If $L>\max_i\norm{B_i}_\op$, then
\begin{equation}\label{eq:instrument-tail}
 \Prb\bigl\{Y\notin[-L,L]^2\bigr\}
 \le\sum_{i=1}^2 2e^{-\left(L-\norm{B_i}_\op\right)^2/(2\eta^2)}.
\end{equation}
\end{lemma}
\begin{proof}
In a $B_1$ eigenbasis, put $\bar\lambda_{ab}=(\lambda_a+\lambda_b)/2$.
The product $g_{\eta,y}(\lambda_a)g_{\eta,y}(\lambda_b)$ equals
$e^{-(\lambda_a-\lambda_b)^2/(8\eta^2)}$ times the density of
$N(\bar\lambda_{ab},\eta^2)$. Its zeroth, first and second moments
therefore give
\begin{align*}
 \int g_{\eta,y}(B_1)A g_{\eta,y}(B_1)\,dy
 &=\mathcal D_\eta(A),\\
 \int y g_{\eta,y}(B_1)A g_{\eta,y}(B_1)\,dy
 &=\tfrac12\{B_1,\mathcal D_\eta(A)\},\\
 \text{and}\quad
 \int y^2 g_{\eta,y}(B_1)A g_{\eta,y}(B_1)\,dy
 &=\eta^2\mathcal D_\eta(A)
   +\tfrac14\{B_1,\{B_1,\mathcal D_\eta(A)\}\}.
\end{align*}
For any Hermitian $B$, spectral integration also gives
\[
 \int y g_{\eta,y}(B)^2\,dy=B
 \quad\text{and}\quad
 \int y^2g_{\eta,y}(B)^2\,dy=B^2+\eta^2\Id.
\]
Integrating the second measurement first and using these identities,
together with $\mathcal D_\eta(\Id)=\Id$, proves the first moments and
$\E Y_1^2$. For the remaining moments, with
$g_1=g_{\eta,y_1}(B_1)$,
\begin{align*}
 \E Y_2^2
 &=\Tr\sigma\int g_1(B_2^2+\eta^2\Id)g_1\,dy_1
   =\Tr\sigma\mathcal D_\eta(B_2^2)+\eta^2,\\
 \text{and}\quad \E Y_1Y_2
 &=\Tr\sigma\int y_1g_1B_2g_1\,dy_1
   =\tfrac12\Tr\sigma\{B_1,\mathcal D_\eta(B_2)\}.
\end{align*}
All polynomial moments are absolutely integrable: finite spectral
expansions reduce them to Gaussian integrals, justifying the order of
integration. The first marginal is a mixture of $N(b,\eta^2)$ laws
with $b$ in the spectrum of $B_1$. The second is the analogous mixture
for $B_2$ in the nonselective state $\mathcal D_\eta(\sigma)$, which is
positive and has trace one. Thus, for $L>\norm{B_i}_\op$,
\[
 \Prb\bigl(|Y_i|>L\bigr)
 \le2\exp\left\{-\left(L-\norm{B_i}_\op\right)^2/(2\eta^2)\right\}.
\]
A union bound over $i=1,2$ proves~\eqref{eq:instrument-tail}.
\end{proof}
For Table~\ref{tab:collective-output-main}, diagonalize $B_{n,i}=V_i\Lambda_iV_i^*$ and factor the enlarged state as $\sigma=C_\sigma C_\sigma^*$. Here $V_i$ is unitary, $\Lambda_i=\diag(\lambda_{i,b})$ contains the eigenvalues of $B_{n,i}$, and $C_\sigma$ is a matrix factor of the state. The row $[A]_{b,\cdot}$ is row $b$ of a matrix $A$. The density is evaluated without random sampling as
\[
 p(x,y)=\sum_b g_{\eta,y}(\lambda_{2,b})^2
   \left\|\bigl[V_2^*V_1g_{\eta,x}(\Lambda_1)V_1^*C_\sigma\bigr]_{b,\cdot}\right\|^2.
\]
For the settings in Table~\ref{tab:collective-output-main}, $s=0.5$,
$\eta=0.7$, $r=0.5$ and $W=I_2$. Thus $V_X=(1+s)I_2$ and
$v=\norm{V_X+\eta^2I_2}_\op=1.99$. Set $\delta=10^{-12}$ and
integrate on $[-L,L]^2$ with
\[
 L=\max\left\{\max_i\norm{B_{n,i}}_\op+\eta\sqrt{2\log(4/\delta)},
                 r+\sqrt{2v\log(4/\delta)}\right\}.
\]
Both output and comparison tail probabilities are at most $\delta$. The omitted part of the half-$L^1$ TV integral is at most half their sum. The omitted projected-risk integral is at most $4r^2$ times the output tail bound. 

\section{Proofs of the collective inference corollaries}
\label{app:collective-corollary-proofs}
This appendix collects the proofs of
Corollaries~\ref{cor:collective-shrinkage}--\ref{cor:finite-readout}
from Section~\ref{sec:output-inference}.

\subsection{Proof of Corollary~\ref*{cor:collective-shrinkage} (Finite-radius shrinkage)}
\label{app:collective-shrinkage-proof}
\begin{proof}[Proof of Corollary~\ref{cor:collective-shrinkage}]
Write the comparison output as $G=h+Z_0$, with $\E Z_0=0$ and covariance $V_X+\eta^2I_p$. Projection cannot increase loss to $h\in K_r$, so its Gaussian risk is bounded by
$$c^2A_\eta+(1-c)^2\norm h_W^2\le c^2A_\eta+(1-c)^2B_r.$$
The projected loss is in $[0,4B_r]$, so the TV bound~\eqref{eq:collective-TV} gives the risk-error term in~\eqref{eq:collective-shrinkage}. Completing the square gives
\[
 c^2A_\eta+(1-c)^2B_r
 =(A_\eta+B_r)\left(c-\frac{B_r}{A_\eta+B_r}\right)^2
   +\frac{A_\eta B_r}{A_\eta+B_r}.
\]
The minimizer lies in $[0,1)$. The zero estimator has worst-ball risk exactly $B_r$, attained in a top eigenvector direction of $W$ when $r>0$. Comparing the two known upper bounds proves the selection claim and its strict-improvement condition.
\end{proof}

\subsection{Proof of Corollary~\ref*{cor:collective-rate} (One fixed-dimensional schedule)}
\label{app:collective-rate-proof}
\begin{proof}[Proof of Corollary~\ref{cor:collective-rate}]
For fixed observable norms $w_1,\ldots,w_{2p}$,
\[
 \mathfrak M_k=O(\eta_n^{-k}),\quad
 a_{\eta_n}^{2p}=O(\eta_n^{-p})
 \quad\text{and}\quad R_{{\rm out},n}^p=O((\log n)^{p/2}).
\]
Thus the leading compact-volume term is $O(n^{-1/2}\eta_n^{-(p+3)}(\log n)^{p/2})$. Balancing it with the smoothing cost $\eta_n^2$ amounts to
\[
 \eta_n^{p+5}=n^{-1/2}(\log n)^{p/2},
\]
which gives the choice $\eta_n=n^{-1/[2(p+5)]}(\log n)^{p/[2(p+5)]}$ in Corollary~\ref{cor:collective-rate} and makes the leading term $O(\eta_n^2)$.
The fourth-moment term has ratio $O((\sqrt n\eta_n)^{-1})=o(1)$ to the leading compact-volume bound $O(n^{-1/2}\eta_n^{-(p+3)}(\log n)^{p/2})$; lower moment terms are smaller for $\eta_n\le1$. The Gaussian tail is at most $2pn^{-4}$. This proves the rate in~\eqref{eq:collective-rate} and its consequence for the projected risk in~\eqref{eq:collective-risk}.
\end{proof}

\subsection{Proof of Corollary~\ref*{cor:collective-inference} (Local confidence regions and testing)}
\label{app:collective-inference-proof}
\begin{proof}[Proof of Corollary~\ref{cor:collective-inference}]
Write $P_h=P_{n,h}^{\eta}$ and $Q_h=N(h,V_\eta)$, where
$V_\eta=V_X+\eta^2I_p\succ0$ by~\eqref{eq:real-gram-lower}.
Put $\varepsilon=\varepsilon_n(r,\eta,R_{\rm out})$, and let
$\mathcal B(\R^p)$ denote the Borel sets of $\R^p$.
Equation~\eqref{eq:collective-TV} and the definition of TV give
\[
 \sup_{h\in K_r}\sup_{A\in\mathcal B(\R^p)}
 \bigl|P_h(A)-Q_h(A)\bigr|\le\varepsilon.
\]
For $t\ge0$, define the closed ellipsoid
\[
 A_h(t)=\bigl\{y\in\R^p:(y-h)^TV_\eta^{-1}(y-h)\le t\bigr\}.
\]
Under $Q_h$, the vector $Z=V_\eta^{-1/2}(Y-h)$ has distribution
$N(0,I_p)$, where $V_\eta^{-1/2}$ is the symmetric positive definite
inverse square root. Thus
\[
 (Y-h)^TV_\eta^{-1}(Y-h)
 =\norm Z^2=\sum_{j=1}^p Z_j^2\sim\chi_p^2
 \quad\text{and}\quad Q_h\bigl(A_h(q_{p,1-\alpha})\bigr)=1-\alpha,
\]
using continuity of the chi-square distribution.
For each true $h\in K_r$, injectivity of
$u\mapsto\theta_0+u/\sqrt n$ gives the event identity
\[
 \left\{\theta_0+h/\sqrt n\in\mathcal C_\alpha(Y)\right\}
 =\bigl\{Y\in A_h(q_{p,1-\alpha})\bigr\}.
\]
Consequently,
\[
 \begin{aligned}
 &\left|
 \Prb_{\theta_0+h/\sqrt n}
 \left\{\theta_0+h/\sqrt n\in\mathcal C_\alpha(Y)\right\}
 -(1-\alpha)\right|\\
 &\quad=
 \left|P_h\bigl(A_h(q_{p,1-\alpha})\bigr)
       -Q_h\bigl(A_h(q_{p,1-\alpha})\bigr)\right|
 \le\varepsilon.
 \end{aligned}
\]
Taking the supremum over $h\in K_r$
proves~\eqref{eq:collective-coverage}.
If $\varepsilon<\alpha$, then $0<\alpha-\varepsilon<1$, and the
region $\mathcal C_{\alpha-\varepsilon}(Y)$ uses the quantile
$q_{p,1-\alpha+\varepsilon}$. The same Gaussian calculation and TV
bound give
\[
 P_h\bigl(A_h(q_{p,1-\alpha+\varepsilon})\bigr)
 \ge(1-\alpha+\varepsilon)-\varepsilon=1-\alpha,
\]
uniformly over $h\in K_r$.

For testing, fix $h_0\in K_r$ and define the Borel rejection set
\[
 R_{h_0}=\bigl\{y\in\R^p:
 (y-h_0)^TV_\eta^{-1}(y-h_0)>q_{p,1-\alpha}\bigr\}.
\]
The central calculation above gives $Q_{h_0}(R_{h_0})=\alpha$,
so $P_{h_0}(R_{h_0})\le Q_{h_0}(R_{h_0})+\varepsilon
=\alpha+\varepsilon$, proving the size bound.
At any $h\in K_r$, under $Q_h$ we have
\[
 V_\eta^{-1/2}(Y-h_0)=Z+\delta_h,\quad
 Z\sim N(0,I_p)
 \quad\text{and}\quad \delta_h=V_\eta^{-1/2}(h-h_0).
\]
Hence the Gaussian testing statistic has distribution
 \begin{align*}
 (Y-h_0)^TV_\eta^{-1}(Y-h_0)
 =\sum_{j=1}^p(Z_j+\delta_{h,j})^2\sim\chi_p^2(\lambda),\\
 \lambda=\norm{\delta_h}^2
 =(h-h_0)^TV_\eta^{-1}(h-h_0).
 \end{align*}
Applying the TV bound to $R_{h_0}$ yields
\[
 \left|P_h(R_{h_0})
 -\Prb\bigl\{\chi_p^2(\lambda)>q_{p,1-\alpha}\bigr\}\right|
 =\bigl|P_h(R_{h_0})-Q_h(R_{h_0})\bigr|
 \le\varepsilon,
\]
which proves the asserted power bound for every $h\in K_r$.
\end{proof}

\subsection{Proof of Corollary~\ref*{cor:finite-readout} (Finite readout)}\label{app:finite-readout-proof}
\begin{proof}[Proof of Corollary~\ref{cor:finite-readout}]
Put $R=R_{\rm out}>r$, $A=[-R,R]^p$, $w=\norm W_\op$ and
$\varepsilon=\varepsilon_n(r,\eta,R_{\rm out})$.
Write $P_h=P_{n,h}^{\eta}$ and $Q_h=N(h,V_\eta)$, with
$V_\eta=V_X+\eta^2I_p$ and $v_\eta=\norm{V_\eta}_\op$.
Let $C_1,\ldots,C_m$ be a finite Borel partition of $A$ into
rectangular grid cells of side lengths at most $\Delta$, assigning
shared boundaries to one cell. Let $c_j$ be the midpoint of the
rectangle defining $C_j$, and put $C_0=A^c$.

Integrating the signal POVM in~\eqref{eq:signal-povm} gives
\[
 F_j=M_{n,\eta}^{\rm signal}(C_j)\succeq0\quad(0\le j\le m)
 \quad\text{and}\quad \sum_{j=0}^mF_j=M_{n,\eta}^{\rm signal}(\R^p)=\Id.
\]
Thus $(F_j)_{j=0}^m$ is a finite POVM, with label zero denoting
failure. Its outcome law $P_h^\Delta$ and the same discretization
$Q_h^\Delta$ of the Gaussian law satisfy
\[
 P_h^\Delta\bigl(\{j\}\bigr)=\Tr\rho_{n,h}F_j=P_h(C_j)
 \quad\text{and}\quad Q_h^\Delta\bigl(\{j\}\bigr)=Q_h(C_j).
\]
Writing $p_h,q_h$ for the densities of $P_h,Q_h$, disjointness
of the cells and~\eqref{eq:collective-TV} imply
\[
 \TV(P_h^\Delta,Q_h^\Delta)
 =\tfrac12\sum_{j=0}^m\left|\int_{C_j}(p_h-q_h)\,dy\right|
 \le\TV(P_h,Q_h)\le\varepsilon.
\]
Applying the same reporting map to both finite laws preserves
this bound, since any event for the reported value corresponds
to its inverse image in the finite label space.

For the risk comparison, let $a(y)=\Pi_r^W y$ and define
$a_\Delta(y)=\Pi_r^W c_j$ for $y\in C_j$, $j\ge1$, and
$a_\Delta(y)=0$ for $y\in A^c$. Both estimators take values in
$K_r$. For $y\in C_j$ with $1\le j\le m$, the midpoint error and projection
nonexpansiveness in~\eqref{eq:projection-facts} give
\[
 \norm{c_j-y}\le\frac{\sqrt p\,\Delta}{2}
 \quad\text{and}\quad \norm{a_\Delta(y)-a(y)}_W\le\norm{c_j-y}_W
 \le\frac{\sqrt{wp}\,\Delta}{2}.
\]
Fix $h\in K_r$ and write $\ell_h(z)=\norm{z-h}_W^2$.
For $y\in A$, abbreviate $a=a(y)$ and $b=a_\Delta(y)$.
Since $a,b,h\in K_r$, both $\norm{a-h}_W$ and
$\norm{b-h}_W$ are at most $2r\sqrt w$. The difference of
the two losses therefore obeys
\[
 \bigl|\ell_h(b)-\ell_h(a)\bigr|
 =\bigl|\langle b-a,b+a-2h\rangle_W\bigr|
 \le\frac{\sqrt{wp}\,\Delta}{2}\,4r\sqrt w
 =2rw\sqrt p\,\Delta.
\]
On $A^c$, each loss lies in $[0,4wr^2]$, so their absolute
difference is at most $4wr^2$. Integrating these two bounds
under the same law $P_h$ yields
\[
 \left|\E_{P_h}\ell_h\bigl(a_\Delta(Y)\bigr)-\E_{P_h}\ell_h\bigl(a(Y)\bigr)\right|
 \le2rw\sqrt p\,\Delta+4wr^2P_h(A^c).
\]
The first expectation is the risk of the finite reporting rule,
because its cell probabilities are $P_h(C_j)$.

Under $Q_h$, coordinate $Y_i$ has mean $h_i$ and variance
$(V_\eta)_{ii}\le v_\eta$. Since $|h_i|\le r<R$, the Gaussian
tail bound and a union bound give
\[
 Q_h(A^c)\le\sum_{i=1}^p Q_h\bigl\{|Y_i-h_i|>R-r\bigr\}
 \le2p e^{-(R-r)^2/(2v_\eta)}.
\]
The TV comparison then gives, uniformly over $h\in K_r$,
\[
 P_h(A^c)\le Q_h(A^c)+\varepsilon
 \le2p e^{-(R-r)^2/(2v_\eta)}+\varepsilon.
\]
Substituting this estimate into the risk comparison proves the
stated bound for every $h\in K_r$. Taking suprema of the two
risks proves the claimed increase in worst-case risk.
\end{proof}

\section{Pauli likelihood, observed information and coverage}\label{app:pauli-diagnostics}
The following calculations give both an event-perturbation bound and a direct
reference-point calibration for the Pauli measurement. They include the
empirical-curvature localization argument, exact-binomial confidence regions,
reference and off-center diagnostics, and all associated numerical tables.
\subsection{A perturbation alternative for Wald calibration}\label{app:wald-old}
For the four-outcome Pauli measurement of Section~\ref{sec:transverse-qubit}, let $N_{a,+},N_{a,-}$ be the counts for axis $a\in\{x,y\}$, and put $N_a=N_{a,+}+N_{a,-}$, $S_a=N_{a,+}-N_{a,-}$ and $K=N_x$. Define $T_a=S_a/\sqrt{N_a}$ when $N_a>0$ and $T_a=0$ otherwise. For the global MLE $\widehat\theta$ on the known physical disk $\norm\theta\le b<\sqrt{1-s^2}$, let $W_{\rm obs}=n\widehat\theta^T\widehat F\widehat\theta$, with $\widehat F$ and its fallback as in the setup preceding Proposition~\ref{prop:observed}. The scalar $x_{\rm tail}>0$ below is a tail exponent.
For comparison with~\eqref{eq:wald-new}, a direct event argument gives, at zero and for $n>8x_{\rm tail}$, $16x_{\rm tail}/n<b^2$,
\[
 d_{\rm K}(W_{\rm obs},\chi_2^2)
 \le\min\left\{1,8/\sqrt n+2e^{-n/8}+4e^{-x_{\rm tail}}+16x_{\rm tail}^2/(n-8x_{\rm tail})\right\}.
\]
To see this, condition on $K\in[n/4,3n/4]$. The scalar Berry--Esseen inequality with constant one, together with contraction of Kolmogorov distance under convolution, bounds the distance between $V=T_x^2+T_y^2$ and $\chi_2^2$ by $8/\sqrt n$. Hoeffding's inequality, applied conditionally on $K$, gives $T_x^2,T_y^2\le2x_{\rm tail}$ except with probability $4e^{-x_{\rm tail}}$. On that event the ratio vector $(S_x/N_x,S_y/N_y)$ is inside the disk and $0\le W_{\rm obs}-V\le32x_{\rm tail}^2/(n-8x_{\rm tail})$. The $\chi_2^2$ density is bounded by $1/2$. Apply upper and lower bounds for the conditional cumulative distribution function (CDF), then add the probability that $K\notin[n/4,3n/4]$, $2e^{-n/8}$ once. Table~\ref{tab:direct} reports the transformed-CDF bound~\eqref{eq:wald-new}. The transformed-CDF argument compares the coordinates before
convolution and avoids this event-perturbation cost.

\subsection{Exact observed Pauli curvature}\label{app:pauli-curvature}
Use the observed counts $N_{a,+},N_{a,-}$, axis totals $N_a$ and signed counts $S_a$ from Section~\ref{app:wald-old}. At the reference point zero, the local information coordinate specializes to $z=\sqrt{n/2}\,\theta$. Write $u_a=\sqrt2z_a/\sqrt n=\theta_a$ for its physical component. Direct differentiation gives the diagonal negative Hessian
\begin{equation}\label{eq:pauli-curvature}
 \bigl[-\nabla^2\ell_n(z)\bigr]_{aa}
 =\frac{2N_a}{n}\frac{1+u_a^2}{(1-u_a^2)^2}
 -\frac{4S_a}{n}\frac{u_a}{(1-u_a^2)^2}.
\end{equation}
Put $a_s=1-s^2$, and set $t_H=x+\log12$, $b=\sqrt2r_{\rm M}/\sqrt n<\sqrt{a_s}$, and
\[
 e_N=\sqrt{2t_H/n}+2t_H/(3n),\quad
 e_S=\sqrt{t_H/n}+2t_H/(3n),\quad\text{\normalfont and}\quad f(b)=\frac{1+b^2}{(1-b^2)^2}.
\]
Three scalar Bernstein applications give, except with probability $6e^{-t_H}=e^{-x}/2$,
$|2N_a/n-1|\le e_N$ for both axes and $|S_a/n|\le e_S$ for each axis. Indeed the two count deviations are negatives of one another; their summands have variance one, whereas each signed count summand has variance $1/2$. It follows that
\begin{equation}\label{eq:pauli-sharp}
 \sup_{\norm z\le r_{\rm M}}\norm{\nabla^2\ell_n(z)+I_2}_\op
 \le\delta_n^{\rm sharp}:=e_N f(b)+f(b)-1+\frac{4e_Sb}{(1-b^2)^2}.
\end{equation}
Combine this event with the score event in Theorem~\ref{thm:measured}. All measured conclusions, including observed-information coverage, hold with $\delta_n^{\rm sharp}$ whenever $z_n<(1-\delta_n^{\rm sharp})r_{\rm M}$.

The log likelihood is globally concave on any known closed disk strictly inside the physical parameter domain. If that disk contains the local ball, local positive curvature and the inward boundary derivative produce a stationary interior point. Concavity makes it a global maximizer; strict concavity along a segment near that point excludes any second maximizer. Thus this argument applies to an implementable global MLE, without assuming full-rank curvature of each individual outcome. At an interior solution, $\widehat\theta_a=S_a/N_a$ when $N_x,N_y>0$; otherwise maximize on the known disk. Its observed information is
\[
 \widehat F_{aa}=\frac{N_{a,+}}{n(1+\widehat\theta_a)^2}
 +\frac{N_{a,-}}{n(1-\widehat\theta_a)^2}\quad\text{\normalfont and}\quad \widehat F_{12}=0.
\]
For the generic curvature and score bounds in Table~\ref{tab:measured},
the measured-likelihood constants are $A=\sqrt2$, $B=1$ and
$K=4\sqrt2/(1-b)^3$, with $b=\sqrt2r_{\rm M}/\sqrt n$ as above.
For example, at $n=5000$, the sharp curvature bound $.20346$ in
Table~\ref{tab:measured} certifies localization because its threshold
is about $7.9654$, above the score bound $z_n\approx7.445$; the generic
curvature bound $1.6762$ exceeds one.

\begin{table}[t]
\caption{Pauli curvature and localization, $s=0.5$, $r_{\rm M}=10$, $x=2$. The generic column uses the scalar-net baseline~\eqref{eq:delta}; certification tests $\delta^{\rm sharp}<1$ and $z_n<(1-\delta^{\rm sharp})r_{\rm M}$.
Smaller curvature bounds are better; bold marks the smaller of the generic and sharp bounds in each row.}\label{tab:measured}
\centering\begin{tabular}{rrrrl}
$n$&Generic $\delta_n$&$\delta_n^{\rm sharp}$&$z_n$&Certified\\\hline
500&51.3255&\textbf{4.14059}&7.826&no\\
1000&10.8500&\textbf{1.25373}&7.663&no\\
5000&1.6762&\textbf{0.20346}&7.445&yes\\
10000&0.9737&\textbf{0.10683}&7.394&yes\\
100000&0.2303&\textbf{0.01679}&7.309&yes\\
1000000&0.0669&\textbf{0.00372}&7.282&yes\\
\end{tabular}
\end{table}

\begin{table}[t]
\caption{Observed-information ellipsoid coverage at $\theta=0$, $s=0.5$, nominal level $0.95$. There are 20,000 independent replications per row.
Coverage should be close to $0.95$; at comparable coverage, smaller area is better, and bold marks the smallest mean area across the displayed sample sizes.}\label{tab:coverage}
\centering\begin{tabular}{rrrrr}
$n$&Coverage&Standard error&Mean area&Fallbacks\\\hline
5000&0.95025&0.00154&0.0075269&0\\
10000&0.94990&0.00154&0.0037640&0\\
100000&0.95015&0.00154&\textbf{0.0003764}&0\\
\end{tabular}
\end{table}
Table~\ref{tab:coverage} uses the global MLE on the known disk $\norm\theta\le0.8$ at $s=0.5$.
For example, at $n=5000$, simulated coverage is $.95025$, only $.00025$
from the nominal $.95$ and well within the Monte Carlo standard error $.00154$.
\limitationtext{The generic bound~\eqref{eq:observed} truncates to one in all three rows, whereas the following binomial argument gives useful guarantees for the same statistic.}
Let $F_1$ and $f_1$ be the $\chi_1^2$ distribution function and density, respectively, and
$M_\chi=3^{3/2}e^{-3/2}/\sqrt{2\pi}$. Write $C_{\rm BE}=0.4748$
for the i.i.d. Berry--Esseen constant used below. For an integer $m\ge1$,
let $T_m$ be a normalized sum of $m$ independent Rademacher variables and
define $W_m=T_m^2/(1-T_m^2/m)$, with $W_m=+\infty$ at $T_m^2=m$.

\begin{lemma}[Transformed chi-square reference]\label{lem:chi-transform}
For every $m\ge1$,
\[
 \sup_{t\ge0}\bigl|F_1(t)-F_1\bigl(t/(1+t/m)\bigr)\bigr|\le M_\chi/m.
\]
For the transformed Rademacher sum just defined,
\[
 d_{\rm K}(W_m,\chi_1^2)\le2C_{\rm BE}/\sqrt m+M_\chi/m.
\]
\end{lemma}
\begin{proof}
For $g_a(t)=t/(1+at)$ and $t>0$, differentiation gives
\[
 -\partial_aF_1\bigl(g_a(t)\bigr)=g_a(t)^2f_1\bigl(g_a(t)\bigr).
\]
Since $y^2f_1(y)=y^{3/2}e^{-y/2}/\sqrt{2\pi}$ has maximum $M_\chi$
at $y=3$,
\[
 0\le F_1(t)-F_1\bigl(g_{1/m}(t)\bigr)
 =\int_0^{1/m}g_a(t)^2f_1\bigl(g_a(t)\bigr)\,da
 \le M_\chi/m.
\]
The same bound holds at $t=0$ by continuity.

Let $H_m$ be the CDF of $T_m$. The scalar Berry--Esseen inequality
with the valid i.i.d. constant $C_{\rm BE}=0.4748$ from
\citet{Shevtsova2011} gives
$\sup_u|H_m(u)-\Phi(u)|\le C_{\rm BE}/\sqrt m$.
For $u\ge0$, the left limit at $-\sqrt u$ ensures that any atom at
that endpoint is included:
\[
 \Prb(T_m^2\le u)=H_m\left(\sqrt u\right)-H_m\left(-\sqrt u-\right)
 \quad\text{and}\quad
 F_1(u)=\Phi\left(\sqrt u\right)-\Phi\left(-\sqrt u\right).
\]
Continuity of $\Phi$ extends the same approximation bound to the left
limit, so $|\Prb(T_m^2\le u)-F_1(u)|\le2C_{\rm BE}/\sqrt m$.
For finite $t\ge0$, monotonicity of the transformation gives
$\{W_m\le t\}=\{T_m^2\le g_{1/m}(t)\}$, excluding the outcomes
assigned $+\infty$. Hence
\[
 \begin{aligned}
 \bigl|\Prb(W_m\le t)-F_1(t)\bigr|
 &\le\Bigl|\Prb\bigl(T_m^2\le g_{1/m}(t)\bigr)-F_1\bigl(g_{1/m}(t)\bigr)\Bigr|+\Bigl|F_1\bigl(g_{1/m}(t)\bigr)-F_1(t)\Bigr|\\
 &\le\frac{2C_{\rm BE}}{\sqrt m}+\frac{M_\chi}{m}.
 \end{aligned}
\]
Both CDFs vanish for negative $t$, proving the asserted distance bound.
\end{proof}

For the direct calibration, use the global MLE on a known disk of radius
$b<\sqrt{1-s^2}$ and the all-data observed-information rules in the setup
preceding Proposition~\ref{prop:observed}. Retain
$W_{\rm obs}=n\widehat\theta^T\widehat F\widehat\theta$ and define the
error budget
\begin{align}\label{eq:wald-new}
 B_{\rm P}(n,\varepsilon,b)
 &=2C_{\rm BE}\sqrt{2/n}\bigl\{(1-\varepsilon)^{-1/2}+(1+\varepsilon)^{-1/2}\bigr\}\\
 &\quad+\frac{4M_\chi}{n(1-\varepsilon^2)}+2e^{-n\varepsilon^2/2}
 +4e^{-n(1-\varepsilon)b^2/8}.\nonumber
\end{align}

\begin{theorem}[Reference-point Pauli Wald calibration]\label{thm:direct-wald}
For the preceding procedure, at $\theta=0$, every integer $n$ and
$0<\varepsilon<1$ with $n(1-\varepsilon)/2\ge1$ satisfy
\begin{equation}\label{eq:direct-wald}
 \sup_t\bigl|\Prb_0(W_{\rm obs}\le t)-\Prb(\chi_2^2\le t)\bigr|
 \le\min\bigl\{1,B_{\rm P}(n,\varepsilon,b)\bigr\}.
\end{equation}
\end{theorem}
\begin{proof}
At zero, $K=N_x$ is binomial with parameters $n,1/2$. Put
$k_-=n(1-\varepsilon)/2$, $k_+=n(1+\varepsilon)/2$ and
$\mathcal K=\{k_-\le K\le k_+\}$. Hoeffding's inequality gives
$\Prb_0(\mathcal K^c)\le2e^{-n\varepsilon^2/2}$.
Conditional on $K=k$, the two signed counts are independent Rademacher
sums of lengths $k$ and $n-k$. Let $W_k^{(x)}$ and $W_{n-k}^{(y)}$
be their transformations from Lemma~\ref{lem:chi-transform}, and put
$V_k=W_k^{(x)}+W_{n-k}^{(y)}$. Convolution with independent variables
contracts Kolmogorov distance, so
\[
 d_{\rm K}(V_k,\chi_2^2)
 \le2C_{\rm BE}\bigl\{k^{-1/2}+(n-k)^{-1/2}\bigr\}
    +M_\chi\bigl\{k^{-1}+(n-k)^{-1}\bigr\}.
\]
This statement compares CDFs at finite thresholds and also holds when
a transformed variable has mass at $+\infty$.
The two sums are convex and symmetric about $n/2$, so on $[k_-,k_+]$
their maxima occur at the endpoints:
\begin{gather*}
 k^{-1/2}+(n-k)^{-1/2}
 \le\sqrt{2/n}\bigl\{(1-\varepsilon)^{-1/2}+(1+\varepsilon)^{-1/2}\bigr\},\\
 \text{and}\quad
 k^{-1}+(n-k)^{-1}\le\frac4{n(1-\varepsilon^2)}.
\end{gather*}
These give the first two terms of~\eqref{eq:wald-new}.

Write $R=(S_x/k,S_y/(n-k))$. If $\norm R\ge b$, at least one
coordinate has absolute value at least $b/\sqrt2$. Hoeffding's inequality,
applied conditionally on $K=k$, and $k,n-k\ge k_-$ yield
\[
 \Prb_0\left(\norm R\ge b\mid K=k\right)
 \le2e^{-kb^2/4}+2e^{-(n-k)b^2/4}
 \le4e^{-n(1-\varepsilon)b^2/8}.
\]
On $\{\norm R<b\}$, the ratio estimator is the interior global MLE.
Writing $R_a=S_a/N_a$, its observed information is diagonal with
entries $N_a/[n(1-R_a^2)]$, so
\[
 W_{\rm obs}=\sum_{a=x,y}\frac{N_aR_a^2}{1-R_a^2}
 =\sum_{a=x,y}\frac{T_a^2}{1-T_a^2/N_a}=V_k.
\]
Here $N_a\ge1$ and $|R_a|<b<1$, so the statistic uses no fallback.
The conditional CDFs of $W_{\rm obs}$ and $V_k$ therefore differ by
at most the disk-exit probability. Integrating the conditional bounds
over $\mathcal K$ and adding $\Prb_0(\mathcal K^c)$ once proves
\eqref{eq:wald-new}. Truncation at one gives~\eqref{eq:direct-wald}.
\end{proof}
For a nominal $1-\alpha$ ellipse, coverage error is at most the bound in~\eqref{eq:direct-wald}. Let $B$ denote the coverage-error bound in~\eqref{eq:direct-wald}. If $B<\alpha$, the quantile at level $1-\alpha+B$ gives coverage at least $1-\alpha$ at zero. In Table~\ref{tab:direct}, the prescribed choice $\varepsilon_n=\sqrt{4\log n/n}$ already gives $B<0.05$ at $n=5000$.
For example, at $n=100000$, the analytic bound limits the coverage error
at zero to $.008514$, less than one percentage point, while simulated
coverage is $.95015$.

\begin{table}[t]
\caption{Reference-point coverage-error bounds for the observed-information ellipse, $b=0.8$. The analytic transformed bound uses~\eqref{eq:wald-new}; the final column is Monte Carlo coverage. Bounds are rounded upward. Section~\ref{app:wald-old} retains an alternative perturbation calculation.
Smaller coverage-error bounds are better; bold marks the smallest bound across the displayed sample sizes, and simulated coverage should be close to $0.95$.}\label{tab:direct}
\centering\begin{tabular}{rrr}
$n$&Transformed bound&Simulated coverage\\\hline
5000 & 0.038455 & 0.95025\\
10000 & 0.027082 & 0.94990\\
100000 & 0.008514 & 0.95015\\
1000000 & \textbf{0.002688} & ---\\
\end{tabular}
\end{table}

For coverage throughout the disk, use coordinatewise Clopper--Pearson
intervals~\citep{ClopperPearson1934} with Bonferroni allocation.
The two-coordinate region below specializes standard binomial inversion
and union bounds to this Pauli measurement. For a broader quantum
construction of confidence regions from finite data, see \citet{WangScholzRenner2019}. Conditional on axis counts, invert exact equal-tailed binomial tests with error $\alpha/2$ per coordinate for $p_a=(1+\theta_a)/2$. Let $B^{-1}_{a,b}(u)$ denote the $u$-quantile of a beta distribution with positive shape parameters $a,b$. With $k$ successes in $m>0$ trials the endpoints are
\[
 p^- = B^{-1}_{k,m-k+1}(\alpha/4)\quad\text{\normalfont and}\quad
 p^+ = B^{-1}_{k+1,m-k}(1-\alpha/4),
\]
with $p^-=0$ at $k=0$, $p^+=1$ at $k=m$. If $m=0$, use $[0,1]$. Map by $\theta_a=2p_a-1$ and intersect the rectangle with the known disk. Conditional exact binomial coverage and the classical two-coordinate union bound prove coverage at least $1-\alpha$ uniformly throughout that disk. The region may be empty on exceptional data; this does not change the coverage guarantee.

\begin{table}[t]
\caption{Uniformly valid exact-binomial region versus the observed-information ellipse, at zero. Mean areas here are before disk intersection for both methods; no boundary truncation occurred in these runs. The exact region uses at most four beta quantile evaluations per data set.
Smaller mean area is preferable at comparable coverage; bold marks the smaller area in each row, and the methods' coverage guarantees differ.}\label{tab:exact-region}
\centering\begin{tabular}{rrrrr}
$n$&Exact-region coverage&Standard error&Rectangle area&Ellipse area\\\hline
5000 & 0.95210 & 0.00151 & 0.0081699 & \textbf{0.0075269}\\
10000 & 0.95270 & 0.00150 & 0.0040669 & \textbf{0.0037640}\\
100000 & 0.95175 & 0.00152 & 0.0004035 & \textbf{0.0003764}\\
\end{tabular}
\end{table}
The exact region requires at most four beta quantile evaluations, whereas the ellipse uses count ratios and a diagonal determinant. This computational distinction and the larger area in Table~\ref{tab:exact-region} are the cost of the simple uniform finite-sample guarantee.
For example, at $n=5000$, the exact region has simulated coverage $.95210$
and mean area $.0081699$, about $8.5\%$ larger than the ellipse's $.0075269$,
illustrating the area cost of its uniform coverage guarantee.
Table~\ref{tab:offcenter} also examines points away from zero: at the
near-boundary point $\theta=(.79,0)$, simulated Wald coverage is $.96890$
compared with $.95225$ for the exact region, and $4368$ of $20000$ MLEs
are constrained, illustrating the sensitivity of Wald calibration to the boundary.

\begin{table}[t]
\caption{Additional Monte Carlo coverage at $n=5000$, $s=.5$, disk radius $.8$, with 20,000 replications per point and seed 20261119 for the permuted congruential generator PCG64. \limitationtext{The Clopper--Pearson guarantee is uniform; the Wald theorem is calibrated only at zero. Standard errors (s.e.) are Monte Carlo quantities.} The last column counts constrained MLEs.
Coverage closer to $0.95$ is preferable; bold marks the numerically closer simulated value in each row.}\label{tab:offcenter}
\centering\small
\begin{tabular}{lrrrrr}
$\theta$&Wald coverage&s.e.&Exact coverage&s.e.&Constrained\\\hline
$(0.3,0.2)$ & \textbf{0.94890} & 0.00156 & 0.95275 & 0.00150 & 0\\
$(0.6,0)$ & \textbf{0.94890} & 0.00156 & 0.95350 & 0.00149 & 0\\
$(0.79,0)$ & 0.96890 & 0.00123 & \textbf{0.95225} & 0.00151 & 4368\\
\end{tabular}
\end{table}

\end{appendix}

\FloatBarrier

\end{document}